\documentclass[a4paper, 11pt]{article}
\usepackage{graphicx}
\usepackage{lineno,hyperref}
\usepackage{bm}
\usepackage{color}
\usepackage{tcolorbox}
\usepackage{amsmath}
\usepackage{epstopdf}
\usepackage{labelfig}
\usepackage{adjustbox}
\usepackage{newfloat}
\usepackage{epsfig}
\title{Chemical reaction  planar fronts with a  viscoelastic reaction product
}
\author{Svetlana Petrenko$^1$       \and
        Alexander Freidin$^{2}$ \and Eric Charkaluk$^{1}$
}
\date{ $^1$Laboratoire de M\'ecanique des Solides (CNRS UMR 7649), Ecole Polytechnique, Institut Polytechnique de Paris, \\ Route de Saclay 91120 Palaiseau, France \\ 
$^2$ Institute for Problems in Mechanical Engineering of the Russian Academy of Sciences, Bol'shoy pr. 61, V.O.,\\ 199178 St. Petersburg, Russia}
\begin{document}

\maketitle
\begin{abstract}
A stress-affected chemical reaction front propagation  is considered utilizing the concept of a chemical affinity tensor. A reaction between an elastic solid and diffusing constituents, localized at the reaction front, is considered. As a result of the reaction, the elastic constituent transforms into viscoelastic one. The reaction is accompanied by volume expansion that in turn may result in  stresses at the reaction front which affect the front velocity through the normal component of the chemical affinity tensor.
Considering a plane strain problem with a planar chemical reaction front propagation  under uniaxial deformation, we focus on the studies of the reaction front kinetics in dependence on external strains and material parameters with the use of the notion of the equilibrium concentration. Then stress relaxation behind the propagating reaction front is modelled.
A standard linear solid model  is used for the reaction product, and its particular cases are also considered. Analytical solutions are obtained which allow to study in explicit form the influence of strains and material parameters on the front retardation or acceleration and stress relaxation.
%\keywords{Mechanochemistry  \and  Chemical affinity tensor  \and Reaction front kinetics  \and Stress relaxation \and Standard linear solid model}
% \PACS{PACS code1 \and PACS code2 \and more}
% \subclass{MSC code1 \and MSC code2 \and more}
\end{abstract}

\section{Introduction}
%\label{intro}

The  influence of stress-strain state on chemical reactions has been widely considered since the 70s of the last century. It is of primary importance in such fields as energy storage industry, nuclear power, medicine, aircraft industry, civil engineering and the list is far from being exhausted. The oxidation of silicon in nanowires, e.g.  \cite{ButtnerZacharias2006}, 
or in MEMS,  e.g.  \cite{MuhlsteinRitchie2003},  reactions in ceramic composites with inclusions, e.g.  \cite{Nanko-ceramics2005},   lithiation  of silicon in Li-ion batteries, e.g.  \cite{McDowellNix2013}, can be mentioned among the examples that demonstrate the importance of establishing interconnections between stress-affected chemical reactions and the stress-strain state 
(see also  \cite{Huang1987,Mihalyi,Yen} and the references in \cite{Freidin2020}).

The above mentioned and many  other reactions can be described using a two-phase reaction model in which the reaction is localized at the sharp interface -- a reaction front, and  the diffusing reactant is transported to the reaction front through the transformed solid material.
One of the first and most simple models that described such reactions was the model  proposed by Deal and Grove for a planar oxidation front \cite{DealGrove}. This model gave a general scenario of the problem statement but did not consider stress effects. However, previously mentioned
chemical reactions  are accompanied by volumetric expansion that produces  internal stresses, which in turn may affect the  chemical reaction (see, e.g.  \cite{Kobeda89,EerNisse1979,Beaulieu,Nanko-ceramics2005}).  An extension of the Deal-Grove model was proposed by considering a stress-dependent diffusion coefficient and a reaction rate parameter \cite{Kao85,Kao87,Kao88,Rafferty,Sutardja,Krzeminski}. Other alternatives to take into account the influence of stresses on the diffusion and the reaction is to introduce additional terms in the expression of the diffusion flux, e.g. \cite{Knyazeva,Knyazeva2020}, or to consider the influence through a scalar chemical potential \cite{Brassart2012,Brassart2013,Levitas2013,Loeffel2011,Loeffel2013}.

In fact, stresses may affect  the reaction  front propagation via the influence on diffusion flux (diffusion-controlled reactions) or via the direct influence on the reaction rate (reaction rate-controlled reactions), see, e.g. \cite{Cui2013}.
In the present paper we focus on the second case, for which the reaction front propagation is controlled rather by the reaction rate than by the diffusion (see, e.g.,  \cite{ZhaoPhar2012,Jia}). Modeling of the reaction front kinetics is based on the chemical affinity tensor derived initially in  \cite{FrAPM2009}
for  the case of a chemical reaction between diffusing and nonlinear elastic constituents
as  a configurational force conjugated to the  front velocity in the expression of the energy dissipation due to the front propagation, similar to the derivation  made in \cite{Knowles79} for a propagating phase interface.

Then it was derived from fundamental balance laws and the entropy inequality written down for an open system with a chemical reaction between diffusing and solid constituents of arbitrary rheology  in the case of finite strains, and a kinetic equation in the form of the dependence of the reaction front velocity on the normal component of the affinity tensor was formulated (see \cite{Freidin2013,FrVilKor2014,FreidinMTT2015} and a review \cite{Freidin2020}). 
This consideration is consistent with the approach of classical physical chemistry where reaction rate is determined by a scalar chemical affinity (see, e.g. \cite{PrigogineDefay1954}), and the notion of the chemical affinity arises to pioneering works by Gibbs \cite{Gibbs} and de Donder \cite{Donder}.

In the case of solid constituents the tensorial nature of the chemical affinity  follows from the consideration of a chemical reaction on the oriented area element of the reaction front (see a more detailed discussion in \cite{Freidin2020}), as well as the tensorial nature of the chemical potential followed from the fact that a phase equilibrium took place not just in a point but at oriented area elements of the phase interface passing trough the point (see, e.g.  \cite{Grinfeld}).  
In a quasi-static case, the chemical affinity  tensor is represented by the linear combination of the chemical potential tensors, which are the Eshelby stress tensors divided by the reference mass densities. This combination is the same as the combination of scalar chemical potentials which defines the classical chemical affinity. 

The approach based on the chemical affinity tensor has been applied to the statement and solution of
a number of boundary value problems with propagating reaction fronts in formulations  which assumed  solid constituents to be linear elastic \cite{FrVilKor2014,VilchFr13,FrMorPetrVilActaMech2016,FrKor2016}. Then the theory has been used to describe numerically two-phase
lithiation of Si particles used in Li-ion batteries, where the constituent materials undergoing finite elasto-viscoplastic deformations were considered  \cite{Poluektov2018,Poluektov2019}. 

In the present paper we come back to the case of small strains and develop a model for analytical studies of stress relaxation behind the reaction front. Considered chemical reactions are  accompanied by transformation strains which may generate huge stresses, and  reaction products often demonstrate more viscous than elastic  behaviours, e.g.,
\cite{Kao85,Kao87}. This motivates  the relevance of stress relaxation studies.
In addition, since  the total thickness of the transformed layer is observed in experiments with planar reaction fronts, even simple models may  be useful for the estimation of the impacts of the transformation and inelastic (viscous) strains on the thickening during the front propagation. 

The paper is organized as follows. A short summary of the concept of the chemical affinity tensor is given firstly in Section~\ref{sec1}, along with the formulation of a general quasi-static coupled problem involving  mechanics,  diffusion and  chemistry. 
This is followed in Section~\ref{sec2} by the formulation and solution of the problem for a planar reaction front propagation  in a plate made initially of a linear elastic material that becomes visco-elastic   after the reaction. A Standard Linear Solid Model is taken for the reaction product. 
The kinetic equation for the propagating reaction front and detailed studies of the influence of elastic moduli and an energy parameter on the front propagation kinetics and blocking are presented. Then the derivations and solutions of the equations for the   stress relaxation and inelastic strains description behind the propagating front are given and discussed. 
Conclusions and some perspectives are presented in Sections 4.

\section{General framework}\label{sec1}

In this section, a brief summary of the concept of chemical affinity  tensor, that is used in the present paper, is given below. More detailed explanations are given in \cite{Freidin2013,FrVilKor2014,FreidinMTT2015,Freidin2020}.

\subsection{Chemical affinity tensor. Kinetic equation}

A chemical reaction between solid and diffusing constituents of the following type is considered: 
\begin{equation*}
%\label{reaction} 
\nonumber
n_-B_- +n_*B_* \longrightarrow n_+B_+,
\end{equation*}
where $B_-$, $B_*$ and $B_+$ are the chemical formulae of an initial solid constituent, a diffusing constituent and a transformed solid constituent, respectively, $n_-$, $n_*$ and $n_+$ are the stoichiometric coefficients. Further sub- and superscripts ``$-$'', ``$*$'' and ``$+$'' refer  values to materials $B_-$, $B_*$ and $B_+$.

\begin{figure}[b!]
	\begin{center} \
		\centering \SetLabels
		\L (0.45*0.5)  $\Gamma$ \\
		\L (0.51*0.23) $\mathbf{N}$ \\
		\L (0.79*0.82) $\Omega_+$ \\
		\L (0.58*0.62) $B_+$ \\
		\L (0.33*0.2) $B_{-}$ \\
		\L (0.61*0.95) $B_{*}$ \\
		\endSetLabels
		%\ShowGrid\leavevmode
		\vspace{1mm}
		\AffixLabels{\includegraphics[scale=0.38]{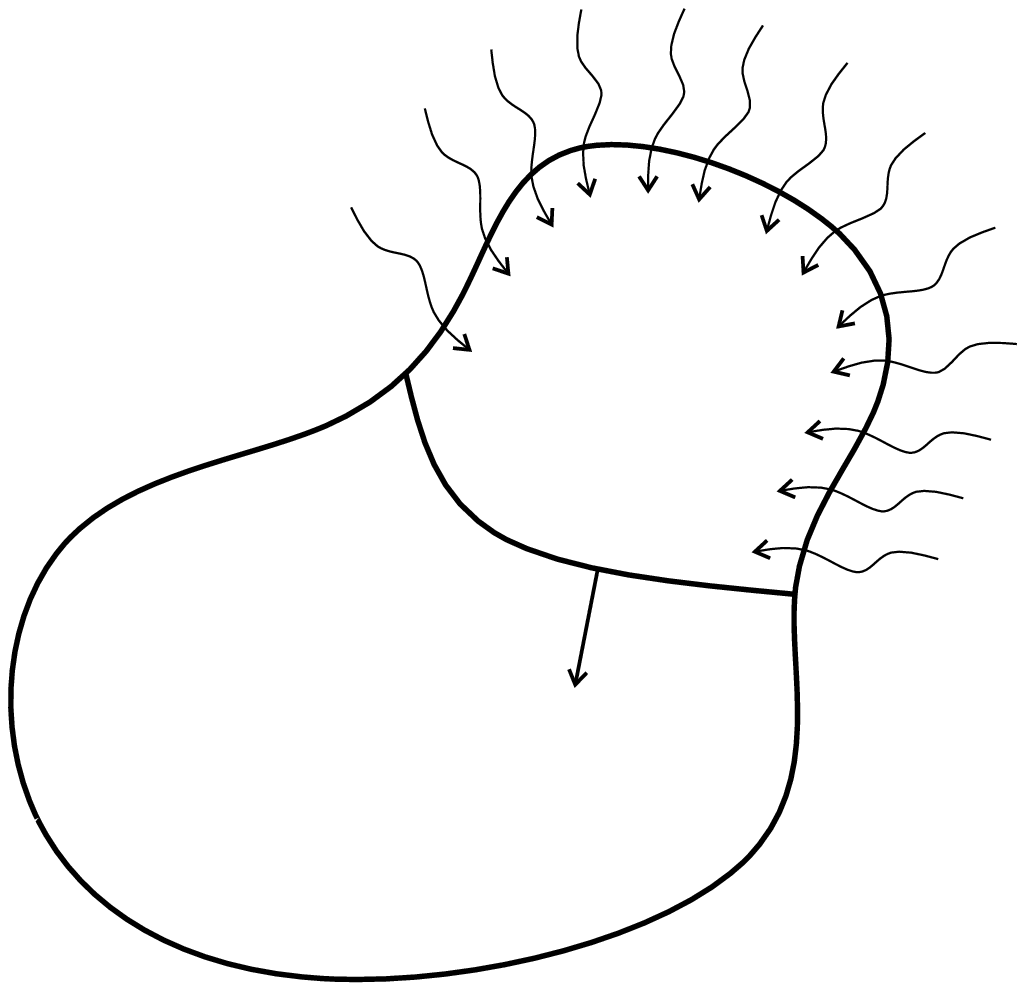}} 
		\caption{Chemical reaction between solid and diffusive constituents, $\Gamma$ is the reaction front\label{body}}
	\end{center}
\end{figure}

The reaction is localized at the reaction front $\Gamma$ that divides the solid constituents $B_-$ and $B_+$, and it is sustained by the diffusion of $B_*$ through the reaction product $B_+$  (see Fig.~\ref{body}). Following \cite{Freidin2013,FrVilKor2014,FreidinMTT2015}, we consider the   constituent $B_+$ as a solid skeleton for the diffusing constituent $B_*$, neglecting the deformations which could be produced in the transformed material by the diffusion. The thermal effects of the chemical reaction are also neglected and  the temperature $T$ is assumed to be a given parameter.

To describe the chemical reaction front kinetics we use an approach based on the concept of chemical affinity tensor  developed in \cite{Freidin2013,FrVilKor2014,FreidinMTT2015} (see also \cite{Freidin2020} and references therein). The normal component of  the chemical affinity tensor appears as a multiplier conjugate to the reaction rate in the expression of the energy dissipation due to the reaction front propagation and acts as a configurational force driving the reaction front. It was shown that the dissipations  per unit area of the reaction front takes the form
\begin{gather}
%\label{Diss}
D=  A_{NN}\omega_N, \nonumber
 %=\frac{\rho_-}{n_-M_-}V_N A_{NN}, 
\end{gather}
where $\omega_N $ is the reaction rate at the reaction front surface element with normal $\mathbf{N}$, 
$A_{NN}=\mathbf{N}\cdot \mathbf{A}\cdot \mathbf{N}$ is the normal component of the chemical affinity tensor 
$ \mathbf{A}$. In 
a quasi-static approach the chemical affinity tensor is defined as:    
\begin{equation}%\label{aff_tens}
\mathbf{A}=n_-M_- \mathbf{M}_- +n_*M_*\mu_*\mathbf{I} - n_+M_+\mathbf{M}_+, \nonumber
\end{equation}
where $\mathbf{M}_-$ and $\mathbf{M}_+$ are the chemical potential tensors which are equal to the Eshelby energy-momentum tensors, divided by the reference mass densities $\rho_-$ and $\rho_+$, and $\mu_*$ is the chemical potential of the diffusing constituent; $\mathbf{I}$ is the second-rank identity tensor; $M_{\pm,*}$ are the molar masses of $B_{\pm,*}$, respectively.  The stresses and strains affect the reaction front propagation as they are present in the configurational force.

The substitution of the normal component $A_{NN}=\mathbf{N}\cdot \mathbf{A}\cdot \mathbf{N}$ of the chemical affinity tensor  into a known formula for the reaction rate \cite{PrigogineDefay1954}  instead of a scalar chemical potential gives the following formula for the reaction rate $\omega_N$ at the reaction front surface element with normal $\mathbf{N}$ \cite{Freidin2013}:
\begin{gather} \nonumber
\omega_N=k_*c \left( 1- \exp \left( -\frac{A_{NN}}{RT} \right) \right),
\end{gather}
where $k_*$ is the kinetic constant, $c$ is the molar concentration of the diffusive constituent per unit volume. Note that the value $k_*c$ represents the partial rate of a direct reaction between diffusing and solid constituents.  Then, since the normal component  $V_N$ of the reaction front velocity is related to the reaction rate as
\begin{equation}
%\label{VN} 
\nonumber
V_N=\dfrac{n_-M_-}{\rho_-}\omega_N,
\end{equation}
where $\rho_-$ is the mass density of the initial material $B_-$, we come to the following dependence of the normal component of the reaction front velocity   on the normal component of the affinity tensor \cite{Freidin2013,FrVilKor2014}:
\begin{gather}
%\label{vel}
\nonumber
V_N=\frac{n_-M_-}{\rho_-}k_*c \left( 1- \exp \left( -\frac{A_{NN}}{RT} \right) \right).
\end{gather}

It can be shown that in the case of small strains, with a chemical potential of the diffusing constituent  taken as
\begin{equation}\label{mustar}
M_* \mu_* = f_* +RT \ln{\cfrac{c}{c_*}},
\end{equation}
where $f_*$ and $c_*$ are the reference chemical energy and volume concentration of the diffusing constituent,  the normal component of the chemical affinity tensor takes the form 
\cite{FrVilKor2014,FreidinMTT2015,FrKor2016}:
\begin{gather}
A_{NN}\!=\!\frac{n_- M_-}{\rho_-} (\gamma+ w_-\!\!- w_+\!+\bm{\sigma}_\pm\!:[\![\bm{\varepsilon}]\!] )  \label{A-small}
\!+n_*RT\ln\frac{c}{c_*}
\end{gather}
which can be rewritten as
\begin{gather} \label{A-small-chi}
A_{NN}\!=\!\frac{n_- M_-}{\rho_-} (\gamma-\chi ) 
+n_*RT\ln\frac{c}{c_*}, 
\end{gather}
where 
$$\gamma=f_0^--f_0^++\dfrac{\rho_-}{n_-M_-}f_*$$ 
is the temperature-dependent chemical energy parameter equal to the combination of the chemical energies  $f_0^-$, $f_0^+$ of the solid constituents and the reference energy $f_*$ of the diffusing constituent, $\gamma$ is taken as a  parameter at given temperature;    $w_\pm$ are the strain energies of the solid constituents per unit volume,  ${[\![\bm{\varepsilon}]\!]=\bm{\varepsilon}_+-\bm{\varepsilon}_-}$ where  $\bm{\varepsilon}_{\pm}$ are the strains at the reaction front,
\begin{gather}
\chi=w_+ -w_- - \bm{\sigma}_\pm\!:[\![\bm{\varepsilon}]\!]  \label{chi}
\end{gather}
characterizes the input of stresses and strains. Note that we neglect the influence  of the pressure produced by the diffusing constituent on the stresses. From the displacement and traction continuity it follows that the stresses $\bm{\sigma}_\pm$ on any side of the front can be substituted into Eq.~\eqref{A-small}. Indeed, since the jump of strain tensor is represented in the form
\begin{equation*}
[\![\bm{\varepsilon}]\!]=\dfrac12(\mathbf{a}\mathbf{N}+\mathbf{N}\mathbf{a}),
\end{equation*} 
the following equalities are valid:
\begin{equation*}
\bm{\sigma}_-\!:[\![\bm{\varepsilon}]\!]
=\mathbf{a}\cdot\bm{\sigma}_-\cdot\mathbf{N}=\mathbf{a}\cdot\bm{\sigma}_+\cdot\mathbf{N}=\bm{\sigma}_+\!:[\![\bm{\varepsilon}]\!].
\end{equation*} 

The equilibrium concentration $c_{eq}$ can be introduced such that \cite{Freidin2013,FrVilKor2014}
\begin{equation}\label{Ann_ceq}
A_{NN}(c=c_{eq})=0.
\end{equation} 
Then the normal component of the affinity tensor can be expressed via the equilibrium concentration $c_{eq}$ and chemical potential of the diffusing constituent calculated at the current concentration
$c(\Gamma)$  and the equilibrium concentration $c_{eq}$  found from \eqref{Ann_ceq} for stresses and strains at the reaction front as
\begin{equation}
%\label{A-ceq}
\nonumber
A_{NN}=n_*M_* \left(\mu_*(c(\Gamma))- \mu_*(c_{eq})\right)
\end{equation}

In a solid skeleton approach $\chi$ does not depend on the concentration. Then from \eqref{A-small-chi} and \eqref{Ann_ceq} it follows that 
\begin{equation}\label{ceq-chi}
\dfrac{c_{eq}}{c_*}=\exp\left\{-\frac{n_- M_-}{\rho_-}\dfrac {(\gamma- \chi )}{n_*RT} \right\}
\end{equation}

During further analysis the stoichiometric coefficients  are normalized by  $n_*$: 
$$n_-\rightarrow n_-/n_*, \quad{n_+ \rightarrow n_+/n_*},\quad  {n_*\rightarrow 1}.$$ Then, if the chemical potential of the diffusing constituent is given by Eq.~\eqref{mustar}, the reaction rate at the front and the reaction front velocity are expressed directly through the current concentrations of the diffusing constituent at the front and the equilibrium concentration corresponding to the stresses at the front: 
\begin{gather}\label{eq_kin_conc}
\omega_N=k_*\left( c(\Gamma)-c_{eq}\right), \quad
V_N=\cfrac{n_-M_-}{\rho_-}k_*\left( c(\Gamma)-c_{eq}\right).
\end{gather}
In such a representation stresses and strains affect the reaction rate via the equilibrium concentration, and one can see that the front may propagate only if at the front $c>c_{eq}$. 

\subsection{Problem statement}

To find the reaction front velocity one has to find stresses and strains at the reaction front, to solve the diffusion problem and to calculate $A_{NN}$ (or find $c_{eq}$ corresponding to stresses and strains at the front). Note the chemo-mechanical coupling: the front velocity depends on stress-strain state and the concentration while stress-state and the concentration depend on the front kinetics and position.

To find the stresses in quasistatic case,  in the absence of body forces, one has to solve the equilibrium equation 
\begin{equation} \label{mech_eq}
\nabla\cdot \bm{\sigma}=0,
\end{equation}
where $\bm{\sigma}$ is the Cauchy stress tensor. The equation Eq.~\eqref{mech_eq} is to be solved within domains $\upsilon_-$ and $ \upsilon_+$, which are occupied by materials $B_-$ and $B_+$, respectively, with boundary conditions at the outer surface of the body (i.e. forces and/or displacements), and with displacement and traction continuity conditions at the reaction front. 

To find the  concentration $c$ at the reaction front we assume that the diffusion flux $\mathbf{j}_*$ is given by Fick's law
\begin{equation} \nonumber
\mathbf{j}_*=-D\nabla c, 
%\label{diffusion}
\end{equation}
where $D$ is the diffusion coefficient of the reactant $B_*$ through $B_+$. Further we assume that $D$ is a constant, the diffusion process happening on much faster time scale than the chemical reaction. We neglect the initial stage of the diffusion prior to the start of the reaction at the outer boundary of the body. Considering the front propagation controlled by the reaction rate rather than by the diffusion rate, we also assume that the  diffusion process is fast enough to consider a steady-state diffusion. Then the diffusion problem  is described by the Laplace equation 
\begin{equation}\nonumber
%\label{Laplace}
\Delta c =0 
\end{equation}
with the boundary conditions
\begin{gather} 
%\label{bond_cond_con}
\left.D \cfrac{\partial c}{\partial N}\right|_{\Omega_+} - \alpha\left( c_* - c (\Omega)\right)=0, \qquad
\left.D\cfrac{\partial c}{\partial N}\right|_{\Gamma} + \omega_N=0, \nonumber
\end{gather}
where $\Omega_+$ is the part of the outer surface of the body corresponding to the transformed material,  $c_*$ is the solubility of $B_*$ in the material $B_+$,
$\alpha$ is the mass transfer coefficient, $\omega_N$ is the reaction rate at the surface element of the reaction front with the normal $\mathbf{N}$. Without loss of generality, one may take $c_*$ also as the reference volume density in Eq.~\eqref{mustar}.

The first boundary condition states that the diffusion flux through the outer boundary of the body becomes zero if the saturation $c_*$ is reached. The second condition means that all the diffusing  reactant is fully consumed at the reaction front and with the use of Eq.~\eqref{eq_kin_conc}$_1$
can be rewritten as  
\begin{gather} 
%\label{bond_cond_conc}
\left.D\cfrac{\partial c}{\partial N}\right|_{\Gamma} + k_* \left( c (\Gamma)- c_{eq} \right)=0. \nonumber
\end{gather}
Finally we come to the coupled problems for a solid with internal unknown propagating interfaces which velocity depends on mechanical stresses and the concentration of a diffusing matter, while the stresses and concentration depend on the position of the interface. 
All the equations are summarized in the following   
 set of equations (see the Box below). Note that this set of equations is valid for any constitutive models of solid constituents.
  	\begin{tcolorbox}[colback=white] 
\text{Equilibrium equation:} 
    \begin{gather*}
        \nabla \cdot \bm{\sigma}=0 \quad +\quad  \text{B.C.}\quad + \quad
        \left.[\![\bm{\sigma}]\!]\right|_{\Gamma}\cdot\mathbf{N}=0\quad
        \text{+\quad Constitutive equations}. 
    \end{gather*} 
\text{Diffusion problem:}
  \begin{gather*}
        \Delta c =0, \\
        \left.D\cfrac{\partial c}{\partial N}\right|_{\Omega}-\alpha\left( c_*-c(\Omega)\right)=0, \\
        \left.D\cfrac{\partial c}{\partial N}\right|_{\Gamma} + k_* \left( c(\Gamma)-c_{eq} \right)=0.
  \end{gather*}
\text{Chemical reaction front kinetics:}
\begin{gather*}
V_N=\cfrac{n_-M_-}{\rho_-}k_* \left( c(\Gamma) -c_{eq} \right)=0\\
%\mathbf{A}=n_-M_-\mathbf{M}_-+n_*M_*\mu_*\bm{ \mathrm{I} } +n_+M_+\mathbf{M}_+ \\
c_{eq}: \ \left.A_{NN}\right|_{c=c_{eq}}=0,\quad 
A_{NN}=
%\mathbf{N}\cdot \mathbf{A}\cdot \mathbf{N}=
\frac{n_- M_-}{\rho_-} (\gamma+ w_-\!\!- w_+\!+\bm{\sigma}_\pm\!:[\![\bm{\varepsilon}]\!] ).
\end{gather*}
%\end{boxtext}
\end{tcolorbox}

\section{A chemical reaction front propagation in the case of a linear viscoelastic reaction product} \label{sec2}

\subsection{Reaction front kinetics}

\begin{figure}[h!]
\begin{center} \
\centering \SetLabels
\L (0.025*1.0)  $0$ \\
\L (1.01*0.93) $x$ \\
\L (0.02*-0.09) $y$ \\
\L (0.12*0.5) $H$ \\
\L (0.31*0.8) $h$ \\
\L (0.45*0.39) $L$ \\
\L (0.80*0.57) $C_-$ \\
\L (0.80*.795) $C_+$ \\
\endSetLabels
%\ShowGrid\leavevmode
\vspace{1mm}
\AffixLabels{\includegraphics[scale=0.8]{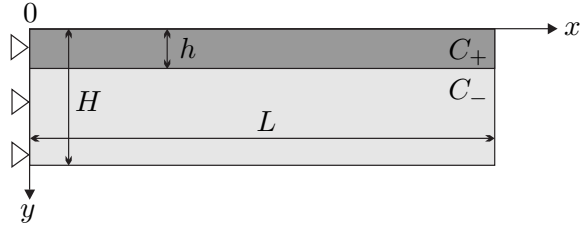}}
\caption{The planar reaction front\label{plate}}
\end{center}
\end{figure}

To demonstrate the influence of the viscosity on the reaction front kinetics, we consider in this section the simple plane strain problem for a chemical reaction in a plate of thickness $H$ and length $L>>H$ with a planar reaction front (see Fig.~\ref{plate}). The  reaction   starts at the outer surface $y=0$ of an initially elastic plate. The  planar reaction front propagates in the $y$-direction,   the reaction front position is given by $y=h$. 
The lower $y=0$ and upper $y=H$ faces of the plate are traction free.
Displacement $u_0$  at the edges $x=\pm L$ prescribes the strain $\varepsilon_0 =u_0/L$ in $x$-direction. Therefore, the strains have to satisfy the following conditions:
\begin{equation}
%\label{strains}
\varepsilon_z= \varepsilon_{xz}=\varepsilon_{yz}=0, \quad \varepsilon_x=\varepsilon_0. \nonumber
\end{equation}

The diffusion problem is reduced to  the diffusion equation
$$
\frac{d^2 c}{dy^2}=0, \quad  y\in[0,h]
$$
with boundary conditions
$$
D\left.\frac{d c}{dy }\right|_{y=0}= \alpha(c(0)-c_*)    , \quad  D\left.\frac{d c}{dy }\right|_{y=h}=-k_*(c(h)-c_{eq}).    
$$
From the solution it follows that the concentration of the diffusing constituent $B_*$ at the reaction front is equal to
\begin{gather}
\nonumber
c(h)=\dfrac{c_*+k_*\left(\dfrac{h}{D} + \dfrac{1}{ \alpha}\right)c_{eq} }{1+k_*\left(\dfrac{h}{D} + \dfrac{1}{ \alpha}\right)}.
\end{gather}
Then,  by  Eq.~\eqref{eq_kin_conc}, the reaction front velocity can be calculated as
\begin{gather} \label{kin_planar}
V=\dfrac{n_- M_-}{\rho_-} \dfrac{c_*- c_{eq}}{\dfrac{1}{k_*}+ \dfrac{1}{\alpha}+\dfrac{h}{D} }, 
\end{gather}
where the equilibrium concentration $c_{eq}$, defined by Eq.~\eqref{Ann_ceq}, depends on stresses and strains at the reaction front.

Formula \eqref{kin_planar} can be rewritten as
\begin{gather} \nonumber
\dot{\xi}=\dfrac{n_- M_-}{\rho_-} \dfrac{c_*- c_{eq}}{T_{ch}+T_D\xi }, 
\end{gather}
where $\xi=\dfrac{h}{H}$, the dot denotes the time derivative, the characteristic times of diffusion and chemical reaction, $T_D$  and  $T_{ch}$, are defined by formulae 
\begin{gather}\label{character-times}
T_D=\frac{H^2}{D}, \quad  T_{ch}={H}\left(\dfrac{1}{k_*}+\dfrac{1}{\alpha}  \right).\end{gather}

In the stress problem, the equilibrium equations and boundary conditions  are satisfied if one takes
\begin{gather}\label{SV}
\sigma_y=0, \qquad \sigma_{xy}=0.
\end{gather}

From continuity of the displacement it follows that at the reaction front
\begin{gather}
%\label{compat}
[\![ \varepsilon_x ]\!] =0. \nonumber
\end{gather}
Then, from Eq.~\eqref{SV} and plane strains conditions  it follows that $\bm{\sigma}_- : [\![ \bm{\varepsilon} ]\!]=0 $ in the expression \eqref{A-small} of the normal component of the chemical affinity tensor.

We assume that the initial material ``$-$'' is isotropic linear elastic. Then 
due to the plane strains conditions, by Hooke's law,  non-zero stresses  in the elastic layer $y\in[h,H]$ are the stresses
\begin{gather}
%\label{hooks}
\sigma^-_x=\frac{4\mu_- \left( 3k_- +\mu_-\right)}{3k_- +4\mu_-} \varepsilon_0, \quad \sigma^-_z=\frac{2\mu_- \left(3k_- - 2\mu_-\right)}{3k_- +4\mu_-}\varepsilon_0, \nonumber
\end{gather}
where $k_-$ and $\mu_-$ are the bulk and shear modules of the material $B_-$. Then the strain energy density of the material $B_-$ is
\begin{gather}\label{w-minus}
w_-=\dfrac{2\mu_- \left(3k_- + \mu_- \right)}{3k_- +4\mu_-} \varepsilon_0^2.
\end{gather}

For the inelastic material ``$+$'' we assume that volumetric strains are elastic, and inelastic behaviors are represented by  rheological models  formulated as relationships between deviatoric parts of stress and strain tensors. Then we use the following decompositions
\begin{gather} \label{stress_and_strain_tensors}
\bm{\sigma}=\sigma \mathbf{I} +\mathbf{s}, \qquad \bm{\varepsilon} =\cfrac{\vartheta}{3} \mathbf{I} +\mathbf{e}, 
\end{gather}
where $\sigma=\cfrac{1}{3}\mathrm{tr}\,  \bm{\sigma}$ and  $\vartheta= \mathrm{tr}  \, \bm{\varepsilon}$ denote the hydrostatic parts of the stress tensor and volume strain, and $\mathbf{s}$ and $\mathbf{e}$ are the deviatoric stress and strain,  respectively.  

We assume that the transformation strain is  spherical: ${\bm{\varepsilon}^{tr}}=(\vartheta^{tr}/3){\mathbf{I}}$. Then the hydrostatic parts of the stress tensor and volume strain are related in constituent  ``$+$''   as
\begin{equation}\label{spherical}
\sigma^+=k_+  ( \vartheta^+ -\vartheta^{tr} ). 
\end{equation}  

\begin{figure}[b!]
	\begin{center} \
		\centering \SetLabels
		\L (0.065*0.865)  $1$ \\
		\L (0.225*0.865) $2$ \\
		\L (0.098*0.61) $\mu_1$ \\
		\L (0.255*0.51) $\mu_2$ \\
		\L (0.098*0.36) $\eta$ \\
		\L (0.418*0.63) $\mu_+$ \\
		\L (0.418*0.23) $\eta$ \\
		\L (0.587*0.525) $\eta$ \\
		\L (0.745*0.515) $\mu_+$ \\
		\L (0.93*0.525) $\eta$ \\
		\L (0.145*-0.07) $a)$ \\
		\L (0.39*-0.07) $b)$ \\
		\L (0.635*-0.07) $c)$ \\
		\L (0.898*-0.07) $d)$ \\
		\endSetLabels
		%\ShowGrid\leavevmode
		\vspace{1mm}
		\AffixLabels{\includegraphics[scale=0.4]{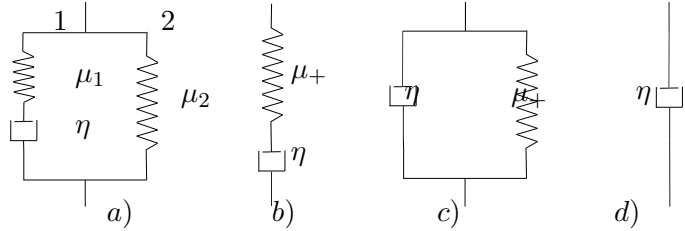}}
			\vspace{3mm}
		\caption{Rheological viscoelastic models: $(a)$ standard linear solid model, $(b)$ Maxwell model, $(c)$ Kelvin-Voigt model, $(d)$ linear viscous model}\label{models}
	\end{center}
\end{figure}

To study how the viscosity and the specified choice of the viscoelastic rheology of the transformed material affects the stress redistribution  due to the chemical reaction, we take at first the standard linear solid model (SLSM)  (Fig.~\ref{models}$a$) that was also referred as the Poynting-Thomson viscoelastic material \cite{Reiner} (see also \cite{Palmov}). Then we examine particular cases of SLSM.

The constitutive equation which relates the deviatoric tensors ${\mathbf{s}^+}$ and $\mathbf{e}^+$ in the material ``$+$'' is derived from the following relationships (see Fig.~\ref{models}):
\begin{gather} \nonumber
\mathbf{s}^+=\mathbf{s}_{1} +\mathbf{s}_{2}, \quad  \mathbf{e^+}=\mathbf{e}_{1} =\mathbf{e}_{2},\\
\nonumber
\mathbf{s}_{1}=\mathbf{s}^e_1=\mathbf{s}^{\eta}, \quad {\mathbf{e}}_1={\mathbf{e}}^e_1+{\mathbf{e}}^\eta,\\
\label{s1s2seta}
\mathbf{s}^e_1=\mathbf{s}_1=2\mu_1\mathbf{e}^e_1, \quad  \mathbf{s}^{\eta}=\mathbf{s}_1=2\eta \dot{\mathbf{e}}^{\eta},\quad
\mathbf{s}_{2}=2\mu_2\mathbf{e}_2 =2\mu_2\mathbf{e}^+,
\end{gather}
where $\mu_1$ and $\mu_2$ are the shear moduli of the elastic elements, $\eta$ is the viscosity, $\mathbf{e}$ and $\mathbf{s}$ with various indices denote deviatoric strains and stresses in corresponding rheological elements, e.g., $\mathbf{s}^e_1$ is the deviatoric stress in the first elastic element, $\mathbf{e}^{\eta}$ is the viscous deviatoric deformation. 
Finally, the constitutive equation takes the known form
\begin{gather}\label{se}
\left(1+\dfrac{\mu_2}{\mu_1}\right)\dot{\mathbf{e}}^+ +\dfrac{\mu_2}{\eta}\mathbf{e}^+=\dfrac{1}{2\mu_1}\dot{\mathbf{s}}^+ +
\dfrac{1}{2\eta}\mathbf{s}^+.
\end{gather}

The strain energy of the constituents ``$+$'' is defined as
\begin{gather}\nonumber
w_+=\dfrac12 k_+(\vartheta^+-\vartheta^{tr})^2
+\mu_1\mathbf{e}_1^e:\mathbf{e}_1^e+\mu_2\mathbf{e}^+:\mathbf{e}^+,
\end{gather}
where it is taken into account that $\mathbf{e}_2 =\mathbf{e}^+$.

To find strain energy $w_+$ at the reaction front there is no need to solve complete viscoelastic problem. Indeed, the viscous strains cannot occur instantaneously at a point when the front passes through this point, while the transformation  and elastic strains appear instantaneously. Therefore: 
\begin{gather} \label{efr}
\mathbf{e}^{\eta}(y,t_y)=0,
\end{gather}
where $t_y$ is the time at which the reaction front passed through the position $y \in [0,h]$. The dependence $t_y=t_y(y)$ is determined by the kinetics of the front propagation: 
\begin{equation}\label{ty}
\int\limits_0^{t_y}V_N(t)dt=y. 
\end{equation}
The condition \eqref{efr} will serve as initial condition in the stress relaxation analysis, but now it is enough to know that from \eqref{efr} it follows that at the reaction front $\mathbf{e}^e_1=\mathbf{e}^+$ and 
\begin{gather}
\nonumber
\bm{\sigma}^+ =k_+ \left(\vartheta^+  -\vartheta^{tr}\right)\mathbf{I}
+2\mu_+\mathbf{e}^+,
\\
\nonumber
w_+=\dfrac12 k_+(\vartheta^+-\vartheta^{tr})^2
+ \mu_+\mathbf{e}^+ :\mathbf{e}^+,
\end{gather}
where 
$\mu_+=\mu_1+\mu_2$,  
%$e^+_{x}, e^+_{y}, e^+_{z}$ are the components of   
deviatoric strain  ${\mathbf{e}^+}$ is taken at the reaction front. 

Due to the plane strain restriction, $\varepsilon_y^+=\vartheta ^+-\varepsilon_0$, 
\begin{equation}
\label{ee}
\mathbf{e}^+ :\mathbf{e}^+ ={\bm{\varepsilon}}^+:{\bm{\varepsilon}}^+-\dfrac{(\vartheta^+)^2}{3}=2\left(\varepsilon_0^2-
\varepsilon_0\vartheta^++\frac{(\vartheta^+)^2}{3}\right).
\end{equation}
Thus, to calculate strain energy $w_+$ at the reaction front it is enough to find the volume strain $\vartheta^+$.
From the relationships:
\begin{gather}
\nonumber
\sigma^+_y=k_+  ( \vartheta^+ -\vartheta^{tr} ) +2\mu_+ e_y^+=0,\\
\nonumber
e_x^+= \varepsilon_0-\dfrac{\vartheta^+}{3}, \quad e_z^+=-\dfrac{\vartheta^+}{3},\quad e_y^+=-(e_x^++e_z^+)=\dfrac{2\vartheta^+}{3}-\varepsilon_0,
\end{gather}
 it immediately follows that at the reaction front:
\begin{gather}
\label{thetaplus}
 \vartheta^+=
\dfrac{3(2 \mu_+ \varepsilon_0+k_+ \vartheta^{tr})}{3k_+ +   {4} \mu_+}, \\
\label{exeyez}
e^+_{x} =\dfrac{(3k_++2\mu_+)\varepsilon_0-k_+ \vartheta^{tr}}{3k_+ + 4 \mu_+},\quad
e^+_{y} =\dfrac{k_+(2\vartheta^{tr}- 3\varepsilon_0)}{3k_+ + 4 \mu_+}, \quad
e^+_{z} =-\dfrac{2\mu_+\varepsilon_0   +   k_+  \vartheta^{tr}}{3k_+ + 4 \mu_+}.
\end{gather}
The relationships \eqref{exeyez} will be also used further in the stress relaxation analysis.

With the use of \eqref{thetaplus} and \eqref{ee}, strain energy $w_+$ becomes a function of $\varepsilon_0$ 
and material parameters. Then the substitution of \eqref{w-minus} for  $w_-$  and obtained expressions of  $w_+$ and $\vartheta^+$
into \eqref{chi} gives $\chi$ as the quadratic function of  external and transformation strains and elastic moduli of the constituents:
\begin{equation}\label{chi-eps}
\chi(\varepsilon_0)=2 (G_+-G_- ) \varepsilon_0^2
-3S\vartheta^{tr} \varepsilon_0 +
S(\vartheta^{tr})^2,
\end{equation}
where
\begin{gather} \label{GS}
G_\pm=\dfrac{\mu_\pm \left(3k_\pm +\mu_\pm \right)}{3k_\pm+4\mu_\pm}=\frac{E_\pm}{4(1-\nu_\pm^2)} , \quad
S=\dfrac{2k_+  \mu_+  }{3k_+ +4  \mu_+ }=\frac{E_+}{9(1-\nu_+)} 
%=\frac{E_+}{9(1-\nu_+)},
\end{gather}
$E_\pm$ and $\nu_\pm$ are the Young moduli and Poisson's ratios.
%We fix above the fact of the dependence of $\chi$ on $\varepsilon_0$ and for the sake of brevity do not write down the corresponding formula. 
Substitution  of \eqref{chi-eps}
into Eq.~\eqref{A-small-chi} and \eqref{ceq-chi} leads to the explicit dependencies    of  $A_{NN}$ and $c_{eq}$ at the reaction front on external and transformation strains, elastic moduli of the constituents and the chemical energies. In particular,
\begin{equation}\label{ceq-chi-pl}
\frac{c_{eq}}{c_*}=\exp\left\{-\frac{n_- M_-}{\rho_-}\dfrac {(\gamma- \chi(\varepsilon_0) )}{ RT} \right\}.
\end{equation}
Note that at given $\varepsilon_0$ the equilibrium concentration does not depend on the front position.
Then the integration of the equation \eqref{kin_planar} leads to the kinetic equation in  the form of the parabolic law:
\begin{gather}
\label{kinetics}
\dfrac {T_D}{2}\xi^2 + T_{ch}\xi = Q t, 
\end{gather}
where 
$Q=\dfrac{n_- M_-}{\rho_-}c_*(1- \phi)$,
\  $\phi=\dfrac{c_{eq}}{c_*}=\exp\left\{-\dfrac{n_- M_-}{\rho_-}\dfrac {(\gamma- \chi(\varepsilon_0) )}{ RT} \right\}$ (cf. with \cite{Morozov2020IMC}).

By Eq.~\eqref{kinetics}, the dependence \eqref{ty} for $t_y(y)$, which is further substituted into \eqref{efr}, can be presented in the explicit form:
\begin{gather}
\label{ty-explicit}
t_y=\dfrac1Q\left(\frac{T_D}{2}\left(\dfrac{y}{H}\right)^2 + {T_{ch}} \left(\dfrac{y}{H}\right) \right).
\end{gather}

\begin{table}
	% table caption is above the table
	\caption{Material parameters used in the simulations for the case $G_+>G_-$}
	\label{Table1}       % Give a unique label
	% For LaTeX tables use
	\begin{tabular}{lcccccc}
		\hline\noalign{\smallskip}
		\textbf{Parameter} & $k_-$[GPa]  & $\mu_-$[GPa]  & $k_+$[GPa] & $\mu_+$[GPa]& $\eta_0$[GPa$\cdot$s]  & $\varepsilon^{tr}$       \\
		%\noalign{\smallskip}\hline\noalign{\smallskip}
		\textbf{Value}& $27.3$  & $25.9$ & $90.9$  & $67.7$ & $15.9$& $0.01$     \\
		\noalign{\smallskip}\hline
		%\hline\noalign{\smallskip}
		\textbf{Parameter}  & $\gamma $[MJ/m$^3$] & $\gamma_0$[MJ/m$^3$]  & $\gamma_*$[MJ/m$^3$]  & $\alpha$[m/s]  & $k_*$[m/s] & $D$[m$^2$/s]  \\
		%\noalign{\smallskip}\hline\noalign{\smallskip}
		\textbf{Value}& $42.9$ &$20.4$ &$14.03$  & $2.3 \cdot 10^{-7}$ &  $1.27\cdot10^{-6}$    & $8 \cdot 10^{-10}$  \\
		\noalign{\smallskip}\hline
			\textbf{Parameter}  & $M_-$[g/mol] &  $\rho_-$[g/cm$^3$]  & $T^{\circ} C$   &   $c_*$  & $H$[m]  & $n_-$
			\\
		\textbf{Value}& $28.1$ & $2.2$  & $920$ & $1$ & $10^{-3}$&  $1$\\
		\noalign{\smallskip}\hline
	\end{tabular}
\end{table}

\begin{table}
% table caption is above the table
\caption{Material parameters used in the simulations for the case $G_+<G_-$}
\label{Table2}       % Give a unique label
% For LaTeX tables use
\begin{tabular}{lccccccc}
\hline\noalign{\smallskip}
\textbf{Parameter}  & $k_-$[GPa]  & $\mu_-$[GPa] & $k_+$[GPa] & $\mu_+$[GPa] & $\gamma$ [MJ/m$^3$]  &  $\gamma_0$[MJ/m$^3$]  & $\gamma_*$[MJ/m$^3$] \\
%\noalign{\smallskip}\hline\noalign{\smallskip}
\textbf{Value}   & $62.3$  & $33.9$ & $27.3$  & $25.9$& $21.9$ & $6.86$ & $14.08$\\
\noalign{\smallskip}\hline
\end{tabular}
\end{table}

\subsection{Equilibrium concentration, kinetics of the reaction front   and  blocking effect}

In this Section,  the dependencies of the reaction front position on time  and of the reaction front velocity on the front position,  which are deduced from Eq~\eqref{kinetics} and \eqref{kin_planar}, will be presented finally at various values of the external strain $\varepsilon_0$, energy parameter $\gamma$ and elastic moduli. 
Since, by \eqref{kin_planar}, the reaction front velocity increases if   $c_{eq}/c_*$ decreases and, respectively, the velocity decreases if  $c_{eq}/c_*$ increases,   the influence of various parameters on the reaction front behavior     can be  predicted qualitatively  
if one knows how the parameters affect the equilibrium concentration. So, we start with such predictions. 

By  \eqref{kin_planar}, the reaction front can propagate only if the stress-strain state at the front and the energy parameter are such that $c_{eq}<c_*$. We study further how the condition $c_{eq}<c_*$ is affected by the parameters. By \eqref{ceq-chi-pl}, this is possible only if the transformation strain, external strains,  elasticity parameters  and the  energy parameter are such that 
$\chi<\gamma$  \cite{Freidin2013,FrVilKor2014}.
By \eqref{chi-eps}, in the considered case this condition 
%\eqref{chi-gamma} 
takes the form
\begin{equation}\nonumber
\chi(\varepsilon_0)-\gamma=2 (G_+-G_- ) \varepsilon_0^2
-3S\vartheta^{tr} \varepsilon_0 -
\left(\gamma-\gamma_0\right)<0,
\end{equation}
where
\begin{gather} \label{gammastar}
\gamma_0= {S(\vartheta^{tr})^2}
%=\frac{E_+(\vartheta^{tr}/3)^2}{(1-\nu_+)} 
\end{gather}
is the critical value of the parameter $\gamma$ in the sense that  the reaction front may propagate
at the external strain $\varepsilon_0=0$ only if
\begin{equation*}% \label{gammastar}
\gamma>\gamma_0.
\end{equation*}
Formula \eqref{gammastar} for the critical value of $\gamma$ for the plane transformation strain with slightly different $S$ was presented in \cite{FrKor2016,FrVilKor2014} for the case of elastic reaction constituents.  The same formula appears here because of the pure elastic behavior of both constituents at the front at the transformation moment~$t_y$.
 
\begin{figure}[tb]
	\begin{center} \
		\centering \SetLabels
		\L (0.145*0.95)  $c_{eq}/c_*$ \\
		\L (0.765*0.95) $c_{eq}/c_*$ \\
		\L (0.19*0.7) $1$ \\
		\L (0.815*0.7) $1$ \\
		\L (0.13*0.03) $\varepsilon^I$ \\
		\L (0.72*0.03) $\varepsilon^{I}$ \\
		\L (0.34*0.03) $\varepsilon^{I\!I}$ \\
		\L (0.845*0.03) $\varepsilon^{I\!I}$ \\
		\scriptsize
		\L (0.204*-0.02) $(a)$ \\
		\L (0.78*-0.02) $(b)$ \\
		\endSetLabels
		%\ShowGrid\leavevmode
		%\vspace{1mm}
		\AffixLabels{\includegraphics[scale=0.3]{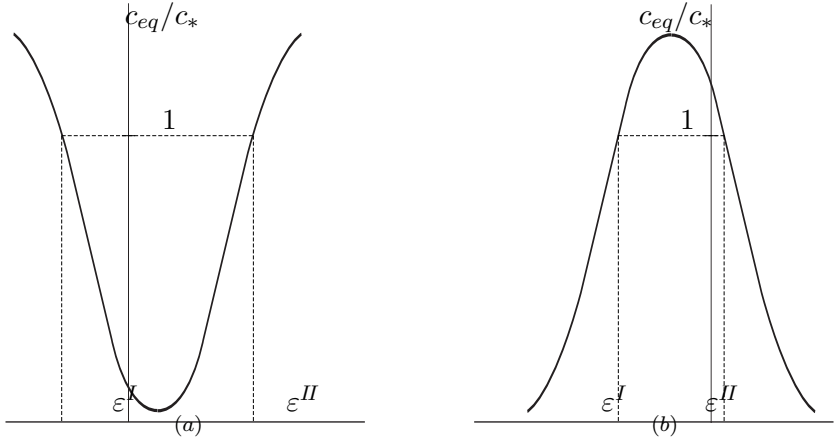}} 
		\vspace{5mm}
		\caption{Dependencies of the equilibrium concentration on the external strain: (a)~${G_+>G_-}$, ${\gamma>\gamma_{*}}$, (b)  ${G_+<G_-}$, ${\gamma_0 <\gamma<\gamma_*}$.} \label{ceq-schem}   
	\end{center}
\end{figure}

The dependencies of $c_{eq}/c_*$ on external strain $\varepsilon_0$ for the planar front are schematically shown in Fig.~\ref{ceq-schem}.
% (see also \cite{FrVilKor2014} where such a dependence was shortly discussed for the case $G_+<G_-$). 
 The extrema is reached at $\varepsilon_0=\varepsilon_0^*$ with $c_{eq}/c_*=c_{eq}^*/c_*$ where 
\begin{gather}\nonumber
\dfrac{c_{eq}^*}{c_*}=\exp\left\{\frac{n_- M_-}{\rho_-}
\dfrac {(\chi(\varepsilon_0^*) -\gamma)}{ RT} \right\},\\ 
\label{extrema}
\varepsilon_0^*=\dfrac{3S\vartheta^{tr}}{4(G_+-G_-)}, \quad 
\quad \chi({\varepsilon_0^*})=
%-\left(\frac{9(H\vartheta^{tr})^2}{8(G_+-G_-)}-\gamma_0\right)=
\left(1-\frac{9S}{8(G_+-G_-)}\right)
%S(\vartheta^{tr})^2
\gamma_0\equiv \gamma_*.
\end{gather}  
Note that $\varepsilon_0^*\neq 0$, since $k_+$ and $\mu_+$ are positive values.  

 The character of the dependence of $c_{eq}/c_*$ on $\varepsilon_0$ and 
 the signs of $\varepsilon_0^*$ and  $\chi(\varepsilon_0^*)$   
  depend  on the relation between $G_+$, $G_-$ and $S$ and the sign of the transformation strain $\vartheta^{tr}$. Without loss of generality, we assume further that  $\vartheta^{tr}>0$.
  As for the elastic modulus, the following three cases can be listed.

 (i) If 
  \begin{equation}\label{GpmS}
  8(G_+-G_-)- {9S}>0,
  \end{equation}
    then $G_+>G_-$ and, therefore,  $\chi(\varepsilon_0^*)>0$ and $\varepsilon_0^*>0$ correspond to the  minimal value of $\chi(\varepsilon_0)$ as it is shown in Fig.~\ref{ceq-schem}$a$.
  Note that the inequality \eqref{GpmS} can be rewritten through elastic moduli as
  \begin{equation} \nonumber
  %\label{mu0-1} 
  \mu_+>\frac{4\mu_-(3k_-+\mu_-)}{3k_-+4\mu_-}\  \Longleftrightarrow\
  \frac{E_+}{1+\nu_+}>\frac{2E_-}{1-\nu_-^2}.
  \end{equation}
     The front can propagate only if the energy parameter, elastic moduli of the solid constituents and transformation strain are  such that  $\gamma>\gamma_*$ and only at strains $\varepsilon_0\in[\varepsilon^I,\varepsilon^{I\!I}]$ where $\varepsilon^I,\varepsilon^{I\!I}$ are the roots of the quadratic equation $\chi(\varepsilon_0)-\gamma=0$. Both roots are positive if $\gamma_*<\gamma<\gamma_0$, and the front cannot propagate without applied tensile strain $\varepsilon_0>\varepsilon^I>0$. 
     
  If the front can propagate at $\varepsilon_{0}=0$ then $\gamma>\gamma_0$ and the roots have different signs,
  $\varepsilon^{I}<0$, $\varepsilon^{I\!I}>0$.
  The front cannot propagate at $\varepsilon_{0}>\varepsilon^{I\!I}$ and at $\varepsilon_{0}<\varepsilon^{I}$, i.e. the propagation is blocked starting from some strains in both tension and compression. 

If the energy parameter is such that $\gamma<\gamma_*$ 
then ${c_{eq}/c_*>1}$ at all $\varepsilon_0$, and the planar front cannot propagate in $y$-direction. 
To avoid misunderstanding, note that this does not mean that the reaction cannot occur  at all. 
In a general case, $\chi$ depends on the geometry  of the front since the stress-strain state at the reaction front depends on the geometry. 
It may happen that other configurations of the reaction front than a configuration with a planar reaction front may develop. On the other hand, a forbidden zone can be constructed in a strain space, formed by the strains at which the reaction front cannot propagate whatever the local normals to the front are  \cite{FrKor2016,FreidinSharipova2018}. Also, 
strictly speaking, even if  the considered solution with a planar propagating front is allowed by kinetic equation at given $\varepsilon_0$, an additional stability analysis would be appropriate \cite{Morozov2019a}.
Consideration of these aspects is beyond the scope of this article. Note only that the use of the semi-inverse approach may give a mathematically consistent solution but other solutions that do not follow a priori assumptions about the geometry of the front may also be of interest.   
  
  (ii)  If the elastic moduli satisfy the inequalities 
   \begin{equation} \nonumber
   %\label{GpmS-ii}
   G_+>G_-\quad \text{but} \quad 8(G_+-G_-)- {9S}<0,
   \end{equation} which can be rewritten as 
  \begin{equation} \nonumber
  %\label{mu0-2}
 \mu_+<\frac{4\mu_-(3k_-+\mu_-)}{3k_-+4\mu_-}<\dfrac{4\mu_+ \left(3k_+ +\mu_+ \right)}{3k_++4\mu_+} \ \Longleftrightarrow \  \frac{E_+}{2(1+\nu_+)}<\frac{E_-}{1-\nu_-^2}<\frac{E_+}{1-\nu_+^2},
  \end{equation}
  then $\varepsilon_0^*>0$ as in the   case (i), but the minimal value of $\chi$ is negative, $\chi(\varepsilon_0^*)<0$. The front may propagate even at negative jump of the chemical energies, $\gamma<0$, but such that $\gamma>-|\gamma_*|$, and at strains $\varepsilon_0$ such that $\chi(\varepsilon_0^*)<\chi(\varepsilon_0)<\gamma<0$.  This would be impossible without accounting for strain energy effects. 
  
  Since $\varepsilon_0^*>0$ at $G_+>G_-$,  the increase of the tensile strain from $\varepsilon_0=0$ until $\varepsilon_0^*$ accelerates the reaction front in both cases (i) and (ii). Further increase of $\varepsilon_0$ retards the front until blocking at $\varepsilon^{I\!I}$.
  On the whole, if 
  $|\varepsilon_0-\varepsilon_0^*|$ increases then the front velocity decreases until zero.

\begin{figure}[h]
	\begin{center} \
		\centering \SetLabels
		\L (0.31*1.02)  $c_{eq}/c_*$ \\
		\L (1.01*0.08) $\varepsilon_0$ \\
			\L (0.716*0.105) $\hat{\varepsilon}$ \\	
			\L (0.53*0.12) $\varepsilon_0^*$ \\
		\scriptsize 
		\L (0.475*-0.01) $(a)$ \\
		%\footnotesize
		\L (0.63*0.82) $ \gamma_0$ \\
		\L (0.63*0.87) $ \gamma_0/2$ \\
		\L (0.63*0.925)  $ \gamma_*$ \\
		\L (0.63*0.76) $2 \gamma_0$ \\
		\L (0.63*0.705) $5 \gamma_0$ \\
		\scriptsize
		\L (0.26*0.2225) $0.9$ \\
		\L (0.26*0.4425) $1.0$ \\
		\L (0.26*0.66575) $1.1$ \\
		\L (0.26*0.89) $1.2$ \\
		\L (0.335*0.04) $0$ \\
		\L (0.41*0.04) $0.01$ \\
		\L (0.185*0.04) $-0.01$ \\
		\L (-0.015*0.04) $-0.03$ \\
		\L (0.605*0.04) $0.03$ \\
			\L (0.805*0.04) $0.05$ \\
		%\footnotesize (0.66*0.38) $\varepsilon^I_*$ \\
		%\L (0.855*0.38) $\varepsilon^{II}_*$ \\
		%\L (0.965*0.138) $\varepsilon^{IV}_*$ \\
		\endSetLabels
		%\ShowGrid\leavevmode
				\begin{minipage}[h]{1\linewidth} 
        \vspace{1mm}
		\AffixLabels{\includegraphics[scale=0.25]{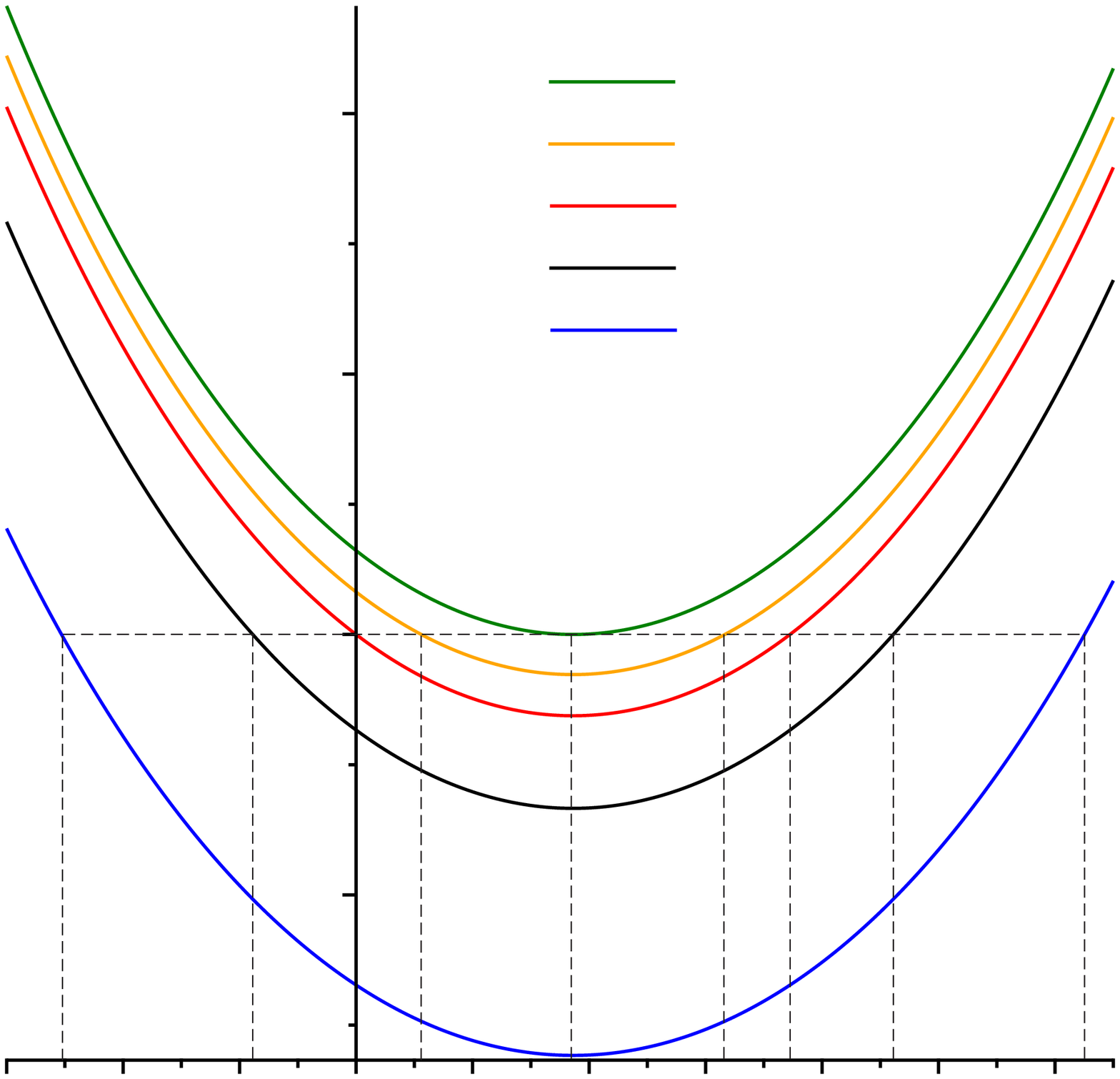} }
        \end{minipage}
        \hfill
        \centering \SetLabels
        \L (0.325*1.0)  $c_{eq}/c_*$ \\
		\L (1.015*0.105) $\varepsilon_0$ \\
		\scriptsize 
		\L (0.49*-0.01) $(b)$ \\
		%\footnotesize
		\L (0.56*0.9) $ 10k_+$ \\
		\L (0.56*0.845) $ k_+$ \\
		\L (0.56*0.79)  $ k_+/10$ \\
		\scriptsize
		\L (0.25*0.189) $0.98$ \\
		\L (0.25*0.32) $1.00$ \\
		\L (0.25*0.445) $1.02$ \\
		\L (0.25*0.57) $1.04$ \\
		\L (0.25*0.6975) $1.06$ \\
		\L (0.25*0.825) $1.08$ \\
		\L (0.355*0.06) $0$ \\
		\L (0.425*0.06) $0.01$ \\
		\L (0.2*0.06) $-0.01$ \\
		\L (0.0*0.06) $-0.03$ \\
		\L (0.625*0.06) $0.03$ \\
		\L (0.815*0.06) $0.05$ \\
		\endSetLabels
		%\ShowGrid\leavevmode
				\begin{minipage}[h]{1\linewidth} 
        \vspace{1mm}
		\AffixLabels{\includegraphics[scale=0.25]{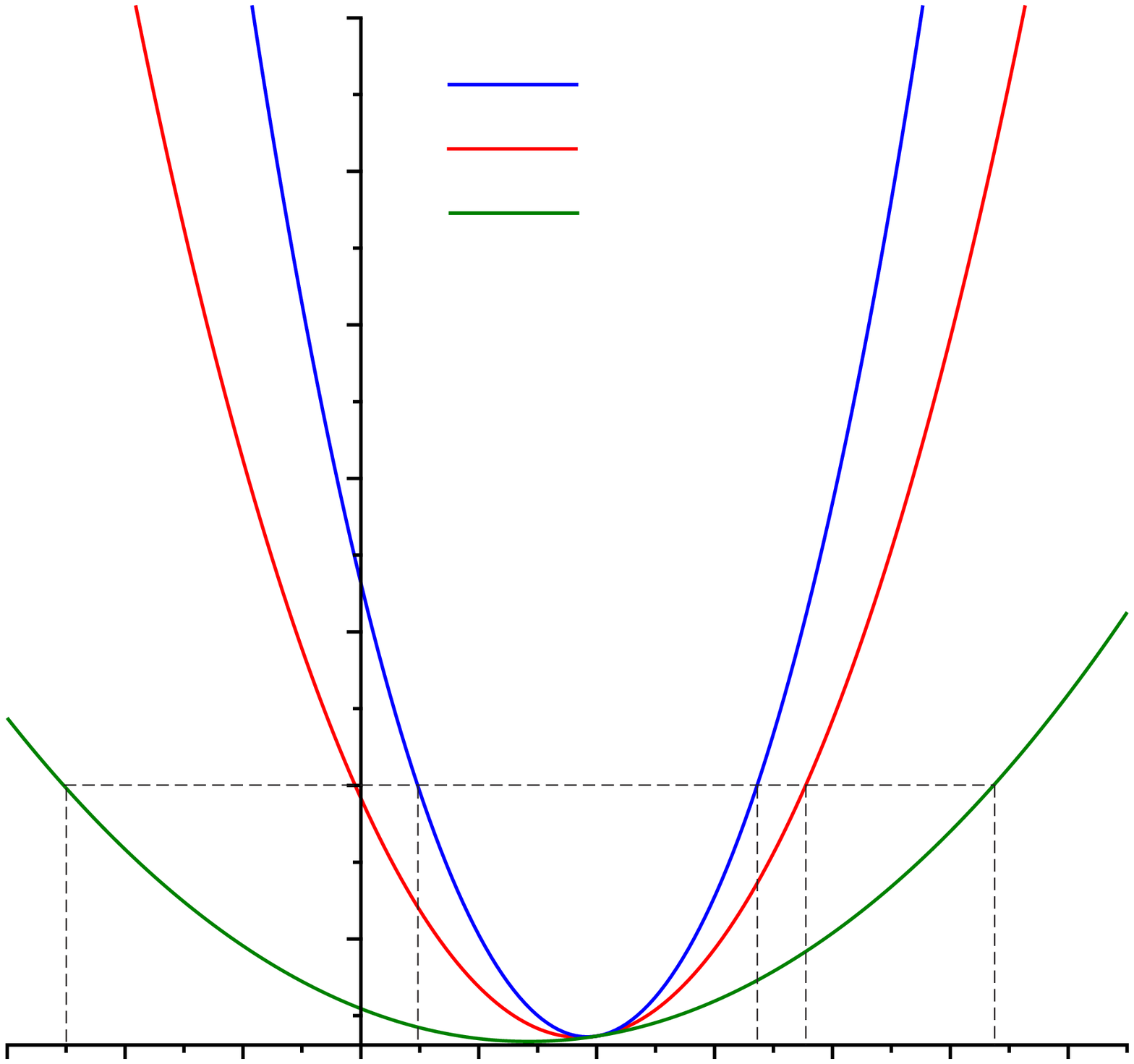}}
        \end{minipage}
        \vfill
        \centering \SetLabels
        \L (0.31*1.01)  $c_{eq}/c_*$ \\
		\L (1.02*0.08) $\varepsilon_0$ \\
		\scriptsize 
		\L (0.49*-0.01) $(c)$ \\
		%\footnotesize
		\L (0.51*0.835) $ \mu_+$ \\
		\L (0.51*0.885) $ 2 \mu_+$ \\
		\L (0.51*0.935)  $ 5 \mu_+$ \\
		\L (0.51*0.78) $\mu_+/2$ \\
		\L (0.51*0.73) $ \mu_+/3$ \\
		%\scriptsize
		\L (0.24*0.175) $0.95$ \\
		\L (0.24*0.4925) $1.05$ \\
		\L (0.24*0.33) $1.00$ \\
		\L (0.24*0.65) $1.10$ \\
		\L (0.24*0.8025) $1.15$ \\
		\L (0.3375*0.04) $0$ \\
		\L (0.4175*0.04) $0.01$ \\
		\L (0.18*0.04) $-0.01$ \\
		\L (-0.015*0.04) $-0.03$ \\
		\L (0.61*0.04) $0.03$ \\
		\L (0.815*0.04) $0.05$ \\
		\endSetLabels
		%\ShowGrid\leavevmode
		 \vspace{3mm}
		\begin{minipage}[h]{1\linewidth} 
        \vspace{1mm}
		\AffixLabels{\includegraphics[scale=0.25]{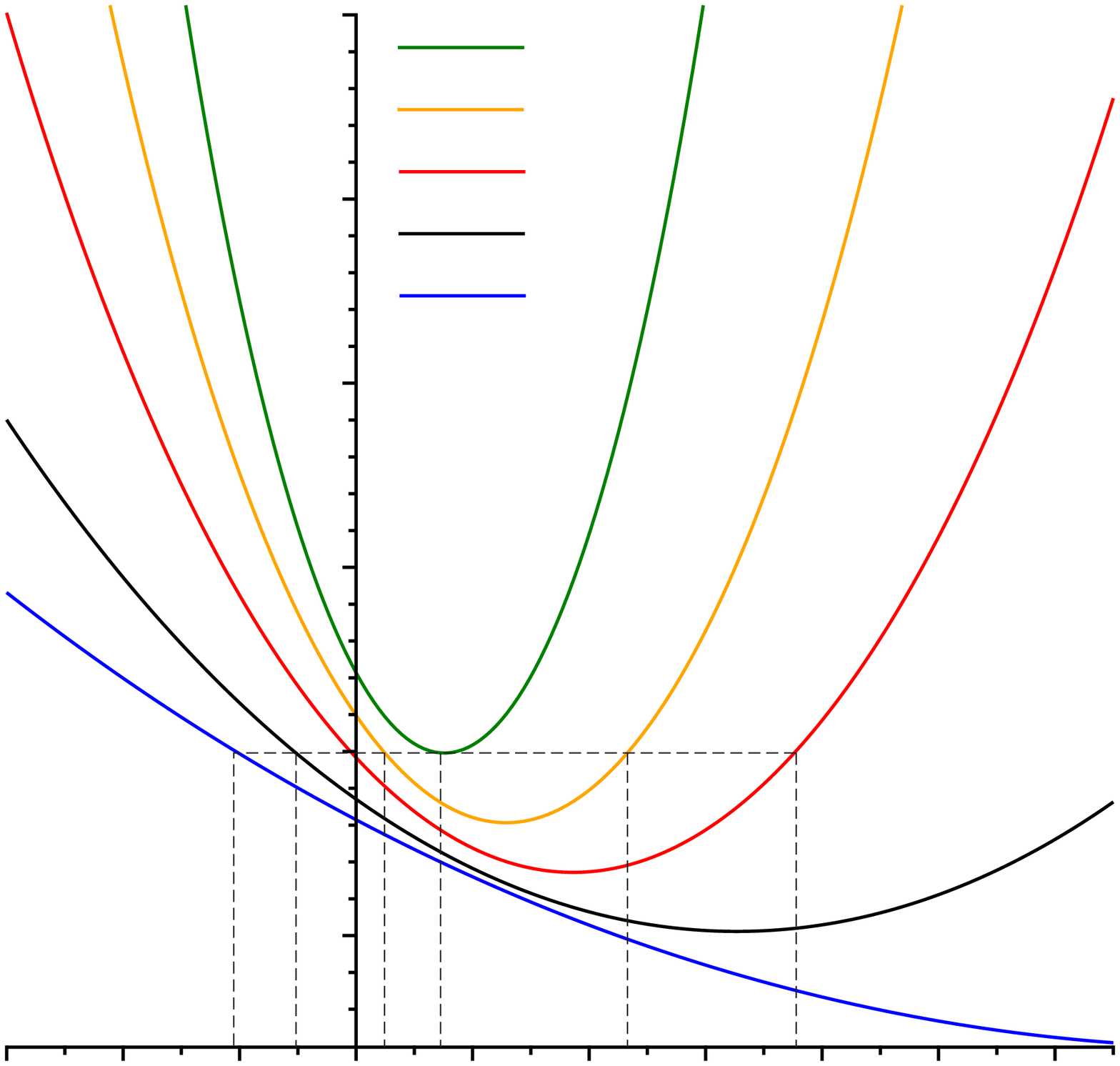}}
        \end{minipage}
    \vspace{2mm}
		\caption{Dependencies of the equilibrium concentration on external strain $\varepsilon_0$ for the case $G_+>G_-$: $(a)$ for different values of the energy parameter $\gamma$; $(b)$ for different values of the bulk modulus $k_+$; $(c)$ for different values of the shear modulus $\mu_+$}
		\label{ceq-gamma-k-mu1}
	\end{center}
\end{figure}

\begin{figure}[h]
	\begin{center} \
	\centering \SetLabels
		\L (0.635*0.99)  $c_{eq}/c_*$ \\
		\L (0.99*0.095) $\varepsilon_0$ \\
			\L (0.24*0.135) $\hat{\varepsilon}$ \\
			\L (0.7*0.135) $\varepsilon^{I\!I}$ \\
			\L (0.15*0.135) $\varepsilon^{I}$\\
				\L (0.45*0.135) $\varepsilon_0^{*}$\\
		\scriptsize 
		\L (0.49*-0.01) $(a)$ \\
		%\footnotesize
		\L (0.855*0.575) $ \gamma_0/2$ \\
		\L (0.855*0.415) $ \gamma_*$ \\
		\L (0.855*0.465) $ 1.5 \gamma_0$ \\
		\L (0.855*0.525)  $ \gamma_{0}$ \\
		\L (0.855*0.355) $2 \gamma_0$ \\
		\L (0.855*0.305) $5 \gamma_0$ \\
		%\scriptsize
		\L (0.53*0.305) $0.85$ \\
		\L (0.53*0.5) $0.90$ \\
		\L (0.53*0.695) $0.95$ \\
		\L (0.53*0.89) $1.00$ \\
		%\L (0.195*0.04) $-0.04$ \\
		\L (0.0*0.055) $-0.06$ \\
		\L (0.39*0.055) $-0.02$ \\
		\L (0.63*0.055) $0$ \\
		\L (0.72*0.055) $0.01$ \\
		\L (0.92*0.055) $0.03$ \\
		%\footnotesize (0.66*0.38) $\varepsilon^I_*$ \\
		%\L (0.855*0.38) $\varepsilon^{II}$ \\
		%\L (0.965*0.138) $\varepsilon^{IV}_*$ \\
		\endSetLabels
		%\ShowGrid\leavevmode
		\begin{minipage}[h]{1\linewidth} 
       % \vspace{1mm}
		\AffixLabels{\includegraphics[scale=0.25]{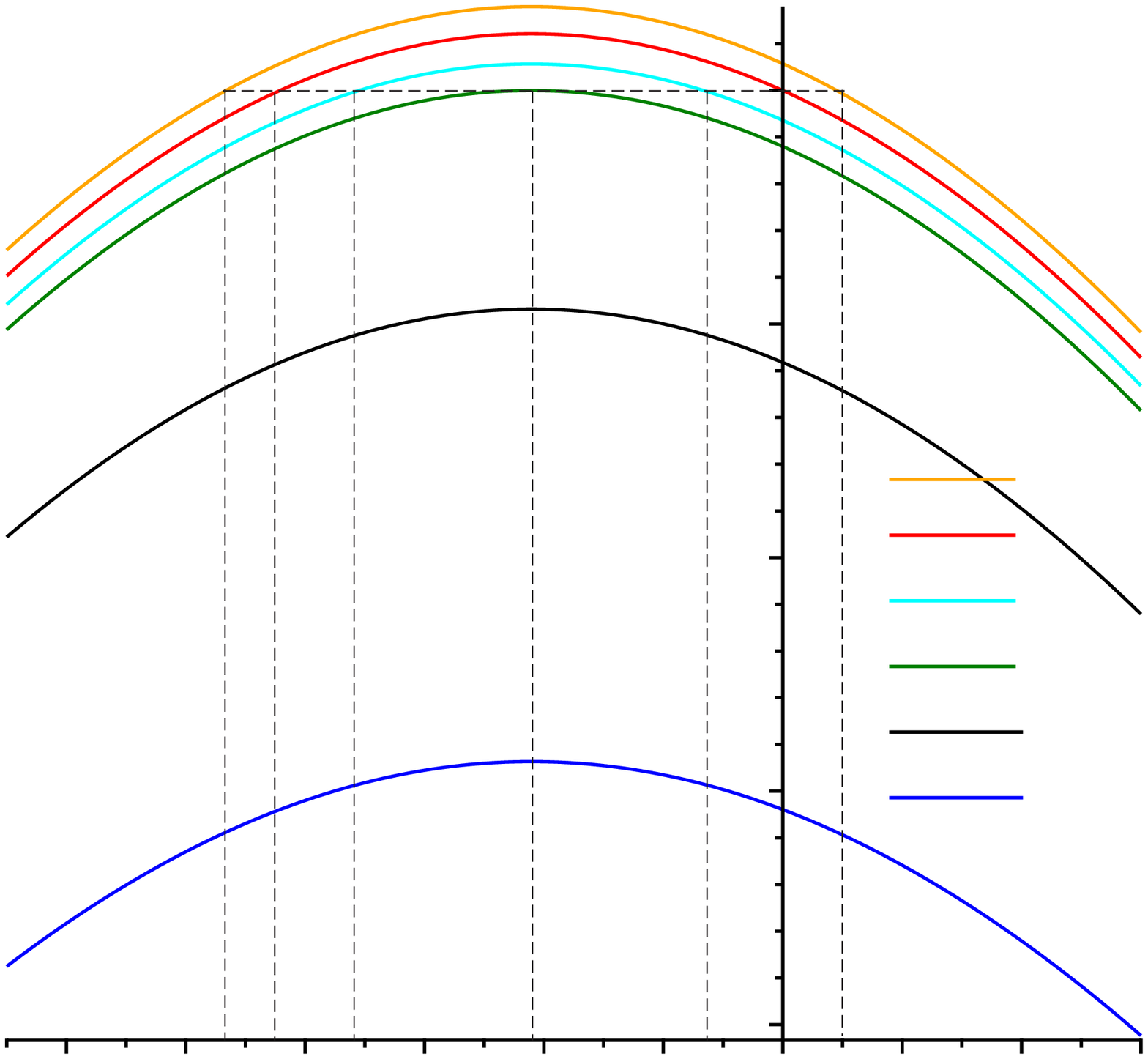} }
        \end{minipage}
	        \hfill
        \centering \SetLabels
        \L (0.585*1.01)  $c_{eq}/c_*$ \\
		\L (1.01*0.105) $\varepsilon_0$ \\
		\scriptsize 
		\L (0.49*-0.01) $(b)$ \\
		%\footnotesize
		\L (0.835*0.915) $ 2k_+$ \\
		\L (0.835*0.86) $ k_+$ \\
		\L (0.835*0.805)  $ k_+/2$ \\
		%\scriptsize
		\L (0.49*0.165) $0.88$ \\
		\L (0.49*0.28) $0.90$ \\
		\L (0.49*0.4) $0.92$ \\
		\L (0.49*0.51) $0.94$ \\
		\L (0.49*0.63) $0.96$ \\
		\L (0.49*0.75) $0.98$ \\
		\L (0.49*0.86) $1.00$ \\
		\L (0.59*0.06) $0$ \\
		\L (0.36*0.06) $-0.02$ \\
		\L (0.18*0.06) $-0.04$ \\
		\L (-0.005*0.06) $-0.06$ \\
		\L (0.74*0.06) $0.02$ \\
		\L (0.915*0.06) $0.04$ \\
		\endSetLabels
		%\ShowGrid\leavevmode
		\begin{minipage}[h]{1\linewidth} 
        \vspace{5mm}
		\AffixLabels{\includegraphics[scale=0.25]{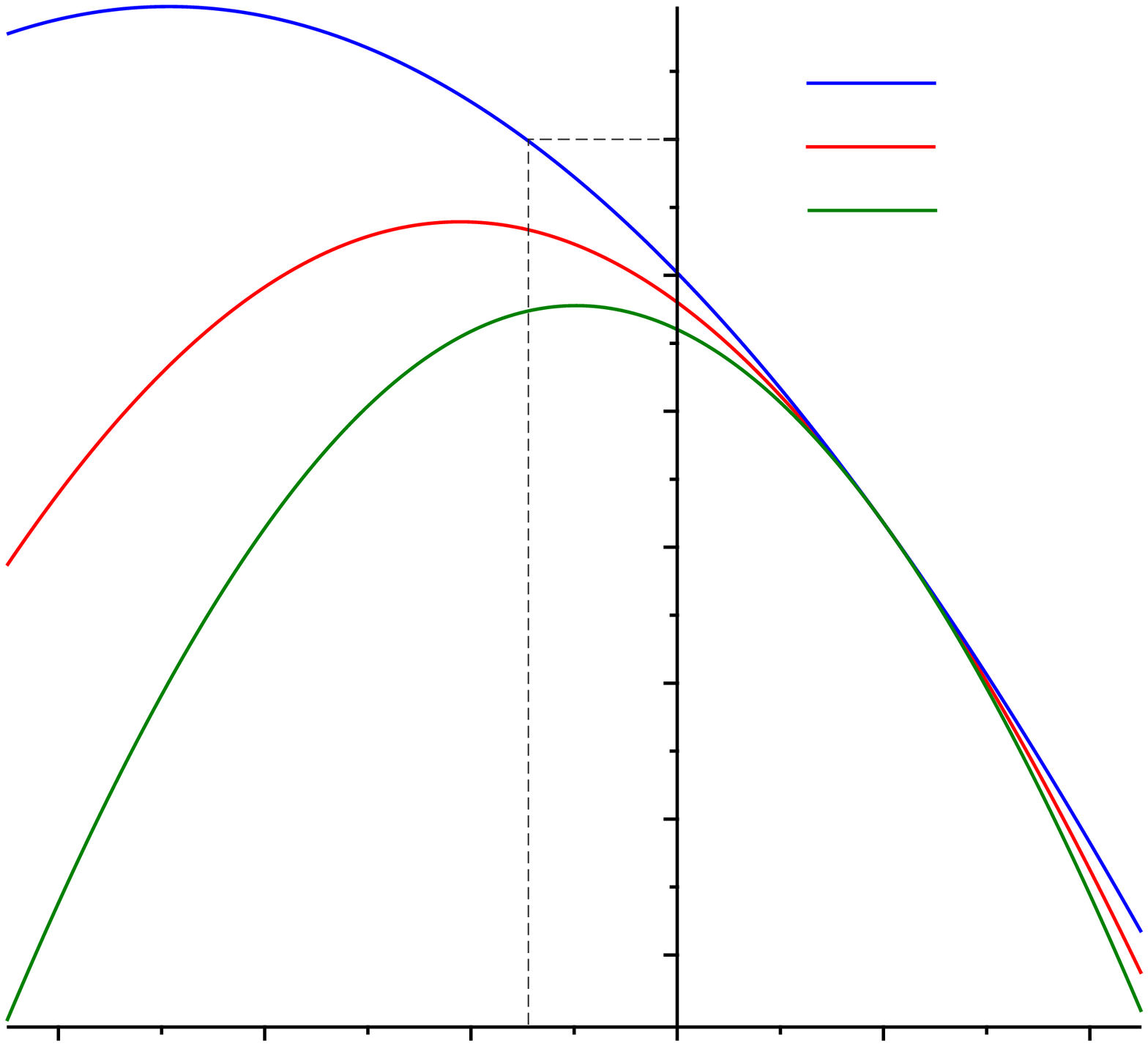}}
        \end{minipage}
        \vfill
        \centering \SetLabels
        \L (0.54*1.01)  $c_{eq}/c_*$ \\
		\L (1.01*0.08) $\varepsilon_0$ \\
		\scriptsize 
		\L (0.49*-0.01) $(c)$ \\
		%\footnotesize
		\L (0.745*0.82) $ \mu_+$ \\
		\L (0.745*0.872) $ 2 \mu_+$ \\
		\L (0.745*0.927)  $ 5 \mu_+$ \\
		\L (0.745*0.77) $\mu_+/2$ \\
		\L (0.745*0.72) $ \mu_+/3$ \\
		%\scriptsize
		\L (0.44*0.52) $0.95$ \\
		\L (0.44*0.315) $0.90$ \\
		\L (0.44*0.735) $1.00$ \\
		\L (0.44*0.94) $1.05$ \\
		\L (0.54*0.04) $0$ \\
		\L (0.39*0.04) $-0.01$ \\
		\L (0.185*0.04) $-0.03$ \\
		\L (-0.015*0.04) $-0.05$ \\
		\L (0.618*0.04) $0.01$ \\
		\L (0.815*0.04) $0.03$ \\
		\endSetLabels
		%\ShowGrid\leavevmode
		\begin{minipage}[h]{1\linewidth} 
        \vspace{4mm}
		\AffixLabels{\includegraphics[scale=0.25]{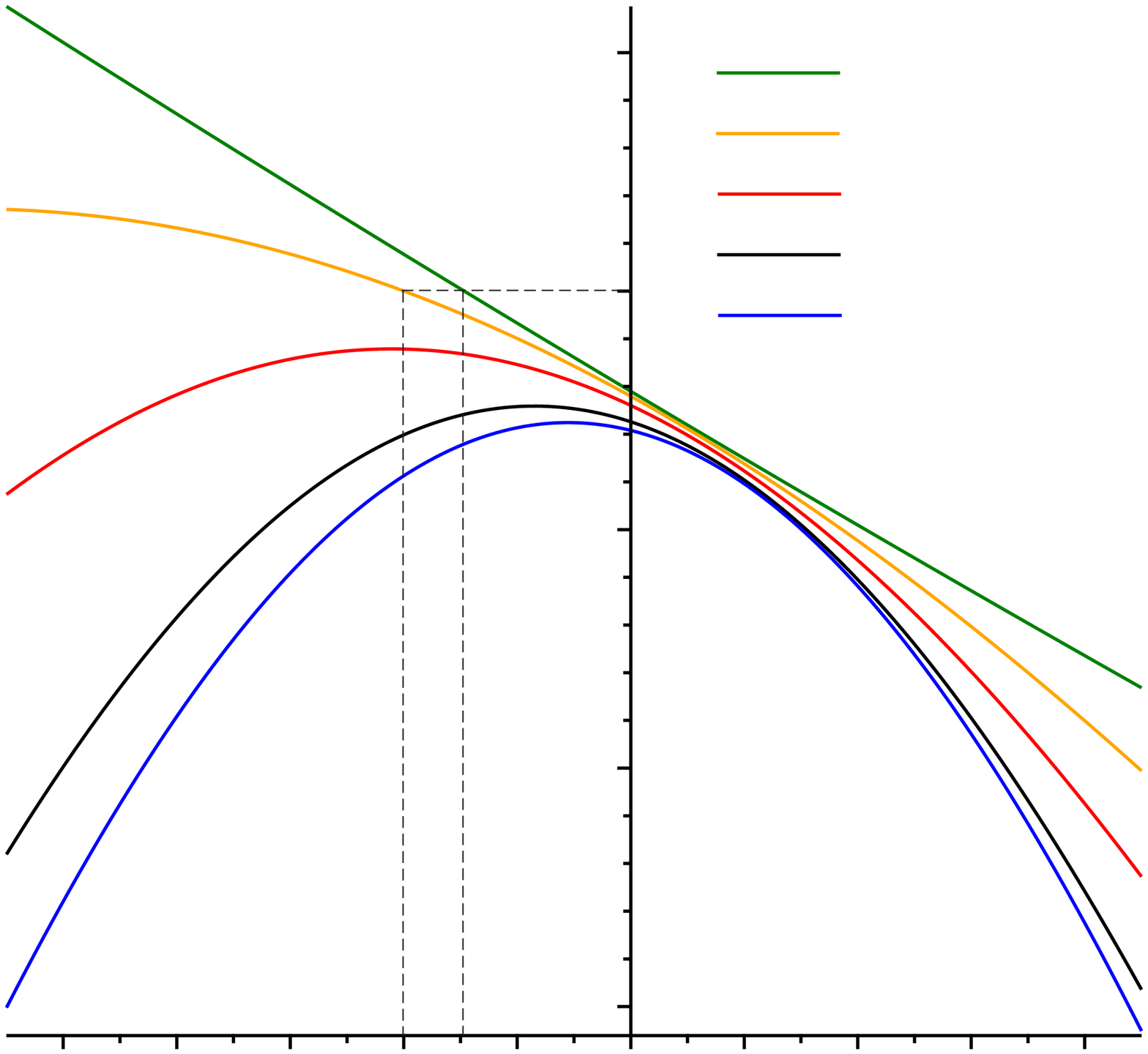}}
        \end{minipage}
    \vspace{3mm}
		\caption{Dependencies of the equilibrium concentration on external strain $\varepsilon_0$ for the case $G_+<G_-$: $(a)$ for different values of the energy parameter $\gamma$; $(b)$ for different values of the bulk modulus $k_+$; $(c)$  for different values of the shear modulus $\mu_+$}
		\label{ceq-gamma-k-mu2}
	\end{center}
\end{figure}
	
(iii) If $G_+< G_-$  then $\gamma_0<\gamma_*$, $\varepsilon_0^*<0$, and $\chi(\varepsilon_0^*)>0$ corresponds to the maximal value on the dependence $\chi(\varepsilon_0)$ 
(Fig.~\ref{ceq-schem}$b$). Such a case was also discussed for an  elastic case with plane transformation strain in \cite{FrVilKor2014}. 
If $\gamma>\gamma_*$ then the front may propagate at any  $\varepsilon_0$. 
If $\gamma<\gamma_*$ then the propagation of the front is blocked at $\varepsilon_0\in[\varepsilon^I,\varepsilon^{I\!I}]$ but
 may start to propagate at proper tension $\varepsilon_0>\varepsilon^{I\!I}$ or compression $\varepsilon_0<\varepsilon^{I}$.   Thus, in this case, in contrast to the previous ones, the front can propagate at any $\gamma$ at some  external strains. 

By \eqref{extrema}, the bulk and shear elastic modules, $k_\pm$ and $\mu_\pm$, affect the dependence $\chi(\varepsilon_0)$ and, thus, the dependencies of $c_{eq}/c_*$ and the reaction front velocity on $\varepsilon_0$ via   parameters $(G_+-G_-)$ and $S$, and the strain $\varepsilon_0^*$ is determined by the dimensionless parameter ${S/(G_+-G_-)}$. 
For example, it is easy to see that $\varepsilon_0^*$ decreases if $\mu_+$ increases and other moduli are fixed.  As another example, one can examine how $k_+$ affects the dependence   $\chi(\varepsilon_{0})$ and the extrema values $\chi(\varepsilon_{0}^*)$ and $\varepsilon_0^*$.   
From \eqref{chi-eps} it follows that
\begin{equation}
\nonumber
\frac{\partial\chi(\varepsilon_{0})}{\partial k_+}=\frac{2\mu_+^2(3\varepsilon_0-2\vartheta^{tr})^2}{(3k_++4\mu_+)^2}\geq 0.
\end{equation}
Then, if the front propagates at a given set of parameters, further increase of $k_+$ increases $\chi(\varepsilon_{0})$ and, therefore, decreases the front velocity. 
Note also that increase of $k_+$ leads to increasing the extrema value $\chi(\varepsilon_{0}^*)$ for both cases $G_+>G_-$ and $G_+<G_-$. 

The dependence of extrema strain $\varepsilon_{0}^*$ on bulk module  $k_+$  is defined by relations between elastic moduli.
Since, by \eqref{extrema}, 
\begin{equation}
\nonumber
\frac{\partial \varepsilon_{0}^*}{\partial k_+}=
\frac{3\mu_+^2\vartheta^{tr}}
{(3k_++4\mu_+)^2}
\frac{8(G_+-G_-)-9S}{4(G_+-G_-)^2},
\end{equation}
one can see that if $8(G_+-G_-)-9S>0$ (the case (i))  then the   point $\varepsilon_{0}^*$ in Fig.~\ref{ceq-schem}$a$ is shifted to the right if $k_+$ increases, and is  shifted to the left leaving  $\varepsilon_{0}^*$ positive if $k_+$ decreases, respectively.

If $G_+>G_-$ but $8(G_+-G_-)-9S<0$  (the case (ii)) or $G_+<G_-$ (the case (iii)), then the extrema point $\varepsilon_{0}^*$ in Fig.~\ref{ceq-schem}$a$ is shifted  to the left or right if $k_+$ increases or decreases, respectively. 

More detailed quantitative  analysis is presented in Fig.~\ref{ceq-gamma-k-mu1} and Fig.~\ref{ceq-gamma-k-mu1}, where  
the dependencies of the relative equilibrium concentration ${c_{eq}^*}/{c_*}$ on the external stain $\varepsilon_0$ at various values of the energy parameter $\gamma$ and the bulk and shear modules $k_+,\mu_+$ of the transformed material are shown for the cases $G_+>G_-$ and $G_+<G_-$, respectively.
 The reference values of the parameters for the cases $G_+>G_-$ and $G_+<G_-$ and corresponding values of 
 %$G_+ $, $G_-$, $S$ ,
  $\gamma_0$ and $\gamma_*$
  %, and $\varepsilon_0^*$
 are given in Tables~\ref{Table1}~and~\ref{Table2}, respectively. Only the parameters differ in two cases are shown in 
 Table~\ref{Table2}. 
 
The choice of the values was made according to the reasons of  the consistency with a small strain approach and better visualisation of the parameters influence.

Fig.~\ref{ceq-gamma-k-mu1}$a$ and Fig.~\ref{ceq-gamma-k-mu2}$a$ reflect the competition between strain and chemical energies at $G_+>G_-$ and $G_+<G_-$, respectively. 
If $\gamma=\gamma_0$ then the dependence of  $c_{eq}/c_*$ on  $\varepsilon_0$ passes through the point $\varepsilon_0=0$, $c_{eq}/c_*=1$. If $G_+>G_-$ and $\gamma=\gamma_0$ then the front may propagate only 
at tension restricted by  the strain $\hat{\varepsilon}=3S\vartheta^{tr}/[2(G_+-G_-)]$, i.e. at strains  $0<\varepsilon_0<\hat{\varepsilon}$  ($\hat{\varepsilon}=0.037$ in Fig.~\ref{ceq-gamma-k-mu1}$a$).
One can see how increasing $\gamma$ results in enlarging the interval of allowed strains $\varepsilon_0$ (see the curves for $\gamma=2\gamma_0$ and  $\gamma=5\gamma_0$) and how the decrease of $\gamma$  shortens and shifts the interval of the  strains at $\gamma_*<\gamma<\gamma_0$. 

If $G_+<G_-$ and $\gamma=\gamma_0$  then 
then the front may propagate only at tension $\varepsilon_0>0$ or compression $\varepsilon_0<\hat{\varepsilon} = -0.042$  (Fig.~\ref{ceq-gamma-k-mu2}$a$).
If $\gamma<\gamma_0$, for example, 
$\gamma=\gamma_0/2$, then the front  can propagate only if additional tension $\varepsilon_0>\varepsilon^{II}>0$ or compression $\varepsilon_0<\varepsilon^{I}<0$ is applied. 
The case $\gamma_0<\gamma < \gamma_* $ is presented by  $\gamma=1.5\gamma_0$, and the front propagation is blocked at $\varepsilon_0\in[-0.036,-0.006]$.  If $\gamma > \gamma_* $ ($\gamma=2\gamma_0$ and  $\gamma=5\gamma_0$ in Fig.~\ref{ceq-gamma-k-mu2}$a$ then the front may propagate at any $\varepsilon_0$.

One can also see in  Fig.~\ref{ceq-schem}$b$,  Fig.~\ref{ceq-gamma-k-mu2}$a$ that if, at $G_+<G_-$, the front can propagate at some $\varepsilon_0$   then further increasing of the absolute value   $|\varepsilon_0|$ decreases ${c_{eq}^*}/{c_*}$ and, thus, increases the front velocity.

Fig.~\ref{ceq-gamma-k-mu1}$b$,$c$ and Fig.~\ref{ceq-gamma-k-mu2}$b$,$c$ characterize quantitatively the role of volume and shear strain energies via the influence of the bulk module  $k_+$ and shear module $\mu_+$ on the dependencies of ${c_{eq}^*}/{c_*}$ on $\varepsilon_0$.
Since $G_+$ decreases if $\mu_+$ decreases and $G_+$ increases if $\mu_+$ increases, one can observe in Fig.~\ref{ceq-gamma-k-mu1}$c$ and Fig.~\ref{ceq-gamma-k-mu2}$c$  how  the dependencies change if $G_+\rightarrow G_-$ due to decreasing $\mu_+$ at $G_+>G_-$ and increasing $\mu_+$ at $G_+<G_-$.

\begin{figure}[tb]
	\begin{center} \
		\centering \SetLabels
		\L (0.145*0.98) $\xi$ \\
		\scriptsize
		\L (0.49*0.005) $(a)$ \\
		%\footnotesize
		\L (0.08*0.96) $1$ \\
		\L (0.04*0.79) $0.8$ \\
		\L (0.04*0.62) $0.6$ \\
		\L (0.04*0.4525) $0.4$ \\
		\L (0.04*0.284) $0.2$ \\
		\L (0.08*0.11) $0$ \\
		\L (0.12*0.065) $0$ \\
		\L (0.375*0.065) $1$ \\
		\L (0.625*0.065) $2$ \\
		\L (0.885*0.065) $3$ \\
		\L (1.01*0.11) $t \cdot 10^{8} [s]$ \\
		%\footnotesize
		\L (0.71*0.475) $\varepsilon_0=0.005$ \\
		\L (0.71*0.42) $\varepsilon_0=0.04$ \\
		\L (0.71*0.37) $\varepsilon_0=-0.005$ \\
		\L (0.71*0.58) $\varepsilon_0=0.019$ \\
		\L (0.71*0.53) $\varepsilon_0=0.009$ \\
		\endSetLabels
		%\ShowGrid\leavevmode
        \begin{minipage}[h]{1\linewidth} 
        \vspace{1mm}
		\AffixLabels{\includegraphics[scale=0.25]{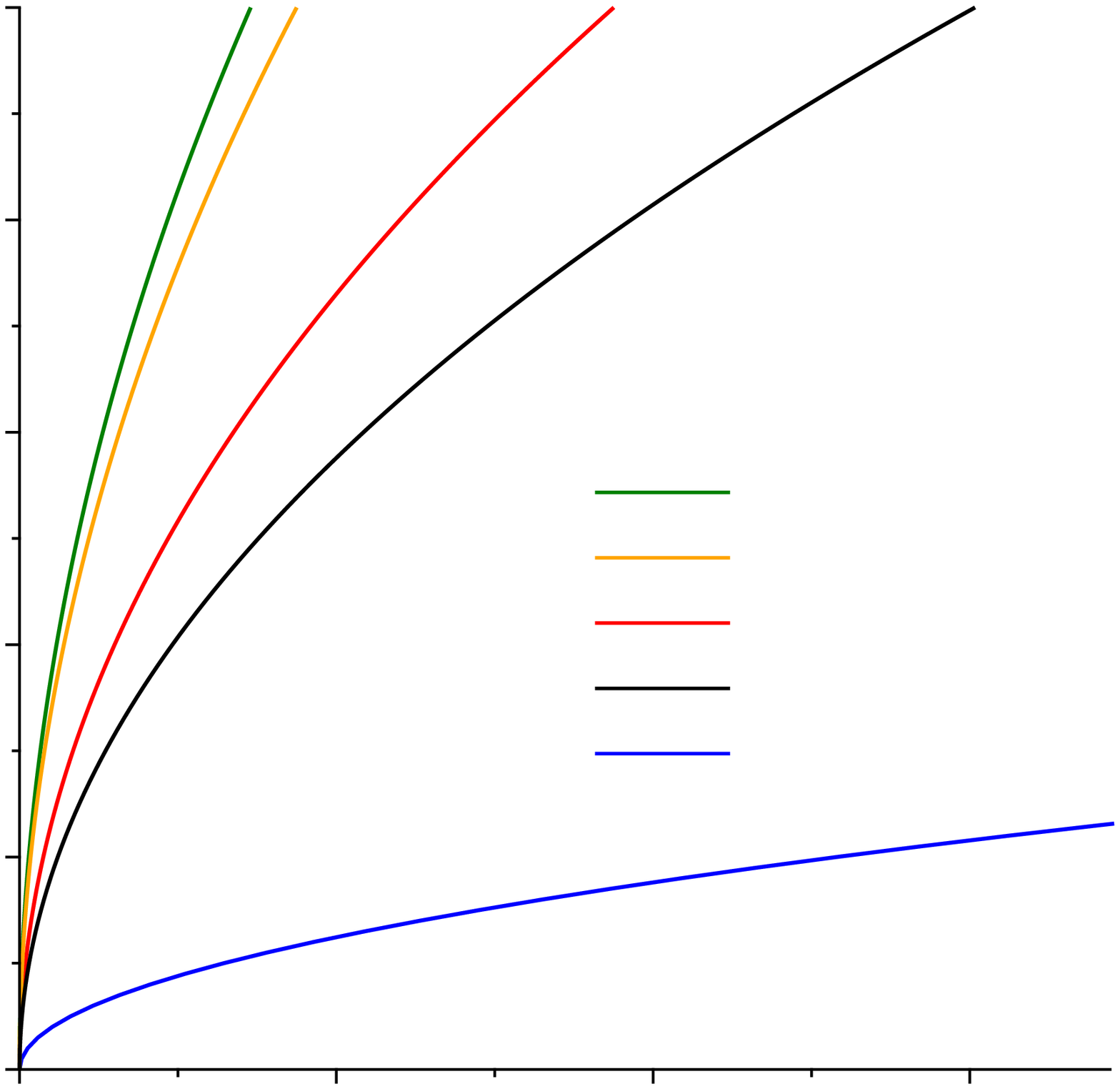}}
        \end{minipage}
        \hfill
	    \centering \SetLabels
		\L (1.014*0.09) $\xi$ \\
		\scriptsize
		\L (0.5*-0.03) $(b)$ \\
		\L (0.24*0.99) $V [nm/s]$ \\
        \L (0.165*0.16) $1$ \\
        \L (0.165*0.42) $2$ \\
        \L (0.165*0.675) $3$ \\
        \L (0.165*0.93) $4$ \\
        \L (0.2*0.035) $0$ \\
        \L (0.345*0.035) $0.2$ \\
        \L (0.50*0.035) $0.4$ \\
        \L (0.655*0.035) $0.6$ \\
        \L (0.81*0.035) $0.8$ \\
        \L (0.98*0.035) $1$ \\
		\endSetLabels
		%\ShowGrid\leavevmode
       \vspace{2mm} \begin{minipage}[h]{1\linewidth} 
        	\AffixLabels{\includegraphics[scale=0.25]{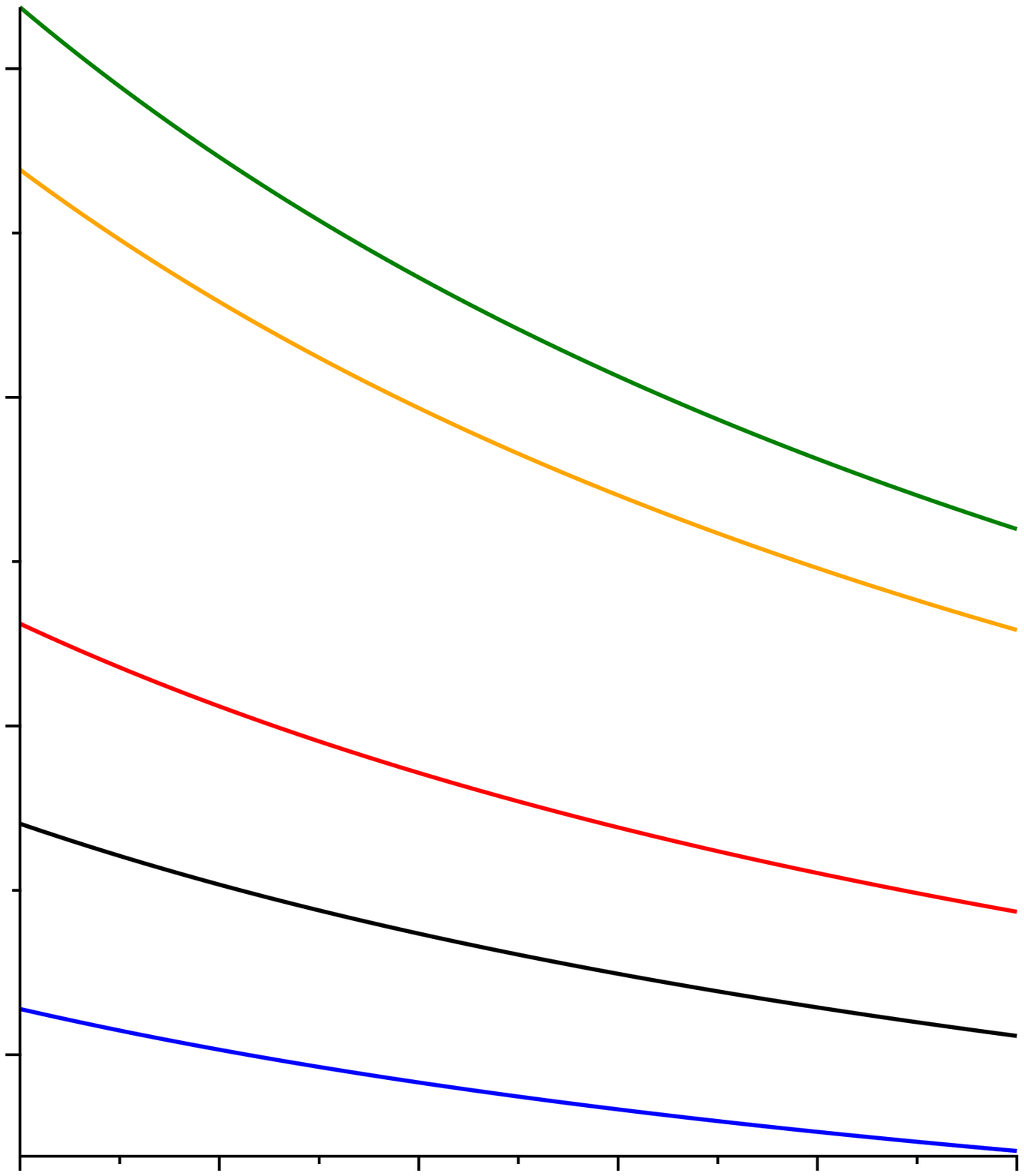}}
        \end{minipage}
        \vfill
        \centering \SetLabels
		\L (0.26*0.98) $h[\mu m]$ \\
		\scriptsize
		\L (0.54*0.005) $(c)$ \\
		%\footnotesize
		\L (0.19*0.93) $5$ \\
		\L (0.19*0.77) $4$ \\
		\L (0.19*0.605) $3$ \\
		\L (0.19*0.44) $2$ \\
		\L (0.19*0.28) $1$ \\
		\L (0.19*0.11) $0$ \\
		\L (0.23*0.065) $0$ \\
		\L (0.405*0.065) $2$ \\
		\L (0.577*0.065) $4$ \\
		\L (0.75*0.065) $6$ \\
		\L (0.92*0.065) $8$ \\
		\L (0.99*0.11) $t \cdot 10^{4} [s]$ \\
		%\footnotesize
		\endSetLabels
		%\ShowGrid\leavevmode
        \begin{minipage}[h]{1\linewidth} 
        \vspace{1mm}
		\AffixLabels{\includegraphics[scale=0.25]{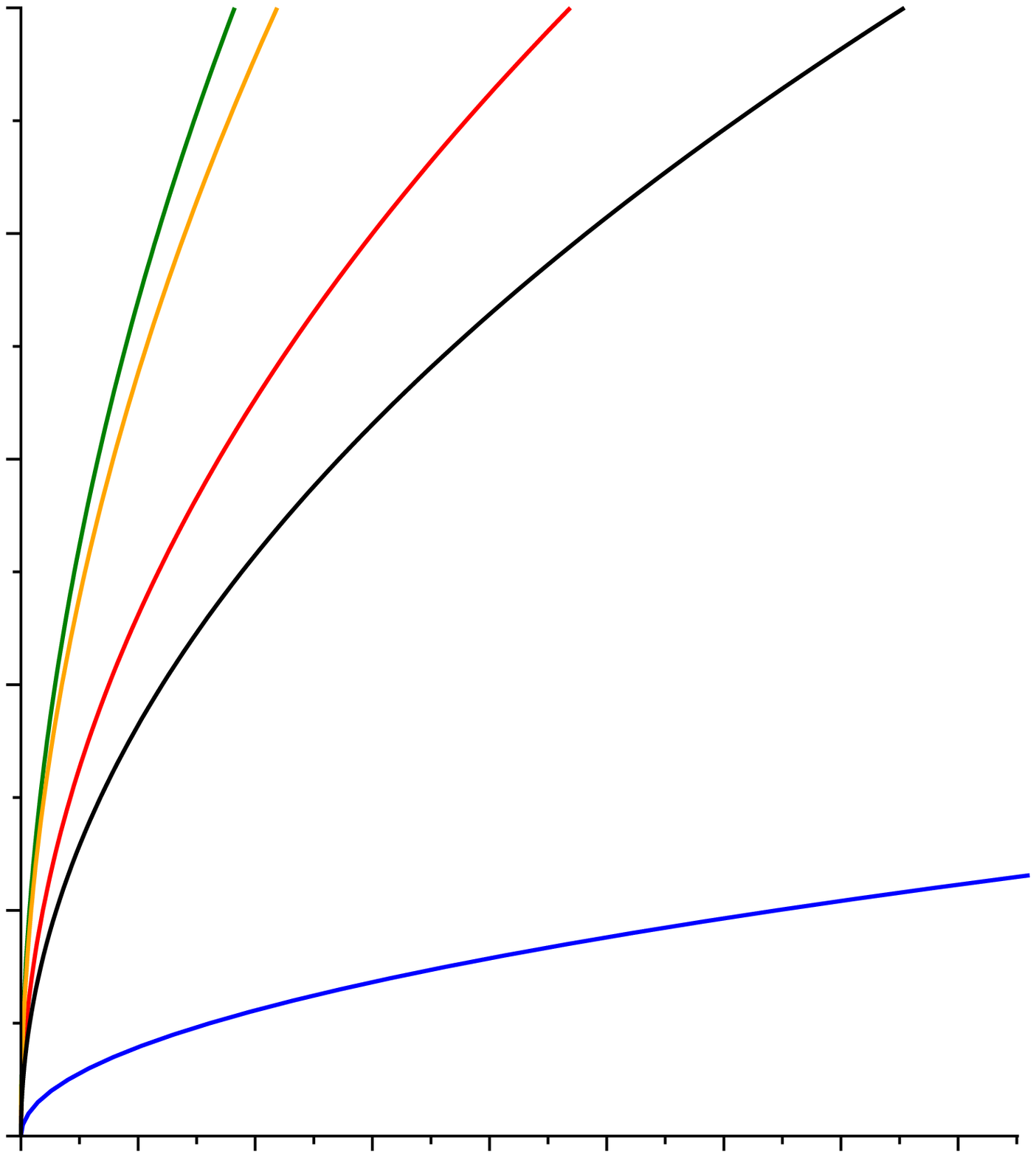}}
        \end{minipage}
     \vspace{1mm}
		\caption{Kinetics of the reaction front at various values of external tension $\varepsilon_0$ for the case $G_+>G_-$. Dependencies of the dimensionless front position on time ($a$),  and the front velocity on the front position (b); (c) -- the front position versus time at the initial stage of the front propagation}\label{kinetics-eps0}
	\end{center}
\end{figure}

Corresponding dependencies of the front position  
on time and the front velocity on the front position under various $\varepsilon_0$ for the case $G_+>G_-$ are shown in Fig.~\ref{kinetics-eps0}. 
The parameters are given in Table~\ref{Table1}; in the   parameters are varied then the values are indicated in figures.
One can see %in  Fig.~\ref{SM-eps0} 
how the strains  retard or accelerate the reaction front. 
The maximal front velocity is observed at tensile strain $\varepsilon_0=\varepsilon_0^*=0.019$. The velocity decreases at both additional  tension  (as at $\varepsilon_0= 0.04$) and at compression relatively to $\varepsilon_0^*$  (as at $\varepsilon_0= 0.005, 0.009, -0.005$), as it has to be in accordance with the increasing $c_{eq}/c_*$ in Fig.~\ref{ceq-gamma-k-mu1}$a$.

\begin{figure}[h!]
	\begin{center} \
		\centering \SetLabels
		\L (0.14*0.98) $\xi$ \\
		\scriptsize
		\L (0.49*0.00) $(a)$ \\
		%\footnotesize
		\L (0.08*0.96) $1$ \\
		\L (0.04*0.79) $0.8$ \\
		\L (0.04*0.62) $0.6$ \\
		\L (0.04*0.45) $0.4$ \\
		\L (0.04*0.28) $0.2$ \\
		\L (0.08*0.11) $0$ \\
		\L (0.12*0.065) $0$ \\
		\L (0.205*0.065) $1$ \\
		\L (0.39*0.065) $3$ \\
		\L (0.565*0.065) $5$ \\
		\L (0.745*0.065) $7$ \\
		\L (0.925*0.065) $9$ \\
		\L (0.98*0.11) $t \cdot 10^{8} [s]$ \\
		%\footnotesize
		\L (0.79*0.35) $\gamma=1.1 \gamma_0$ \\
		\L (0.79*0.297) $\gamma=2 \gamma_0$ \\
		\L (0.79*0.24) $\gamma=5 \gamma_0$ \\
		\endSetLabels
		%\ShowGrid\leavevmode
        \begin{minipage}[h]{1\linewidth} 
        \vspace{1mm}
		\AffixLabels{\includegraphics[scale=0.25]{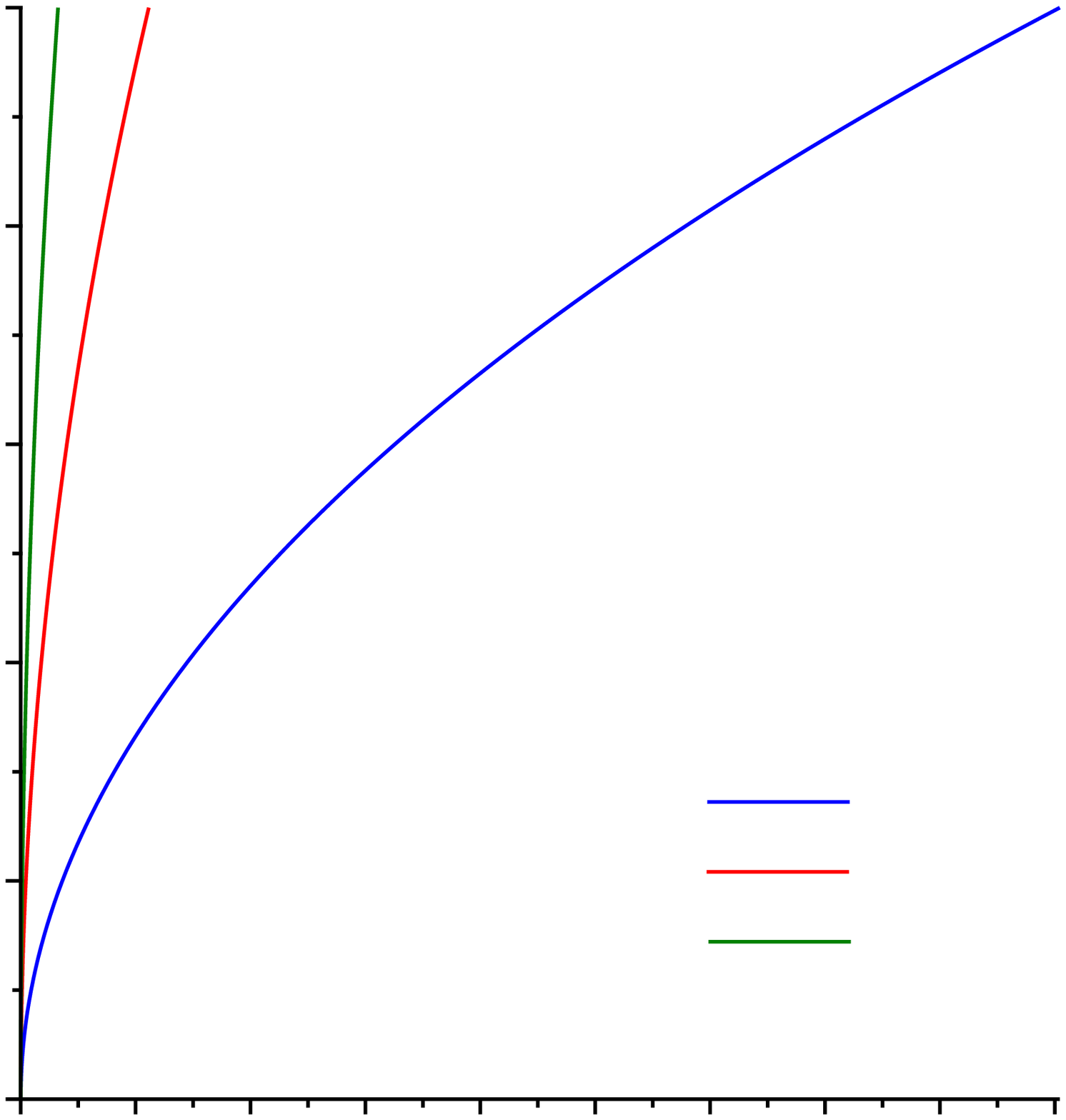}}
        \end{minipage}
        \hfill
        %\hspace{2cm}
	    \centering \SetLabels
		\L (1.01*0.08) $\xi$ \\
		\scriptsize
		\L (0.47*-0.02) $(b)$ \\
		\L (0.23*0.99) $V [nm/s]$ \\
        \L (0.155*0.165) $1$ \\
        \L (0.155*0.355) $3$ \\
        \L (0.155*0.555) $5$ \\
        \L (0.155*0.75) $7$ \\
        \L (0.155*0.94) $9$ \\
        \L (0.195*0.035) $0$ \\
        \L (0.34*0.035) $0.2$ \\
        \L (0.50*0.035) $0.4$ \\
        \L (0.655*0.035) $0.6$ \\
        \L (0.81*0.035) $0.8$ \\
        \L (0.98*0.035) $1$ \\
		\endSetLabels
		%\ShowGrid\leavevmode
        \begin{minipage}[h]{1\linewidth} 
        \vspace{1mm}
		\AffixLabels{\includegraphics[scale=0.25]{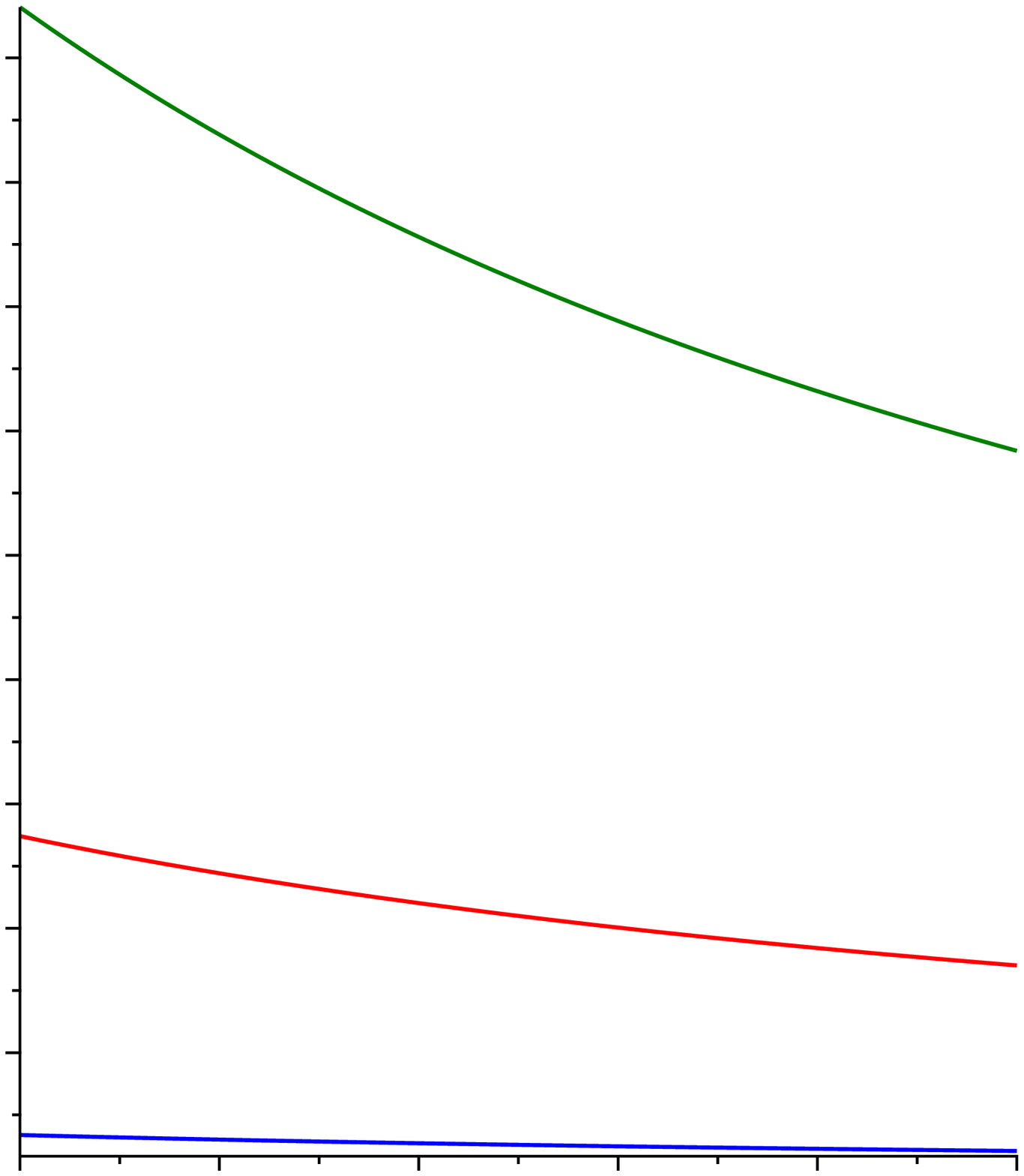}}
        \end{minipage}
		\caption{Dependencies of the front position on time ($a$), and the front velocity on the front position $(b)$ at various values of  
			 energy parameter $\gamma$} \label{kinetics-gamma}
	\end{center}
\end{figure}

\begin{figure}[tb]
	\begin{center} \
		\centering \SetLabels
		\L (0.14*0.98) $\xi$ \\
		\scriptsize
		\L (0.49*0.005) $(a)$ \\
		%\footnotesize
		\L (0.08*0.96) $1$ \\
		\L (0.04*0.79) $0.8$ \\
		\L (0.04*0.62) $0.6$ \\
		\L (0.04*0.45) $0.4$ \\
		\L (0.04*0.28) $0.2$ \\
		\L (0.08*0.11) $0$ \\
		\L (0.12*0.065) $0$ \\
		\L (0.185*0.065) $0.2$ \\
		\L (0.36*0.065) $0.6$ \\
		\L (0.56*0.065) $1$ \\
		\L (0.72*0.065) $1.4$ \\
		\L (0.895*0.065) $1.8$ \\
		\L (0.99*0.11) $t \cdot 10^{8} [s]$ \\
		%\footnotesize
		\L (0.79*0.345) $2 k_+$ \\
		\L (0.79*0.29) $k_+$ \\
		\L (0.79*0.235) $k_+/2$ \\
		\endSetLabels
		%\ShowGrid\leavevmode
		\begin{minipage}[h]{1\linewidth} 
			\vspace{1mm}
			\AffixLabels{\includegraphics[scale=0.25]{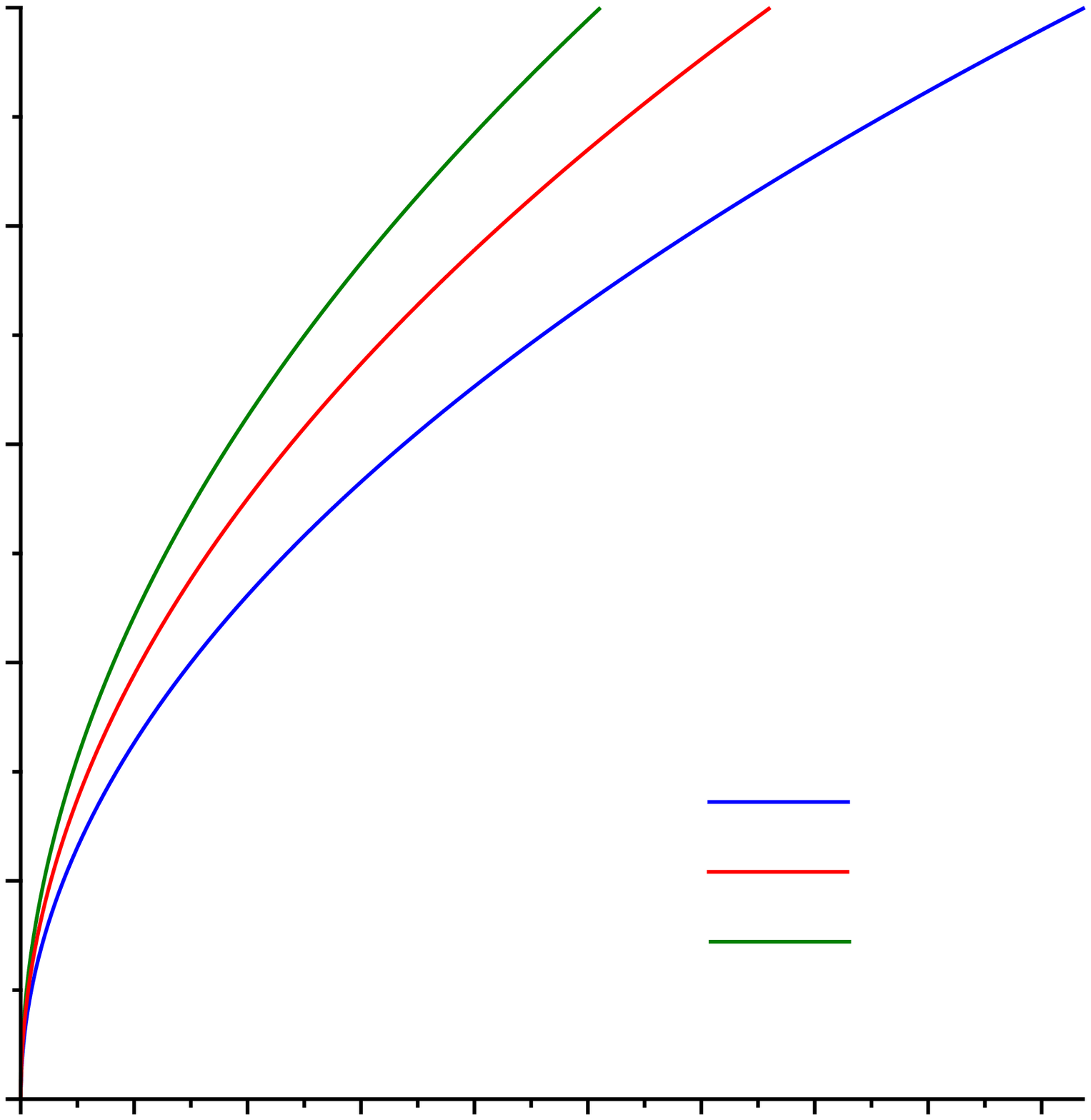}}
		\end{minipage}
	\hfill
		%\hspace{1.5cm}
		\centering \SetLabels
		\L (0.155*0.97) $\xi$ \\
	\scriptsize
	\L (0.49*0.005) $(b)$ \\
	%\footnotesize
	\L (0.09*0.95) $1$ \\
	\L (0.05*0.78) $0.8$ \\
	\L (0.05*0.615) $0.6$ \\
	\L (0.05*0.45) $0.4$ \\
	\L (0.05*0.285) $0.2$ \\
	\L (0.09*0.12) $0$ \\
	\L (0.13*0.07) $0$ \\
	\L (0.385*0.07) $1$ \\
	\L (0.64*0.07) $2$ \\
	\L (0.895*0.07) $3$ \\
	\L (1.01*0.125) $t \cdot 10^{8} [s]$ \\
	%\footnotesize
	\L (0.745*0.475) $\mu_+/5$ \\
	\L (0.745*0.42) $\mu_+/2$ \\
	\L (0.745*0.37) $\mu_+$ \\
	\L (0.745*0.32) $2\mu_+$ \\
	\L (0.745*0.265) $5\mu_+$ \\
	\endSetLabels
		%\ShowGrid\leavevmode
		\begin{minipage}[h]{1\linewidth} 
			\vspace{1mm}
			\AffixLabels{\includegraphics[scale=0.25]{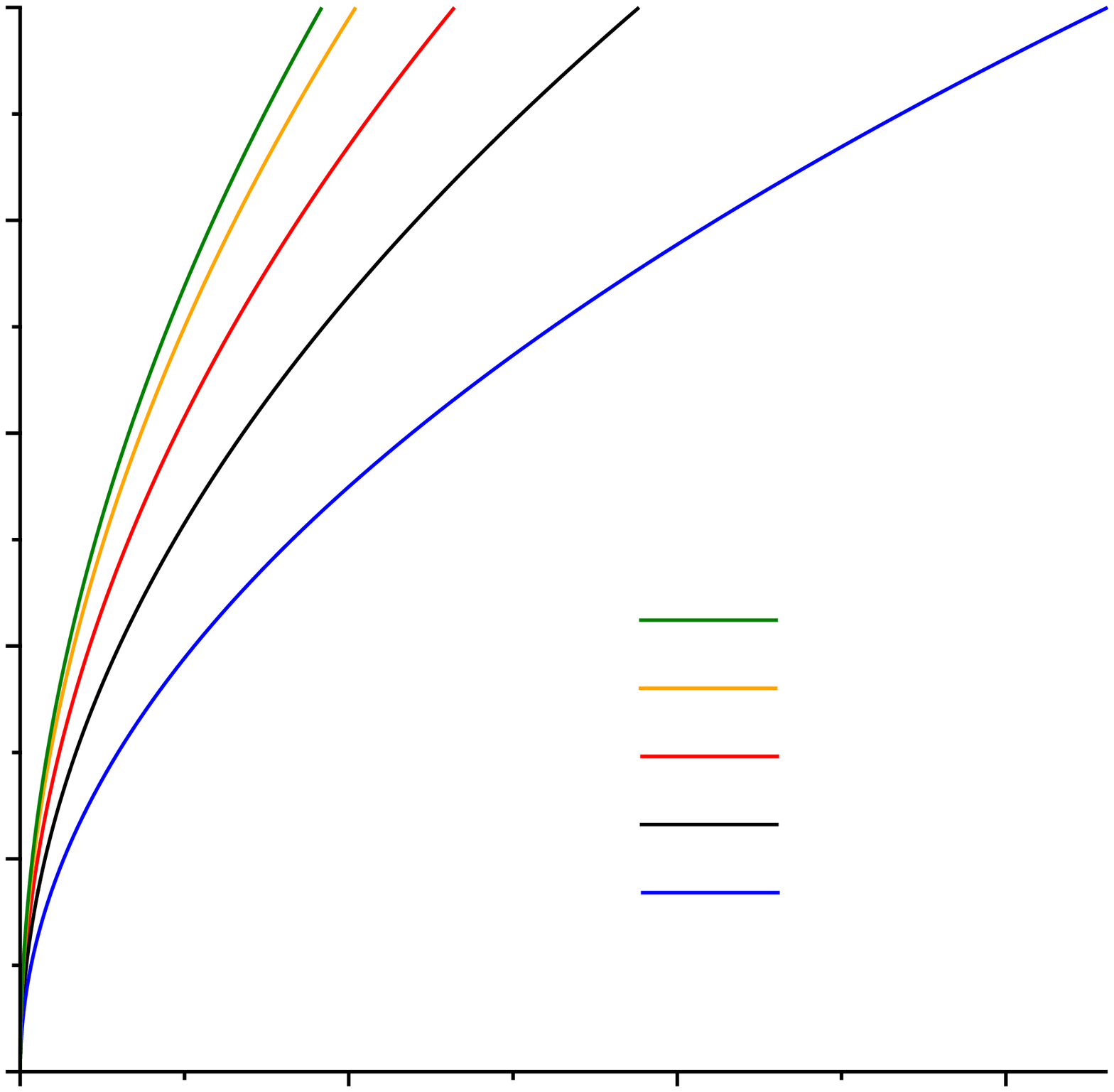}}
		\end{minipage}
		\vspace{1mm} 
		\caption{Dependencies of the front position on time    at various values of the  bulk modulus~$k_+$~(a), and  at   various values of the shear modulus $\mu_+$ (b)  }\label{kinetics-k-mu}
	\end{center}
\end{figure}

Fig.~\ref{kinetics-gamma} and \ref{kinetics-k-mu} demonstrate how   increasing of the energy parameter accelerates the front, and how the values of elastic moduli affect the front kinetics. These dependencies are consistent with the dependencies of  $c_{eq}/c_*$  shown in Fig.~\ref{ceq-gamma-k-mu1}.

\begin{figure}[h!]
	\begin{center} \
		\centering \SetLabels
		\L (0.14*0.98) $\xi$ \\
		\scriptsize
		\L (0.49*0.005) $(a)$ \\
		%\footnotesize
		\L (0.08*0.96) $1$ \\
		\L (0.04*0.79) $0.8$ \\
		\L (0.04*0.62) $0.6$ \\
		\L (0.04*0.45) $0.4$ \\
		\L (0.04*0.28) $0.2$ \\
		\L (0.08*0.11) $0$ \\
		\L (0.117*0.065) $0$ \\
		\L (0.245*0.065) $0.5$ \\
		\L (0.412*0.065) $1$ \\
		\L (0.705*0.065) $2$ \\
		\L (1.01*0.11) $t \cdot 10^{8} [s]$ \\
		%\footnotesize
		\L (0.745*0.475) $H=5\cdot 10^{-4} m$ \\
		\L (0.745*0.42) $H=7.5 \cdot 10^{-4} m$ \\
		\L (0.745*0.365) $H=10^{-3} m$ \\
		\L (0.745*0.315) $H=1.5 \cdot 10^{-3} m$ \\
		\endSetLabels
		%\ShowGrid\leavevmode
        \begin{minipage}[h]{1\linewidth} 
        \vspace{1mm}
		\AffixLabels{\includegraphics[scale=0.25]{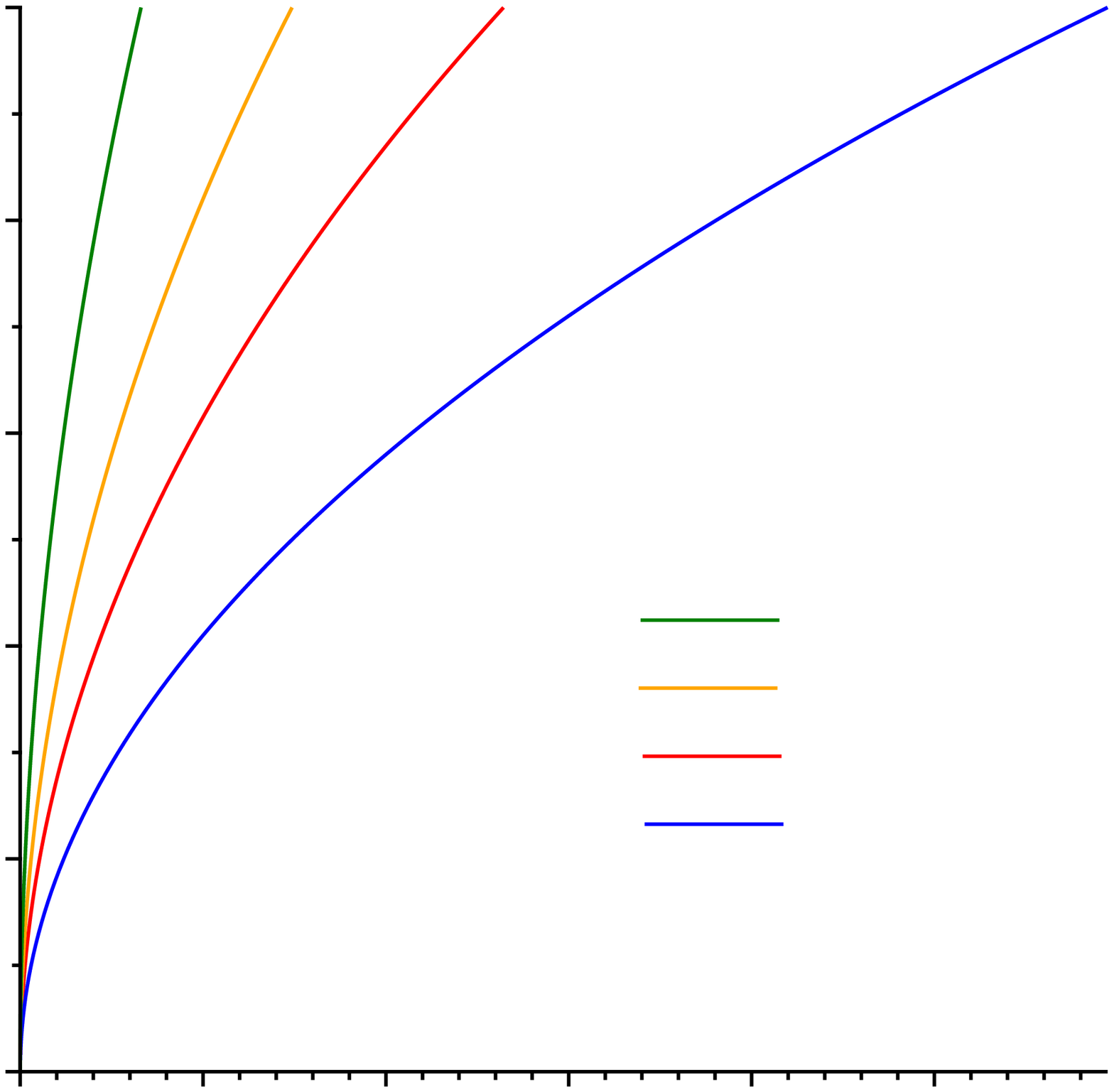}}
        \end{minipage}
        \hfill
        \centering \SetLabels
		\L (1.01*0.08) $\xi$ \\
		\scriptsize
		\L (0.47*-0.02) $(b)$ \\
		\L (0.23*0.99) $V [nm/s]$ \\
        \L (0.16*0.13) $1$ \\
        \L (0.16*0.365) $2$ \\
        \L (0.16*0.6) $3$ \\
        \L (0.16*0.835) $4$ \\
        \L (0.20*0.035) $0$ \\
        \L (0.34*0.035) $0.2$ \\
        \L (0.50*0.035) $0.4$ \\
        \L (0.655*0.035) $0.6$ \\
        \L (0.81*0.035) $0.8$ \\
        \L (0.98*0.035) $1$ \\
		\endSetLabels
		%\ShowGrid\leavevmode
        \begin{minipage}[h]{1\linewidth} 
        %\vspace{1mm}
		\AffixLabels{\includegraphics[scale=0.25]{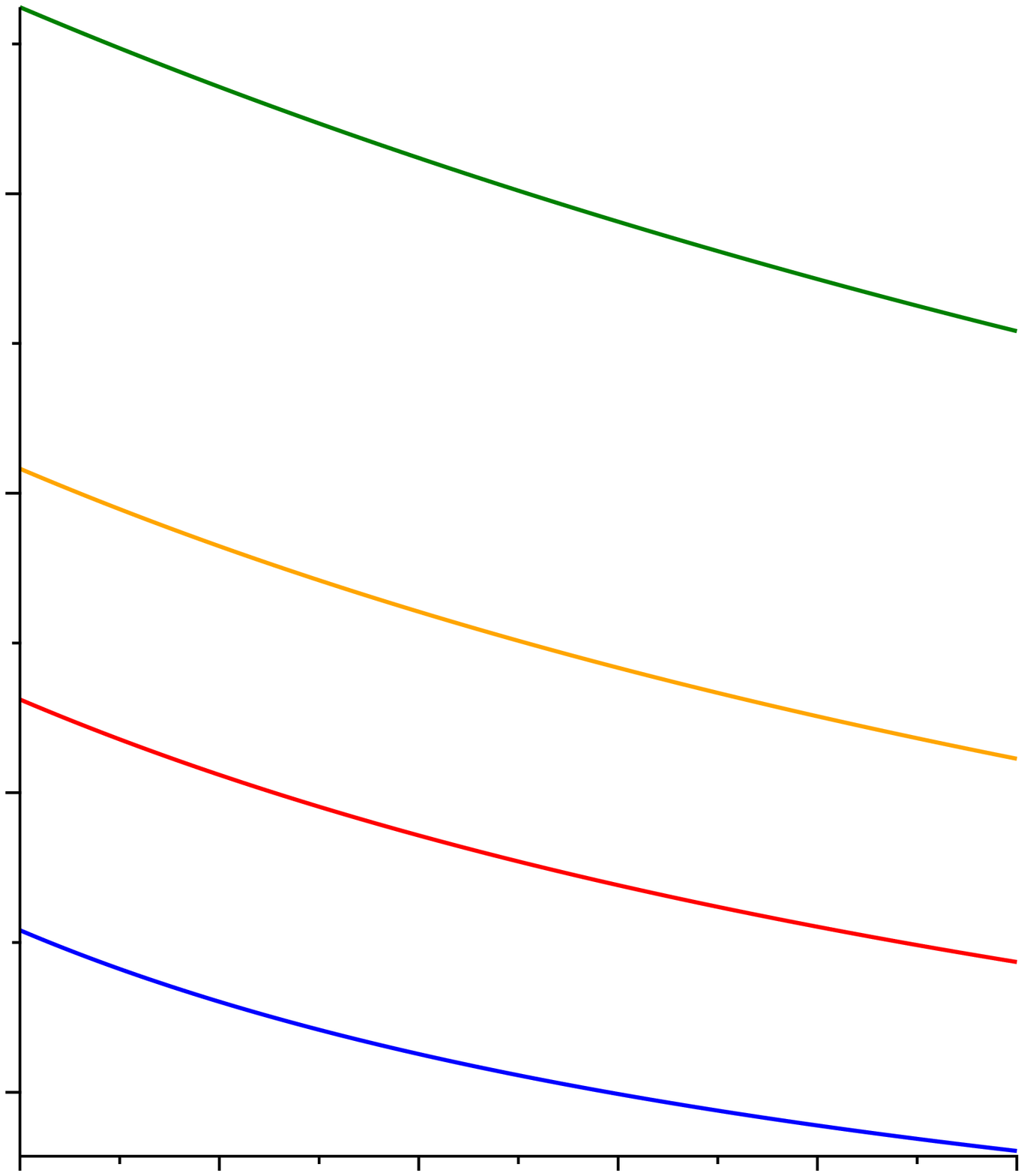}}
        \end{minipage}
		\caption{Dependencies of the front position on time ($a$) and the front velocity on the front position (b)  at   various initial thickness $H$ of the plate }\label{kinetics-H}
	\end{center}
\end{figure}

The initial thickness of the plate also has an effect on the front kinetics 
(Fig.~\ref{kinetics-H})
through characteristic times $T_D$  and $T_{ch}$ of the diffusion supply and  chemical reaction (see Eq.~\eqref{character-times}). Increasing the plate thickness increases the characteristic times and therefore decreases  the relative front velocity. 

\subsection{Stress relaxation behind the reaction front}
 
To calculate stresses  in the transformed material, i.e. behind the reaction front, according to \eqref{stress_and_strain_tensors}, \eqref{spherical} and \eqref{s1s2seta} one has to know time-evolution 
of volume strain $\vartheta^+$ and deviators $\mathbf{e}_2=\mathbf{e}^+$ and~$\mathbf{e}^\eta$ or~$\mathbf{e}^e_1$. 

Substitution of $\bm{\varepsilon}^+-\dfrac{\vartheta^+}{3}{\mathbf{I}}$ and ${\bm{\sigma}}^+-\sigma^+{\mathbf{I}}$ into Eq.~\eqref{se} instead of $\mathbf{s}^+$ and $\mathbf{e}^+$ with the restrictions  $\varepsilon_x^+=\varepsilon_0 $, $\varepsilon_z^+=0$, leads  to the equations
\begin{gather}\label{se-1}
-\left(1+\dfrac{\mu_2}{\mu_1}\right)\dfrac{\dot{\vartheta}^+}{3}+
\dfrac{\mu_2}{\eta}\left(\varepsilon_0-\dfrac{\vartheta^+}{3}\right)=
\dfrac{1}{2\mu_1}\left(\dot{\sigma}^+_x-\dot{\sigma}^+ \right)
+\dfrac{1}{2\eta}(\sigma_x^+-\sigma^+),
\\
\label{se-2}
-\left(1+\dfrac{\mu_2}{\mu_1}\right)\dfrac{\dot{\vartheta}^+}{3}-
\dfrac{\mu_2}{\eta}\dfrac{\vartheta^+}{3}=
\dfrac{1}{2\mu_1}\left(\dot{\sigma}^+_z-\dot{\sigma}^+ \right)+\dfrac{1}{2\eta}(\sigma_z^+-\sigma^+).
\end{gather}

Adding Eq.~\eqref{se-1}  and~ \eqref{se-2} and taking into account Eq.~\eqref{spherical}, we derive the differential equation for $\vartheta^+$:
\begin{gather}\label{eq-theta}
\dot{\vartheta}^+ + \dfrac{\vartheta^+}{\tau_+} - \dfrac{3(k_+ \vartheta^{tr} + 2 \mu_2 \varepsilon_0)}{\tau_1\left(3k_+ +   {4} \mu_+\right)}=0,
\end{gather}
where
\begin{gather}
\tau_1=\dfrac{\eta}{\mu_1}, \quad \tau_+= \dfrac{(3k_+ +  {4}  \mu_+)}{(3k_+ + {4} \mu_2)}\dfrac{\eta}{\mu_1}.  \nonumber
\end{gather}
The initial condition for Eq.~\eqref{eq-theta} is the value $\vartheta^+(t_y)$  at time $t_y$; it is given by \eqref{thetaplus}. Then  the solution of Eq.~\eqref{eq-theta} takes the form:
\begin{gather}\label{thgm}
\vartheta^+(y,t)=
\dfrac{6\mu_1 k_+ ( 3\varepsilon_0-2\vartheta^{tr} )}
{ \left(3k_++ {4} \mu_2 \right)\left(3k_++{4} \mu_+\right)}
\exp \left( - \dfrac{t-t_y}{\tau_+} \right) + \dfrac{3 (k_+ \vartheta^{tr}+ 2 \mu_2 \varepsilon_0)}{3k_++ {4} \mu_2},
\end{gather}
where the dependence $t_y=t_y(y)$ is given by \eqref{ty-explicit}. One can see that the volume strain in points behind the front increases or decreases with time depending on the sign of the difference $( 3\varepsilon_0-2\vartheta^{tr} )$.

Next step is to find $e^+_{x}$ and 
$e_{1x}^e$.
Since 
\begin{equation}
\label{e-x}
e^+_x=\varepsilon_0-\vartheta^+/3,
\end{equation}
from Eq.~\eqref{thgm} it directly follows that 
\begin{multline} \label{ex2gm-res}
e^+_{x}(y,t)=-\dfrac{ 2\mu_1 k_+\left( 3\varepsilon_0 -2\vartheta^{tr} \right)}{\left(3k_++ {4} \mu_2 \right)\left(3k_++ {4} \mu_+\right)} \exp \left( -\cfrac{t-t_y}{\tau_+}\right) \\ + \cfrac{\varepsilon_0 \left( 3k_+ +2\mu_2 \right)-k_+ \vartheta^{tr}}{3k_+ + 4\mu_2}. 
\end{multline}

By constitutive equations~\eqref{s1s2seta}, 
\begin{equation}\nonumber
\mathbf{e}_1^e=\tau_1\dot{\mathbf{e}}^\eta=\tau_1(\dot{\mathbf{e}}^+-\dot{\mathbf{e}}^e_1).
\end{equation}
Then from  \eqref{e-x}
it  follows that ${e}_{1x}^e$ can be found from the equation
\begin{equation}\label{e1e}
\dot{e}_{1x}^e+\dfrac{1}{\tau_1}{e}_{1x}^e=-\dfrac{\dot{\vartheta}^+}{3}
\end{equation}
with the initial condition
\begin{equation}\label{init-ex}
e_{1x}^e(y,t_y)=e_x^+(y,t_y)=\varepsilon_0-\frac{\vartheta^+(y,t_y)}{3}.
\end{equation}
The condition \eqref{init-ex}  follows from   Eq.~\eqref{efr} with $\vartheta^+(y,t_y)$    taken from Eq.~\eqref{thetaplus}.

After calculating the time derivative  $\dot{\vartheta}^+$ from Eq.~\eqref{thgm} and  substituting it into the right hand side of Eq.~\eqref{e1e} we come to the equation for $e_{1x}^e$ that, with the initial condition \eqref{init-ex}, has a solution: 
\begin{gather}
e^e_{1x}(y,t)=\frac{k_+ ( 3\varepsilon_0 -2 \vartheta^{tr})}{2( 3k_+ + 4\mu_+) }
 \exp\left( -\frac{t-t_y}{\tau_+}\right)
 \label{ex1gm-res}
 + \frac{\varepsilon_0}{2} \exp \left( -\frac{t-t_y}{\tau_1} \right).
\end{gather}

Finally, from \eqref{thgm}, \eqref{ex2gm-res} and \eqref{ex1gm-res} it follows that
\begin{multline} 
\sigma^+_x  (y,t)= \dfrac{9\mu_1 k_+^2 \left(3\varepsilon_0 -2 \vartheta^{tr} \right)}{\left(3k_++ {4} \mu_2 \right)\left(3k_++ {4} \mu_+\right)} \exp\left( -\dfrac{t-t_y}{\tau_+}\right)\\ +  \mu_1\varepsilon_0  \exp\left(-\dfrac{t-t_y}{\tau_1}\right)+
  \dfrac{2\mu_2  \left( 2 (3k_+ + \mu_2) \varepsilon_0  -3 k_+ \vartheta^{tr}\right)}{  3k_+ + {4} \mu_2 }. \label{sigx-gm-res}
\end{multline}
Then at the reaction front
\begin{gather}\label{sigx-front}
    \sigma^+_x(y,t_y)=\cfrac{2 \mu_+ \left( 2 \left( 3k_+ +\mu_+ \right) \varepsilon_0 -3 k_+ \vartheta^{tr} \right)}{3k_+ +4\mu_+}.
\end{gather}

Since
%\begin{gather} \nonumber
\begin{multline}
\sigma^+_z (y,t)=3k_+(\vartheta^+(t)-\vartheta^{tr})-\sigma^+_x  (t )=\dfrac{9 \mu_1 {k_+}^2 \left(3 \varepsilon_0 - 2 \vartheta^{tr} \right)}{( 3k_+ + 4\mu_2 )( 3k_+ + 4\mu_+ )} \exp\left( -\dfrac{t-t_y}{\tau_+}\right) \\- \mu_1 \varepsilon_0 \exp \left( -\dfrac{t-t_y}{\tau_1}\right)+  \dfrac{2\mu_2\left( 3 k_+ \left(\varepsilon_0 -\vartheta^{tr} \right) - 2 \mu_2 \varepsilon_0 \right)}{3k_+ +4\mu_2},\label{sigz-gm-res}
\end{multline}
 at the reaction front 
 \begin{gather} \nonumber
     \sigma^+_z(y,t_y)=\cfrac{2\mu_+\left(\left(k^+ - 2\mu_+ \right) \varepsilon_0 - k^+ \vartheta^{tr} \right)}{3k^+ + 4\mu_+}.
 \end{gather}

For completeness, we also write down the formulas for  strain the strains $e_y^\eta$  and  $\varepsilon_y^+$.  To specify $\mathbf{e}^{\eta}$, note that from \eqref{s1s2seta} it follows that 
\begin{gather}\label{e-eta}
\mathbf{e}^{\eta}=
\left(1+\dfrac{\mu_2}{\mu_1}\right)\mathbf{e}^+
%\left({\bm{\varepsilon}}^+-\dfrac{\vartheta^+}{3}\mathbf{I}\right)
-\dfrac{1}{2\mu_1}\left({\bm{\sigma}}^+-\sigma^+{\mathbf{I}}\right).
\end{gather}
 Then, with 
$e^+_y=\dfrac23\vartheta^+-\varepsilon_0$ and  $\sigma_y=0$, from \eqref{e-eta}  it follows that
\begin{equation}
\nonumber
e^\eta_y(y,t)=\dfrac{1}{2\mu_1}\left\{\left(k_++\dfrac43 \mu_+\right)\vartheta^+(y,t)-\left(k_+\vartheta^{tr}+2\mu_+\varepsilon_0\right) \right\},
\end{equation}
where the dependence of $\vartheta^+$   is given by \eqref{thgm}.
The dependence $\varepsilon_y^+(y,t)$ follows from the equality $\varepsilon_y^+=\vartheta^+-\varepsilon_0$ and~\eqref{thgm}.

\begin{figure}[b!]
	\begin{center} \
		\centering \SetLabels
		\L (1.0*0.88) $\xi$ \\
		\scriptsize
		\L (0.55*-0.04) $(a)$ \\
		%\footnotesize
		\L (0.21*0) $\sigma^+_x$ \\
		\L (0.27*0.92) $0$ \\
		\L (0.82*0.92) $\xi_1$ \\
		\L (0.935*0.92) $\xi_2$ \\
		\L (0.82*-0.02) $A$ \\
		\L (0.80*0.38) $B$ \\
		\L (0.53*0.28) $t=t_y(\xi_1)$ \\ 
		\L (0.625*0.735) $t=t_y(\xi_2)$ \\ 
		\endSetLabels
		%\ShowGrid\leavevmode
		\begin{minipage}[h]{1\linewidth} 
			\vspace{1mm}
			\AffixLabels{\includegraphics[scale=0.27]{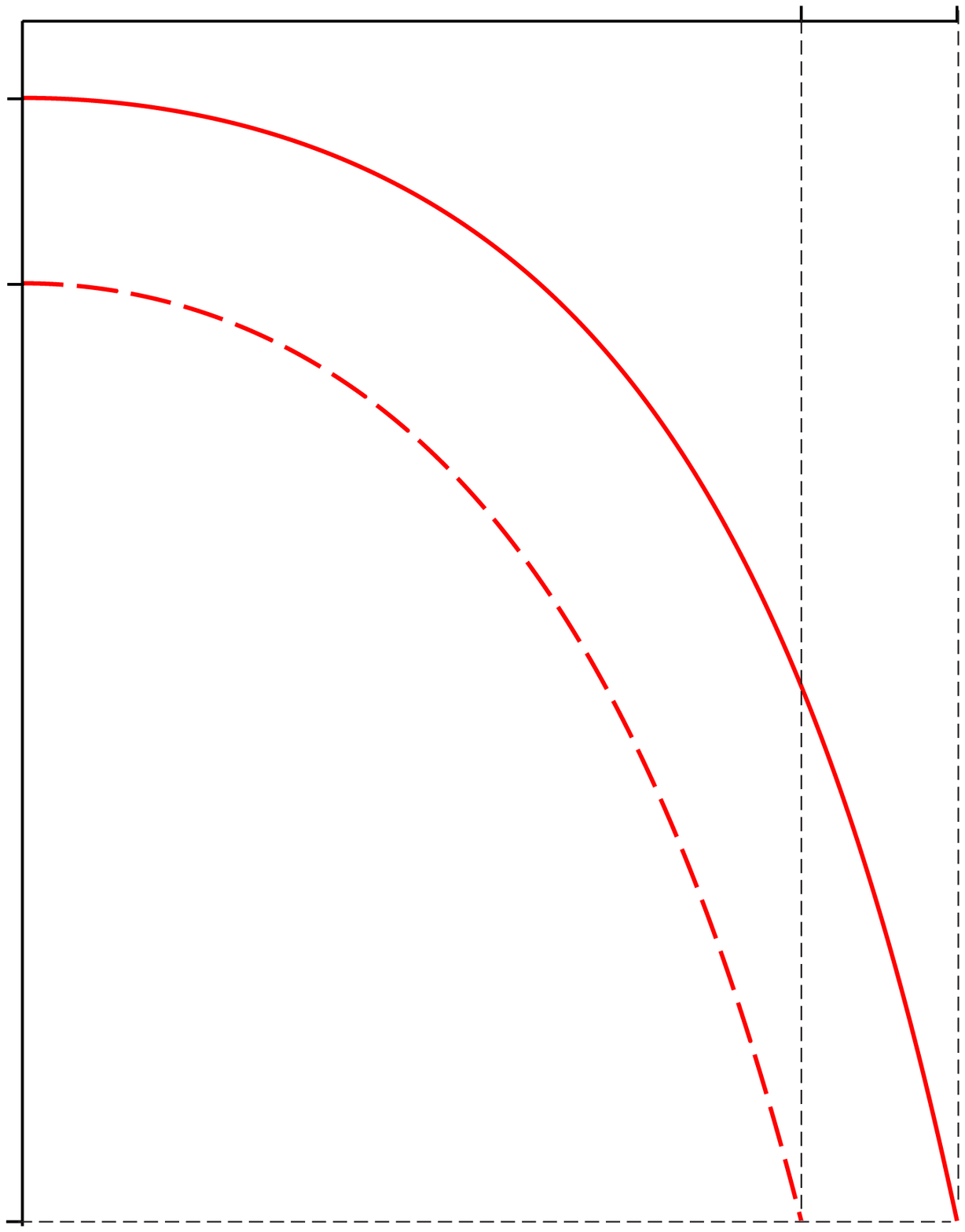}}
		\end{minipage}
		\hfill
	%	\hspace{3cm}
		\centering \SetLabels
		%\L (1.01*0.08) $\xi$ \\
		\scriptsize
		\L (0.54*-0.04) $(b)$ \\
		\L (0.94*0.885) $t$ \\
		\L (0.21*0) $\sigma^+_x$ \\
		\L (0.26*0.91) $0$ \\
		\L (0.4*0.91) $t=t_y(\xi_1)$ \\
		\endSetLabels
		%\ShowGrid\leavevmode
		\begin{minipage}[h]{1\linewidth} 
			\vspace{1mm}
			\AffixLabels{\includegraphics[scale=0.27]{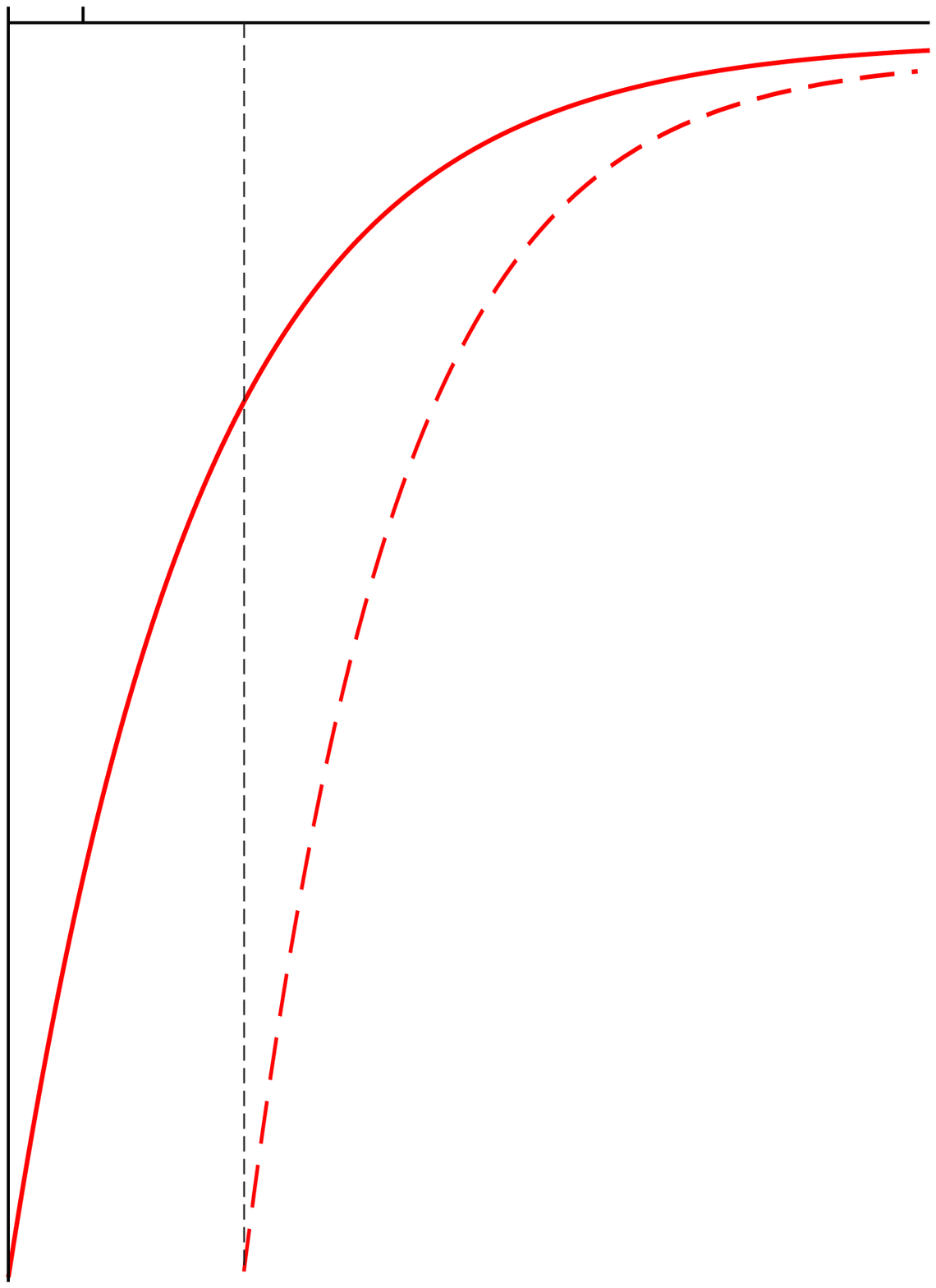}}
		\end{minipage}
		\vspace{3mm}
		\caption{Stress relaxation  behind the reaction front: $(a)$  stress distributions behind the front for two front positions at times $t=t_y(\xi_1)$ and  $t=t_y(\xi_2)$; $(b)$ stress relaxation in points $\xi=0$ and  $\xi=\xi_1$ starting  from the moments $t=t_y(0)=0$ and $t=t_y(\xi_1)$, respectively.
			\label{relax-scheme}}
	\end{center}
\end{figure}
\begin{figure}[tb!]
	\begin{center} \
		\centering \SetLabels
		\L (1.0*0.88) $\xi$ \\
		\scriptsize
		\L (0.485*-0.02) $(a)$ \\
		%\footnotesize
		\L (0.075*0) $\sigma^+_x[\text{GPa}]$ \\
		\L (0.27*0.92) $0$ \\
		\L (0.54*0.92) $0.001$ \\
		\L (0.87*0.92) $0.002$ \\
		\L (0.15*0.73) $-1.4$ \\
		\L (0.13*0.56) $-1.45$ \\
		\L (0.15*0.385) $-1.5$ \\
		\L (0.13*0.21) $-1.55$ \\		
		%\footnotesize
		\L (0.79*0.71) $ \eta_0$ \\
		\L (0.79*0.65) $5\eta_0$ \\
		\L (0.79*0.59) $10\eta_0$ \\
		\L (0.79*0.53) $50 \eta_0$ \\
		\endSetLabels
		%\ShowGrid\leavevmode
		\begin{minipage}[h]{1\linewidth} 
			%\vspace{1mm}
			\AffixLabels{\includegraphics[scale=0.28]{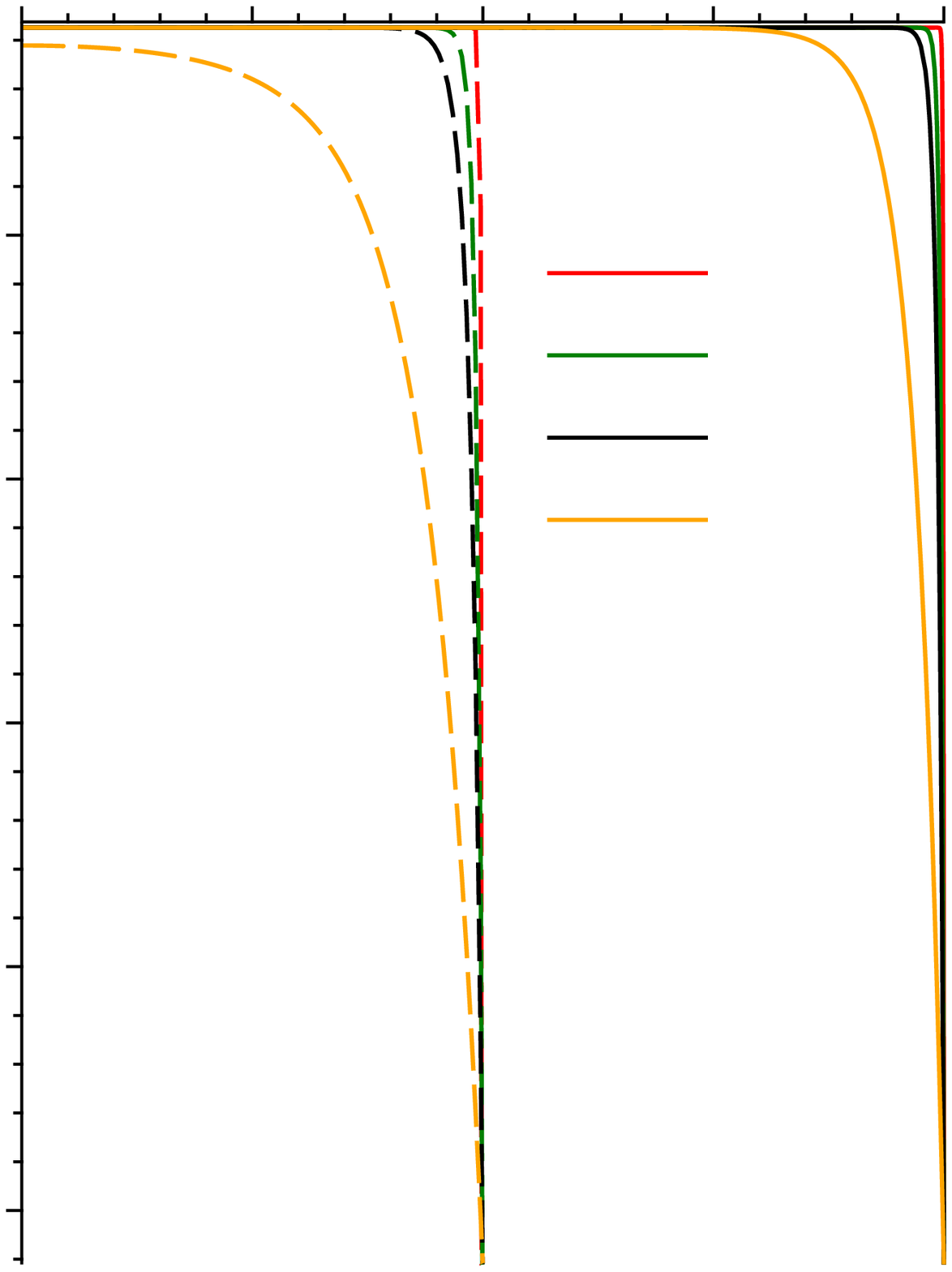}}
		\end{minipage}
		\hfill
		\centering \SetLabels
		%\L (1.01*0.08) $\xi$ \\
		\scriptsize
		\L (0.48*-0.02) $(b)$ \\
		\L (0.98*0.875) $t  [s]$ \\
		\L (0.07*0) $\sigma^+_x[\text{GPa}]$ \\
		\L (0.27*0.92) $0$ \\
		\L (0.43*0.92) $50$ \\
		\L (0.585*0.92) $100$ \\
		\L (0.76*0.92) $150$ \\
		\L (0.92*0.92) $200$ \\
		\L (0.15*0.735) $-1.4$ \\
		\L (0.13*0.56) $-1.45$ \\
		\L (0.15*0.39) $-1.5$ \\
		\L (0.13*0.215) $-1.55$ \\
		\endSetLabels
		%\ShowGrid\leavevmode
		\begin{minipage}[h]{1\linewidth} 
			%\vspace{1mm}
			\AffixLabels{\includegraphics[scale=0.28]{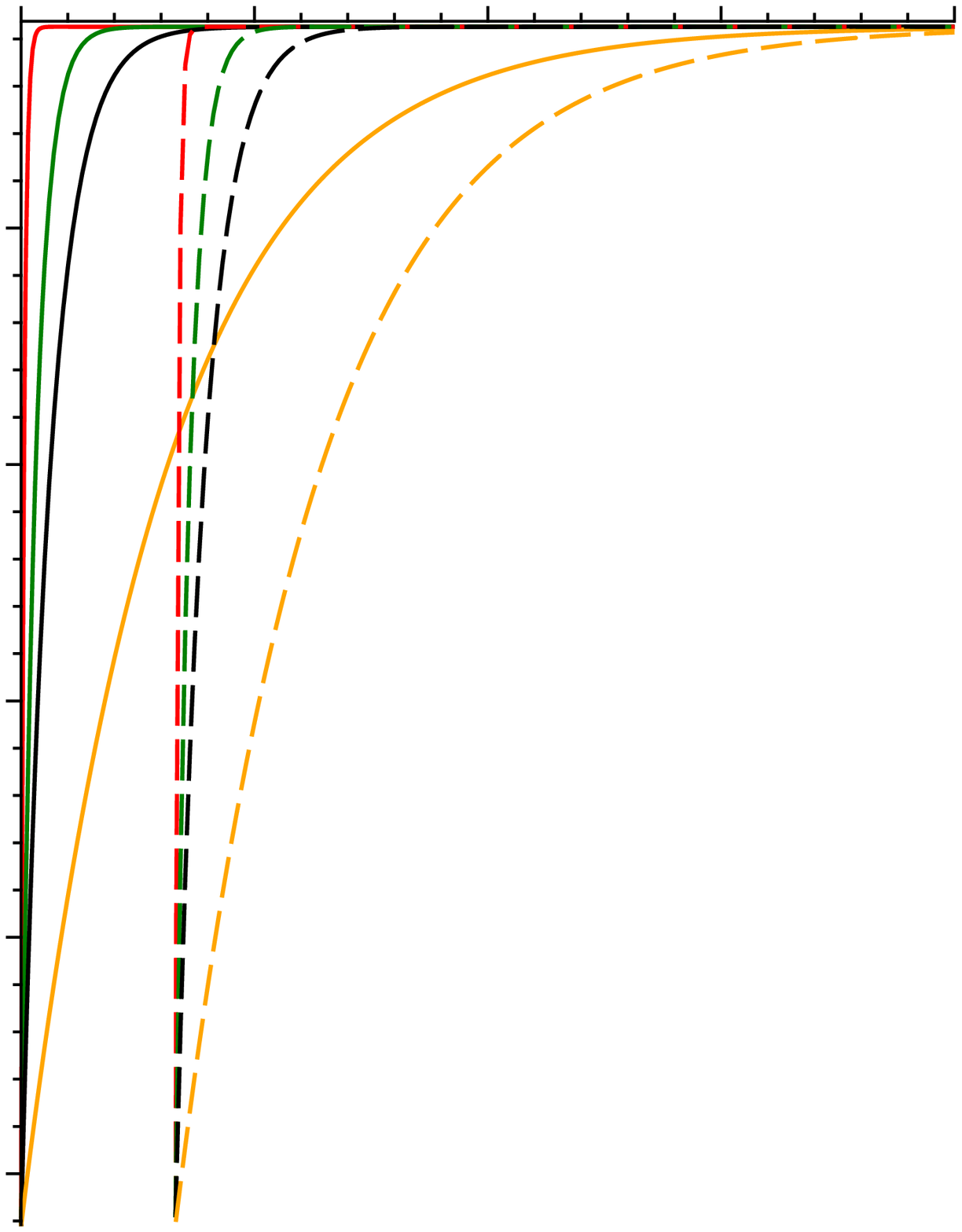}}
		\end{minipage}
		\vspace{3mm}
		\caption{Stress relaxation at various values of viscosity coefficient $\eta$ for the standard linear solid model: $(a)$  stress distributions behind the front for two front positions
			$\xi=0.001$  (dashed lines) and $\xi=0.002$  (solid lines), $(b)$ stress relaxation in  points $\xi=0$ (solid lines) and for $\xi=0.005$  (dashed lines) ; $\varepsilon_0=0$, $\eta_0=15.9 $GPa$\cdot$s. Solid and dashed lines of the same color correspond to the same viscosity coefficient. \label{stress-relax-eta}}
	\end{center}
\end{figure}

\begin{figure}[tb!]
	\begin{center} \
		\centering \SetLabels
		\L (1.01*0.35) $\xi$ \\
		\scriptsize
		\L (0.5*-0.04) $(a)$ \\
		%\footnotesize
		\L (0.035*0) $\sigma^+_x[GPa]$ \\
		\L (0.53*0.32) $0.001$ \\
		\L (0.88*0.32) $0.002$ \\
		\L (0.18*0.935) $4$ \\
		\L (0.18*0.79) $3$ \\
		\L (0.18*0.64) $2$ \\
		\L (0.18*0.5) $1$ \\
		\L (0.18*0.35) $0$ \\
		\L (0.14*0.21) $-1$ \\
		\L (0.14*0.06) $-2$ \\
		%	\footnotesize
		\L (0.785*0.705) $0.04$ \\
		\L (0.785*0.645) $0.019$ \\
		\L (0.785*0.59) $0.009$ \\
		\L (0.785*0.535) $\varepsilon=0$ \\
		\L (0.785*0.475) $-0.005$ \\
		\endSetLabels
		%\ShowGrid\leavevmode
		\begin{minipage}[h]{1\linewidth} 
			\vspace{1mm}
			\AffixLabels{\includegraphics[scale=0.28]{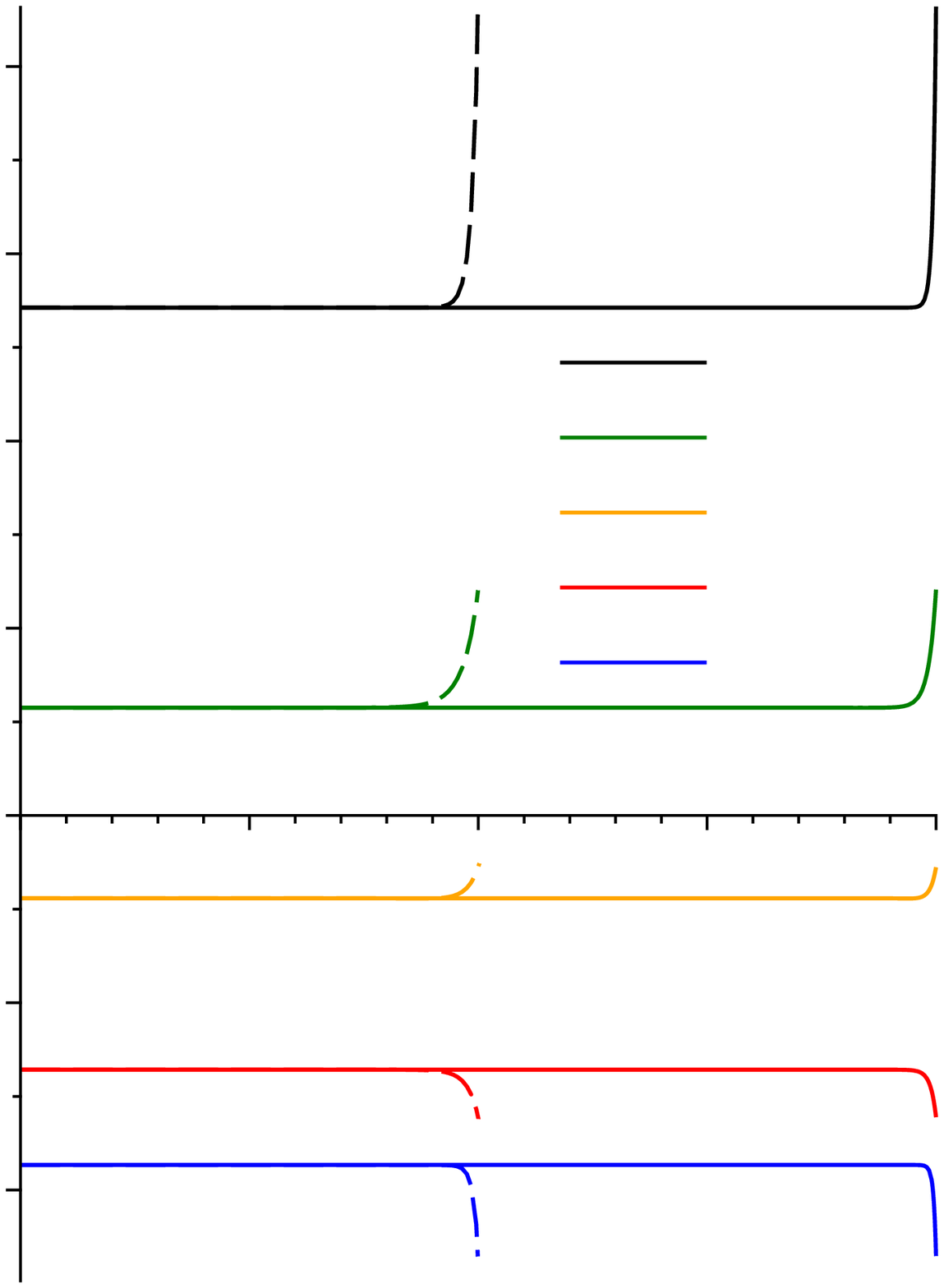}}
		\end{minipage}
		\hfill
		\centering \SetLabels
		%\L (1.01*0.08) $\xi$ \\
		\scriptsize
		\L (0.47*-0.04) $(b)$ \\
		\L (1.0*0.35) $t[s]$ \\
		\L (0.045*0) $\sigma^+_x[Gpa]$ \\
		\L (0.39*0.31) $50$ \\
		\L (0.57*0.31) $100$ \\
		\L (0.745*0.31) $150$ \\
		\L (0.93*0.31) $200$ \\
		\L (0.18*0.935) $4$ \\
		\L (0.18*0.79) $3$ \\
		\L (0.18*0.64) $2$ \\
		\L (0.18*0.5) $1$ \\
		\L (0.18*0.35) $0$ \\
		\L (0.14*0.2) $-1$ \\
		\L (0.14*0.06) $-2$ \\
		\endSetLabels
		%\ShowGrid\leavevmode
		\begin{minipage}[h]{1\linewidth} 
			\vspace{1mm}
			\AffixLabels{\includegraphics[scale=0.28]{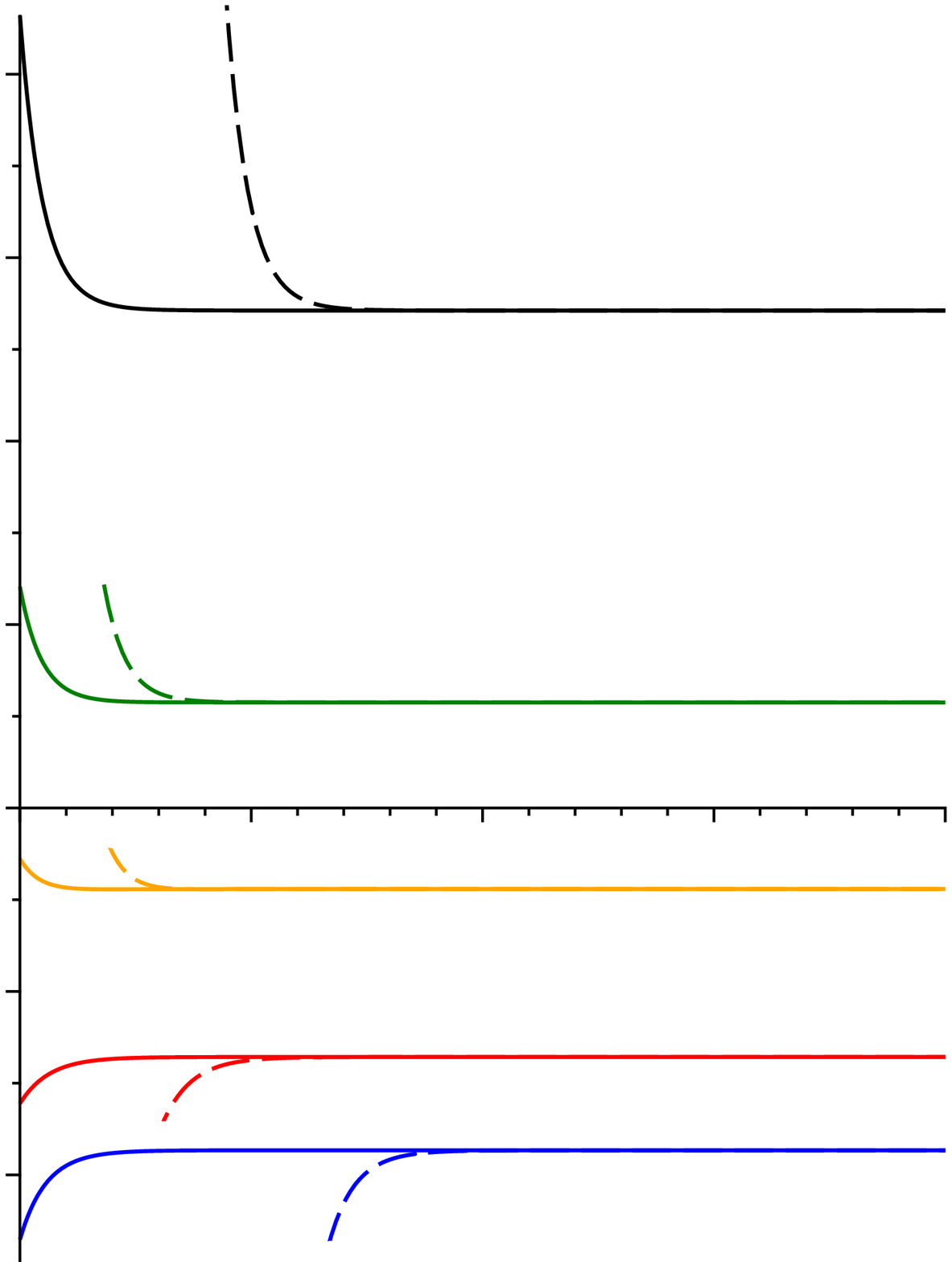}}
		\end{minipage}
	\vspace{4mm}
		\caption{Stress relaxation at various values of external strain $\varepsilon_0$: $(a)$ stresses behind the reaction front; $(b)$ stress relaxation in two points. Solid and dashed lines of the same color correspond to the same external strains.} \label{stress-relax-epsilon}
	\end{center}
\end{figure}

The difference of molar volumes of initial material and transformed material is a source of volume expansion due to chemical reaction. Kinematic  compatibility, i.e. displacement continuity at the reaction front, restricts the transformation strain and produces stresses which can be huge  in the case of an elastic behavior of the reaction product. Viscoelastic assumption allows the strain in the transformed layer to be partly accommodated by the viscous deformation, making it possible that the volume increases due to the increase of the thickness of the transformed layer leading to the stress relaxation (see, e.g., \cite{Courtney,Kobeda89,Kobeda88,EerNisse1977,EerNisse1979}).

The distributions of stress $\sigma_x^+$  behind the reaction front at two moments $t_y(y)$ which correspond to the dimensionless front  positions $\xi=\xi_1$ and  $\xi_2$ are schematically  shown in Fig.~\ref{relax-scheme}$a$. The stress $\sigma_x^+$ at $\xi =\xi_1$ relaxes from  $A$ to  $B$ during the time of the front propagation from  $\xi_1$ to  $\xi_2$. Stress relaxation in two points at $\xi=0$ and  $\xi=\xi_1$ starting  from the moments $t=t_y(0)=0$ and $t=t_y(\xi_1)$, respectively, is shown in Fig.~\ref{relax-scheme}$b$.

\begin{figure}[tb!]
	\begin{center} \
		\centering \SetLabels
		%\L (1.01*0.885) $\xi$ \\
		\scriptsize
		\L (0.5*-0.04) $(a)$ \\
		\L (0.12*0.995) $e^{\eta}_x$ \\
			\L (0.99*0.54) $t[s]$ \\
		%\footnotesize
		\L (0.06*0.85) $0.004$ \\
		\L (0.06*0.685) $0.002$ \\
		\L (0.14*0.53) $0$ \\
		\L (0.02*0.375) $-0.002$ \\
		\L (0.02*0.22) $-0.004$ \\
		\L (0.02*0.06) $-0.006$ \\
		%	\footnotesize
		\L (0.365*0.49) $50$ \\
		\L (0.55*0.49) $100$ \\
		\L (0.74*0.49) $150$ \\
		\L (0.935*0.49) $200$ \\
		\L (0.75*0.44) $\eta_0$ \\
		\L (0.75*0.375) $5 \eta_0$ \\
		\L (0.75*0.315) $10\eta_0$ \\
		\L (0.75*0.255) $50 \eta_0$ \\
		\endSetLabels
		%\ShowGrid\leavevmode
		\begin{minipage}[h]{1\linewidth} 
			\vspace{1mm}
			\AffixLabels{\includegraphics[scale=0.28]{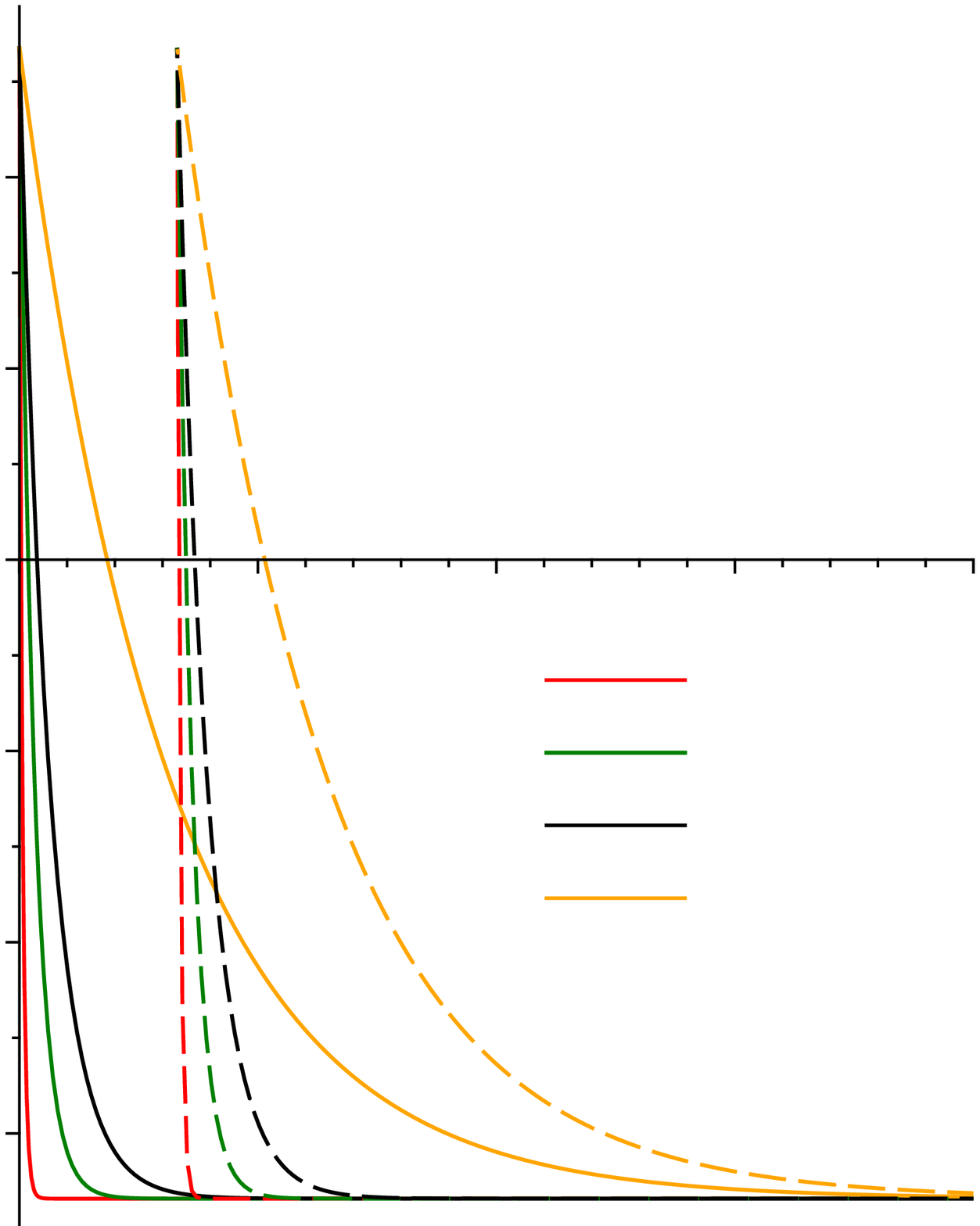}}
		\end{minipage}
		\hfill
		\centering \SetLabels
		%\L (1.01*0.08) $\xi$ \\
		\scriptsize
		\L (0.5*-0.04) $(b)$ \\
		\L (0.12*0.995) $e_x$ \\
			\L (0.99*0.54) $t[s]$ \\
		%\footnotesize
		\L (0.05*0.86) $0.004$ \\
		\L (0.05*0.695) $0.002$ \\
		\L (0.13*0.54) $0$ \\
		\L (0.01*0.38) $-0.002$ \\
		\L (0.01*0.215) $-0.004$ \\
		\L (0.01*0.06) $-0.006$ \\
		%	\footnotesize
		\L (0.325*0.5) $2$ \\
		\L (0.48*0.5) $4$ \\
		\L (0.64*0.5) $6$ \\
		\L (0.8*0.5) $8$ \\
		\L (0.935*0.5) $10$ \\
		\L (0.75*0.38) $e^{\eta}_x$ \\
		\L (0.75*0.32) $e^+_x$ \\
		\L (0.75*0.26) $e^e_{1x}$ \\
		\endSetLabels
		%\ShowGrid\leavevmode
		\begin{minipage}[h]{1\linewidth} 
			\vspace{1mm}
			\AffixLabels{\includegraphics[scale=0.28]{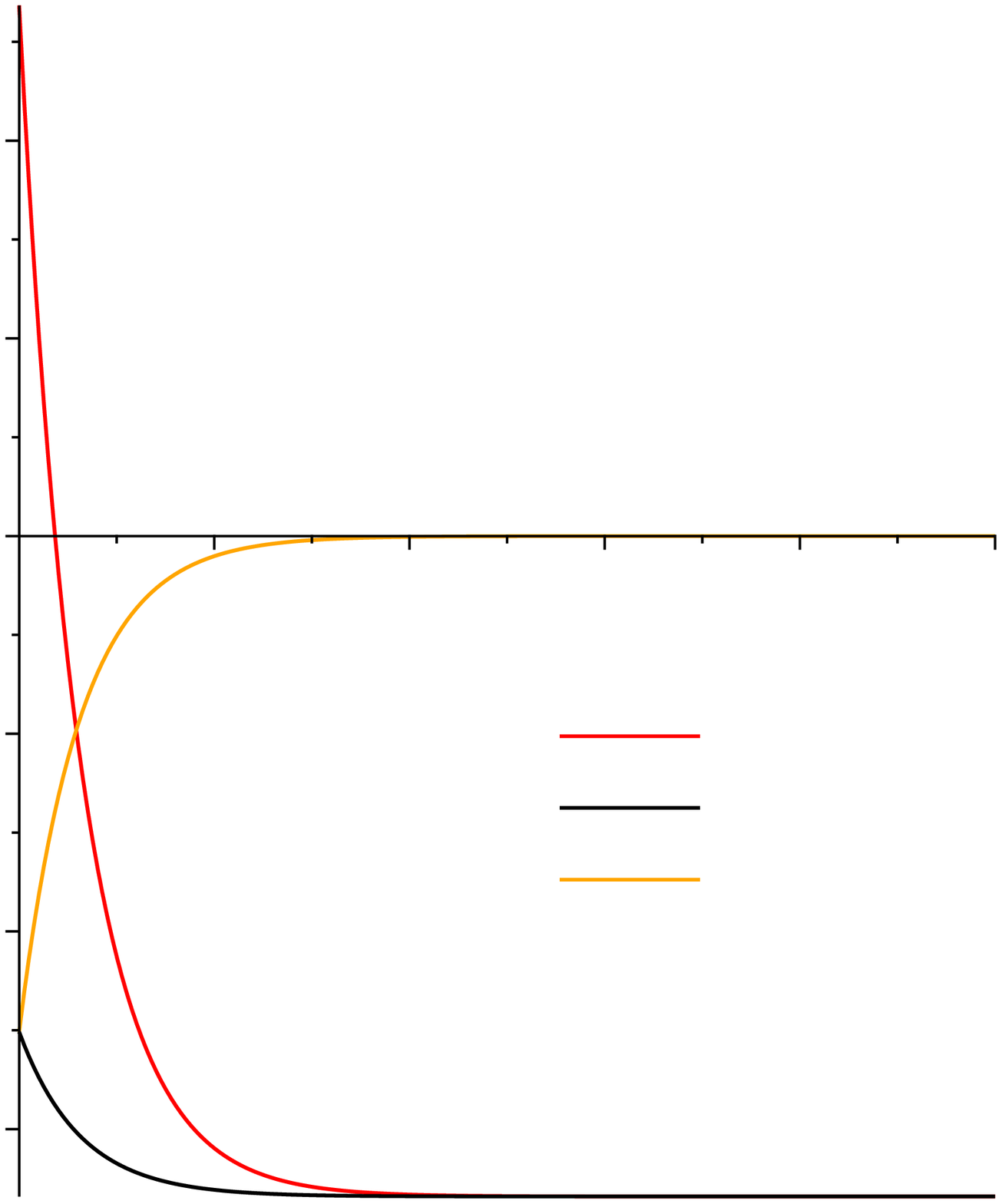}}
		\end{minipage}
	\end{center}
	\vfill
		\begin{center} \
		\centering \SetLabels
		%\L (1.01*0.885) $\xi$ \\
		\scriptsize
		\L (0.5*-0.04) $(c)$ \\
		\L (0.105*0.995) $e^{\eta}_y$ \\
		\L (1.0*0.1) $t[s]$ \\
		%\footnotesize
		\L (0.03*0.83) $0.012$ \\
		\L (0.03*0.71) $0.010$ \\
		\L (0.03*0.58) $0.008$ \\
		\L (0.03*0.455) $0.006$ \\
		\L (0.03*0.33) $0.004$ \\
		\L (0.03*0.21) $0.002$ \\
		\L (0.115*0.085) $0$ \\
		%	\footnotesize
		\L (0.15*0.045) $0$ \\
		\L (0.34*0.045) $50$ \\
		\L (0.53*0.045) $100$ \\
		\L (0.74*0.045) $150$ \\
		\L (0.935*0.045) $200$ \\
		\L (0.75*0.44) $\eta_0$ \\
		\L (0.75*0.375) $5 \eta_0$ \\
		\L (0.75*0.32) $10\eta_0$ \\
		\L (0.75*0.26) $50 \eta_0$ \\
		\endSetLabels
		%\ShowGrid\leavevmode
		%\ShowGrid\leavevmode
		\begin{minipage}[h]{1\linewidth} 
			\vspace{1mm}
			\AffixLabels{\includegraphics[scale=0.28]{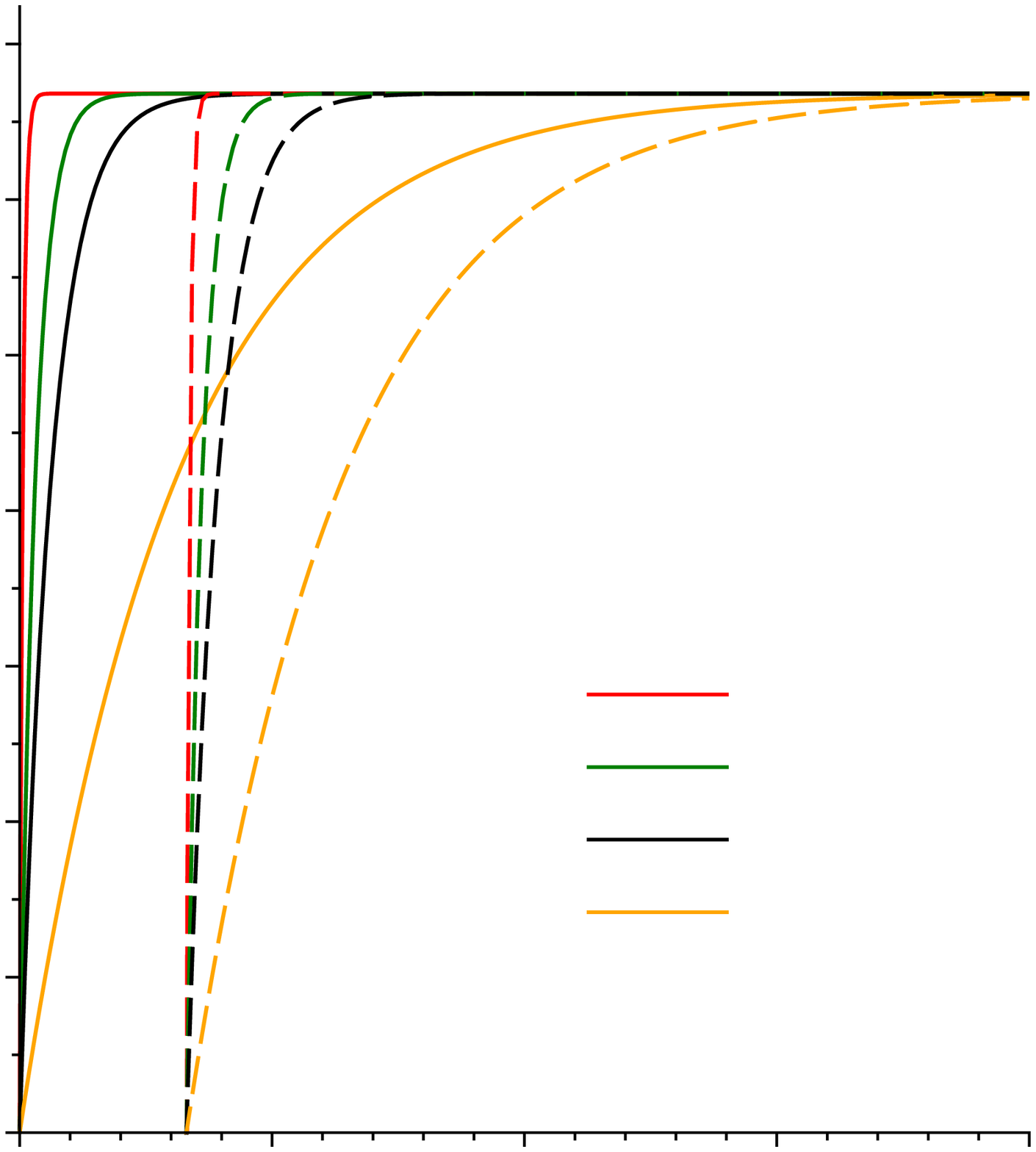}}
		\end{minipage}
		\hfill
		\centering \SetLabels
		%\L (1.01*0.08) $\xi$ \\
		\scriptsize
		\L (0.5*-0.04) $(d)$ \\
		\L (0.095*0.995) $e_y$ \\
		\L (1.0*0.1) $t[s]$ \\
		%\footnotesize
		\L (0.02*0.88) $0.012$ \\
		\L (0.02*0.75) $0.010$ \\
		\L (0.02*0.615) $0.008$ \\
		\L (0.02*0.49) $0.006$ \\
		\L (0.02*0.355) $0.004$ \\
		\L (0.02*0.23) $0.002$ \\
		\L (0.105*0.095) $0$ \\
		%	\footnotesize
		\L (0.14*0.055) $0$ \\
		\L (0.305*0.055) $2$ \\
		\L (0.465*0.055) $4$ \\
		\L (0.63*0.055) $6$ \\
		\L (0.8*0.055) $8$ \\
		\L (0.945*0.055) $10$ \\
		\L (0.75*0.375) $e^{\eta}_y$ \\
		\L (0.75*0.32) $e^+_y$ \\
		\L (0.75*0.26) $e^e_{1y}$ \\
		\endSetLabels
		%\ShowGrid\leavevmode
		\begin{minipage}[h]{1\linewidth} 
			\vspace{1mm}
			\AffixLabels{\includegraphics[scale=0.28]{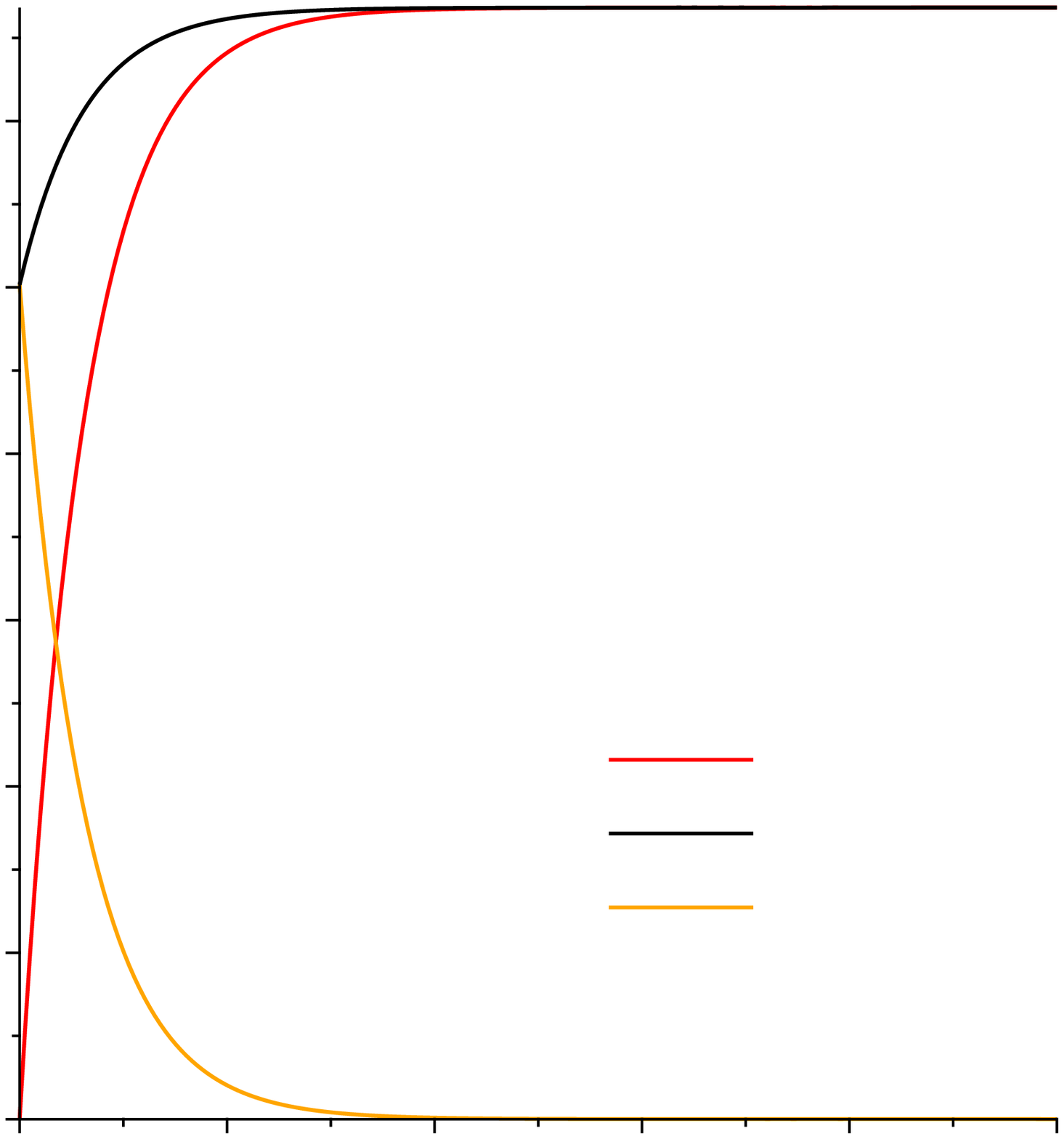}}
		\end{minipage}
		\vspace{2mm}
		\caption{Evolution and redistribution of viscous and elastic strains:
			(a) and (c) -- evolution of viscous strains $e_x^\eta$ and $e_y^\eta$  in  the point  $\xi=0$ from the moment $t=0$  and in the point $y$ that is reached by the reaction front at time  $t_y=200s$   for          
			various viscosity coefficients; (b) and (d) --  relation between the inputs of elastic strains $e^e_{1x}$, $e^e_{1y}$ and viscous strains $e^{\eta}_x$, $e^{\eta}_y$ into total strains $e^+_x$, $e^+_y$.  }\label{strain-evol-time}   
	\end{center}
\end{figure}

The results of quantitative studies of stress relaxation 
and  strains evolution and the redistribution of various modes of strains are shown in 
Fig.~\ref{stress-relax-eta},~\ref{stress-relax-epsilon} and Fig.~\ref{strain-evol-time}, respectively.  Material parameters are given in Table~\ref{Table1}.
Two sets of the stress distributions behind the reaction front at two moments $t_y$ which correspond to the front positions $\xi=0.001$ and  $\xi=0.002$, and two sets of stress relaxation curves for stresses in points $\xi=0$ and $\xi=0.005$ are shown  for various viscosity coefficients $\eta$ and  various external strains $\varepsilon_0$.

The transformation strain may produce huge stresses at the reaction front which would remain  in a pure elastic problem statement and might cause fracture. One can see how narrow the high stresses domain can be and how fast the  stresses may relax due to the viscous behavior of the reaction product at proper viscosities. This in turn demonstrates that, in dependence of the viscosity, the stress relaxation can or cannot prevent damage accumulation and fracture at the reaction front. Fig.~\ref{stress-relax-epsilon} shows how the curves are affected by external strain, in particular, how fast the  limit residual ``elastic'' stress is reached. 

\begin{figure}[tb!]
	\begin{center} \
		\centering \SetLabels
		\L (1.0*0.88) $\xi$ \\
		\scriptsize
		\L (0.485*-0.02) $(a)$ \\
		%\footnotesize
		\L (0.075*0) $\sigma^+_x[\text{GPa}]$ \\
		\L (0.265*0.93) $0$ \\
		\L (0.545*0.93) $0.001$ \\
		\L (0.89*0.93) $0.002$ \\
		\L (0.145*0.735) $-1.4$ \\
		\L (0.125*0.56) $-1.45$ \\
		\L (0.145*0.385) $-1.5$ \\
		\L (0.125*0.21) $-1.55$ \\		
		%\footnotesize
		\L (0.785*0.71) $1.1 \gamma_0$ \\
		\L (0.785*0.65) $2 \gamma_0$ \\
		\L (0.785*0.59) $5 \gamma_0$ \\
		\endSetLabels
		%\ShowGrid\leavevmode
		\begin{minipage}[h]{1\linewidth} 
			%\vspace{1mm}
			\AffixLabels{\includegraphics[scale=0.28]{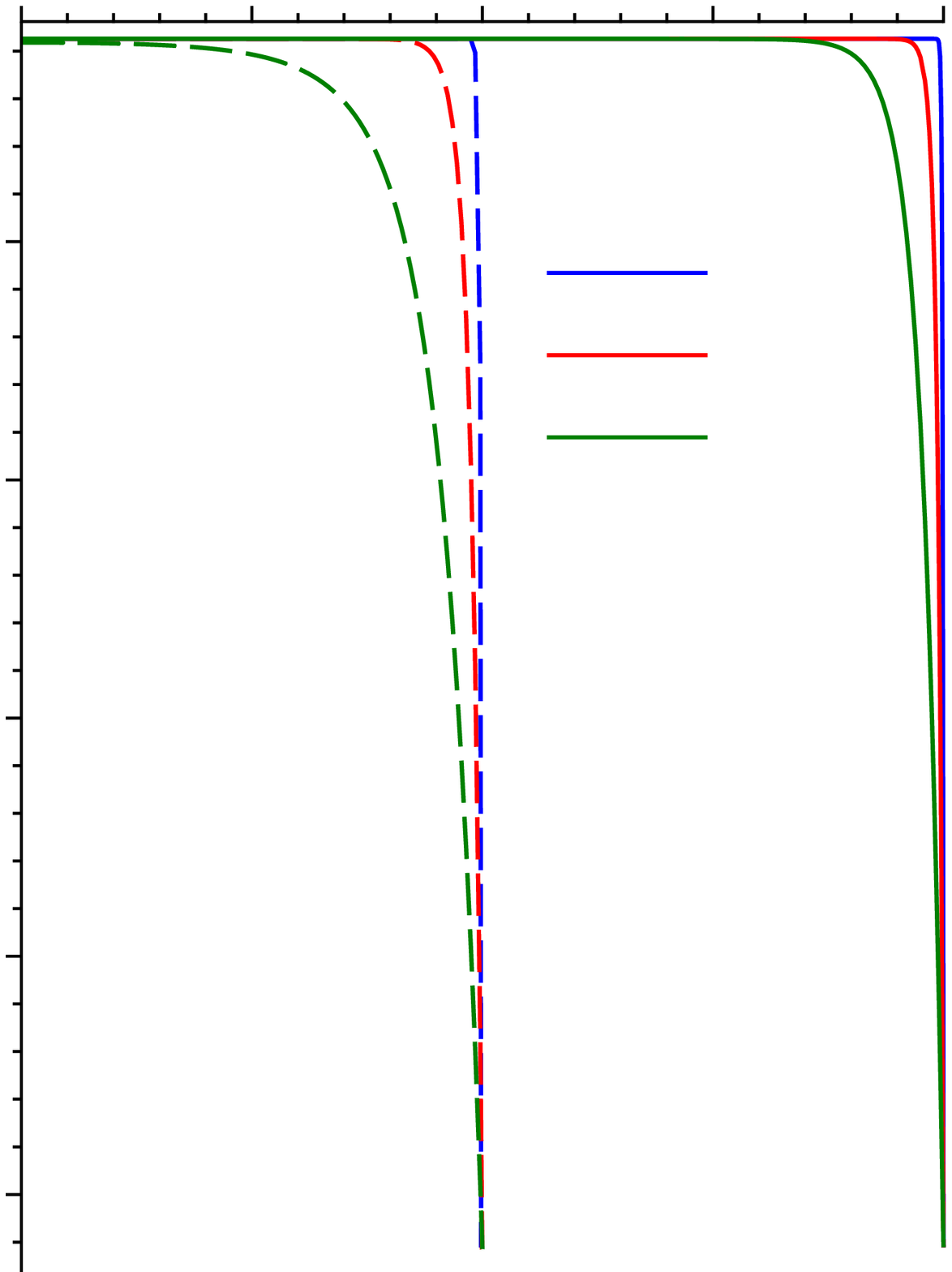}}
		\end{minipage}
		\hfill
		%\hspace{2.5cm}
		\centering \SetLabels
		%\L (1.01*0.08) $\xi$ \\
		\scriptsize
		\L (0.48*-0.02) $(b)$ \\
		\L (0.99*0.875) $t  [s]$ \\
		\L (0.07*0) $\sigma^+_x[\text{GPa}]$ \\
		\L (0.265*0.92) $0$ \\
		\L (0.38*0.92) $100$ \\
		\L (0.51*0.92) $200$ \\
		\L (0.65*0.92) $300$ \\
		\L (0.785*0.92) $400$ \\
		\L (0.92*0.92) $500$ \\
		\L (0.15*0.73) $-1.4$ \\
		\L (0.13*0.56) $-1.45$ \\
		\L (0.15*0.3857) $-1.5$ \\
		\L (0.13*0.22) $-1.55$ \\
		\endSetLabels
		%\ShowGrid\leavevmode
		\begin{minipage}[h]{1\linewidth} 
			%\vspace{1mm}
			\AffixLabels{\includegraphics[scale=0.28]{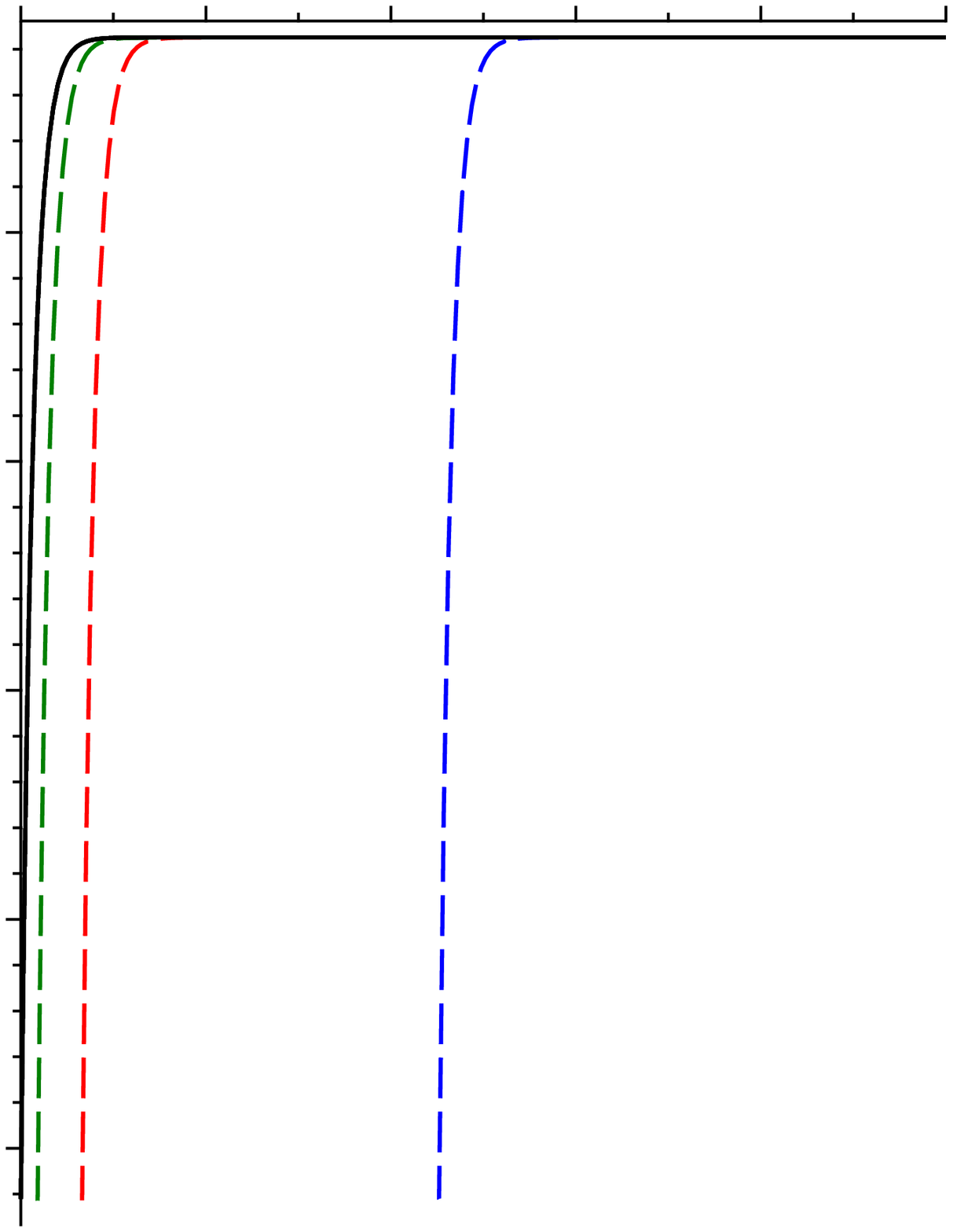}}
		\end{minipage}
		\vspace{2mm}
		\caption{Stress relaxation at various values of energy parameter $\gamma$ for the standard linear solid model: $(a)$  stresses behind the reaction front, $(b)$ stress relaxation in two points $\xi=0$ (solid lines) and for $\xi=0.005$  (dashed lines). Solid and dashed lines of the same color correspond to the same energy parameter. \label{stress-relax-gamma}}
	\end{center}
\end{figure}

Relaxation times $\tau_+$  and  $\tau_1$ do not depend on energy parameter $\gamma$. But $\gamma$ affects the front velocity and, thus, the front kinetics. Increasing $\gamma$ increases the front velocity and decreases the time $t_y$, in other words,  increasing $\gamma$ ``compresses'' the time in Eq.~\eqref{kinetics}  due to the increase of parameter $Q$. That is why stresses found for the same two front positions but at various  $\gamma$ have less time for the relaxation if  $\gamma$ increases.
This tendency is reflected by the stress distributions shown in Fig.~\ref{stress-relax-gamma}$b$.
Note that the energy parameter depends on temperature, and the temperature may also affect stress relaxation via the viscosity coefficient.

Note that in experiments the thickness of the layer of the transformed
material is usually observed which does not coincide with the front position predicted by the model. Fig.~ \ref{strain-evol-time}(b) demonstrates how residual strain $e_y^+=\lim\limits_{t\rightarrow \infty}e_y^\eta$ is formed at various viscosity coefficients. 

\subsection{Particular cases of viscoelastic behaviors}
In this subsection we specify the equations for stresses and strains behind propagating reaction front for three rheological  models which can be considered as particular cases of Standard Linear Solid Model and discuss the applicability of these models in the statement of mechanochemistry problems.  

\subsubsection{Maxwell material} We obtain  the Maxwell material, Fig.~\ref{models}\,$b$, from SLSM  setting $\mu_2=0$ and $\mu_1=\mu_+$.
The formula \eqref{chi-eps} for $\chi_0$ remains the same with new $\mu_+$.
The formulae \eqref{thgm},   \eqref{ex2gm-res},   \eqref{ex1gm-res},   \eqref{sigx-gm-res} \eqref{sigz-gm-res} 
for  strains $\vartheta^+$, $e_x^+$,  $e_{1x}^e$ and stresses  $\sigma_x^+$ and $\sigma_z^+$ behind the front become
\begin{gather} \nonumber
\vartheta^+(y,t)=\vartheta^{tr}+
\dfrac{2\mu_+ \left( 3\varepsilon_0-2\vartheta^{tr}\right)}
{  3k_++{4} \mu_+}
\exp \left( - \dfrac{t-t_y}{\tau_+} \right),
\\ \nonumber
e^+_{x}(y,t)= \varepsilon_0 -\frac{\vartheta^{tr}}{3} -\dfrac{ 2\mu_+ \left( 3\varepsilon_0 -2\vartheta^{tr} \right)}{3\left(3k_++ {4} \mu_+\right)} \exp \left( -\cfrac{t-t_y}{\tau_+}\right), \\ \nonumber
e^e_{x}(y,t)=\frac{k_+ ( 3\varepsilon_0 -2 \vartheta^{tr})}{2( 3k_+ + 4\mu_+) }
\exp\left( -\frac{t-t_y}{\tau_+}\right)+ \frac{\varepsilon_0}{2} \exp \left( -\frac{t-t_y}{\tau_1} \right),
\end{gather}
and
\begin{gather*}
\sigma^+_x  (y,t)= \dfrac{3k_+ \mu_+ \left(3\varepsilon_0 -2 \vartheta^{tr} \right)}{ 3k_++ {4} \mu_+} \exp\left( -\dfrac{t-t_y}{\tau_+}\right) +  \mu_+\varepsilon_0  \exp\left(-\dfrac{t-t_y}{\tau_1}\right),\\ \nonumber
\sigma^+_z (y,t)= \dfrac{3 k_+\mu_+   \left(3 \varepsilon_0 - 2 \vartheta^{tr} \right)}{3k_+ + 4\mu_+ } \exp\left( -\dfrac{t-t_y}{\tau_+}\right) - \mu_+ \varepsilon_0 \exp \left( -\dfrac{t-t_y}{\tau_1}\right),
\end{gather*}
where
\begin{equation}
\nonumber
\tau_1= \dfrac{\eta}{\mu_+},\quad\tau_+= \dfrac{(3k_+ +  {4}  \mu_+)}{3k_+  }\dfrac{\eta}{\mu_+}.
\end{equation}

\subsubsection{Kelvin-Voigt material} Another particular case is the Kelvin-Voigt material  (Fig.~\ref{models}$c$).
In this case
\begin{gather}\label{K-V-mu} 	
\mu_1\rightarrow\infty,\   \mu_2\rightarrow\mu_+,	
\\ 
\label{K-V-s-e}
\mathbf{e}^+=\mathbf{e}^\eta=\mathbf{e}^e, \quad
\bm{\sigma}^+ =k_+ \left(\vartheta^+  -\vartheta^{tr}\right)\mathbf{I}
+2\mu_+\mathbf{e}^+ + 2\eta\dot{\mathbf{e}}^+.
\end{gather}
Since the dash-pot 
element cannot deform simultaneously, at the reaction front 
$\mathbf{e}^\eta=\mathbf{e}^+=0$. 
Then this degenerative  case can be realized only if $\varepsilon_0=0$. 
Then $w_-=0$ and at the reaction front
\begin{equation}\label{K-V-initial}
\vartheta =\vartheta(y,t_y)=0. 
\end{equation}
Then 
$ 
\chi=w_+=\dfrac12k_+(\vartheta^+)^2
$. Mechanics just subtracts  
$\dfrac12k_+(\vartheta^+)^2$ from $\gamma$ in the expression of$A_{NN}$. Of course, this also directly follows from \eqref{chi-eps} and \eqref{GS} if one takes \eqref{K-V-mu}.

The equation \eqref{eq-theta} for $\vartheta$ behind the front takes the form
$$
\dot{\vartheta}^++\dfrac{\vartheta^+}{\tau}-\dfrac{3k_+}{4\eta}\vartheta^{tr}=0,
\quad  \tau=\frac{4\eta}{3k_++4\mu_+}.
$$
The solution, satisfying the initial condition \eqref{K-V-initial}, is
\begin{equation}
\label{K-V-vartheta}
\vartheta^+(y,t)=\frac{3k_+\vartheta^{tr}}{3k_++4\mu_+}\left(
1-\exp\left(-\frac{t-t_y}{\tau}\right)
\right).
\end{equation}
The volume strain in points behind the front increases with time if $\vartheta^{tr}>0$. Since $e_x^+=e_z^+=-\vartheta/3$, from  \eqref{K-V-s-e} and \eqref{K-V-vartheta} it follows that the stresses behind the front can be expressed via $\vartheta$ and relax as
\begin{gather}\nonumber
\sigma_x^+(y,t)=\sigma_z^+(y,t)=
-\dfrac23\eta\dot{\vartheta}^+ +\left(k^+-\dfrac23\mu_+\right)\vartheta^+-k_+\vartheta^{tr}\qquad\qquad\qquad\qquad\\
\qquad\qquad\qquad\qquad
=-\dfrac{6 k_+\mu_+  \vartheta^{tr}}{3k_+ + 4\mu_+ }\left(1+\dfrac{3k_+}{4\mu_+} \exp\left( -\dfrac{t-t_y}{\tau}\right) \right).\label{K-V-sig}
\end{gather}
At the reaction front
\begin{gather}\label{K-V-sig-front}
\sigma_x^+=\sigma_z^+=-\dfrac{3}{2}k_+\vartheta^{tr}.
\end{gather}
Of course, Eq.~\eqref{K-V-vartheta},\eqref{K-V-sig-front} and \eqref{K-V-sig-front} directly follow from \eqref{thgm}, \eqref{sigx-gm-res} and \eqref{sigx-front} if to take  $\mu_1$ and $\mu_2$ from \eqref{K-V-mu}, but, since this case may be of a special interest, we presented the short derivations \eqref{K-V-s-e}--\eqref{K-V-sig}. 

\subsubsection{Pure linear-viscous material}
The linear-viscous material  (Fig.~\ref{models}$d$) can be obtained by setting   $\mu_+=0$ in above formulae. It can be considered only at the same restriction $\varepsilon_0=0$ as above. 
Then at the reaction front $w_-=0$, $\chi=w_+=\dfrac12k_+(\vartheta^+)^2$,
\begin{gather}\nonumber
\vartheta^+=0, \quad \sigma_x^+=\sigma_z^+=-\dfrac{3}{2}k_+\vartheta^{tr}
\end{gather}
It is easy to see that the volume strain in points behind the front increases up to $\vartheta^{tr}>0$ (decreases if $\vartheta^{tr}<0$) with time as
\begin{equation}
\nonumber
\vartheta^+(y,t)=\vartheta^{tr}\left(
1-\exp\left(-\frac{t-t_y}{\tau}\right)
\right), \quad \tau=\dfrac{4\eta}{3k_+}
\end{equation}
and stresses relax as
\begin{gather}\nonumber
\sigma_x^+(y,t)=\sigma_z^+(y,t)=
-\dfrac{3}{2}k_+\vartheta^{tr} \exp\left( -\dfrac{t-t_y}{\tau}\right) \end{gather}

Note that the restriction $\varepsilon_0=0$  makes the Kelvin-Voigt and pure viscous materials  rather unsuitable than suitable as rheological models,  as opposed the Standard Linear Solid Model and the Maxwell material. This indicates that not every common rheological model can be used for reaction constituents  in the  considerations  of coupled problems of mechanochemistry. 

\section{Conclusions}

The stress-affected chemical reaction front propagation in deformable solid in the case of a planar reaction front has been considered basing on the concept of the chemical affinity tensor. The influence of strains and material parameters on the kinetics of the front propagation was studied in detail with the use of the notion of the equilibrium concentration. Two types of the dependencies of the equilibrium concentration and, thus, front velocity on strain are demonstrated, depending on the relations between the combinations of elastic moduli of solid reactants. In the first case the front can propagate only if strains belong to some interval, and it cannot propagate at all if the energy parameter is less than the critical value defined by the elastic moduli and transformation strain. In the second case the front can propagate at any energy parameter at proper strains which are outside of a corresponding interval. Different cases also correspond to different effects of strains on the front acceleration or retardation.
 
The changing of the rheology of a solid constituent due to the localized chemical reaction was taken into account with the use of the Standard Linear Solid Model and its particular cases. The SLSM and Maxwell model allowed to obtain  analytical solutions which gave us possibilities to study the specific effects of material parameters on stress relaxation behind the reaction front. On the other hand, the Kelvin-Voigt  and pure viscous materials  can hardly be considered as proper candidates for modeling the reaction products.
Results show that viscous deformations of the reaction product do not affect directly the kinetics of the front in the case of the SLSM if the  external strain acts in the plane of the interface, since they do not have time to appear at the moment of the transformation. But they enable the possibility for a stress relaxation phenomenon behind the reaction front. Depending on the viscous and elastic parameters, this relaxation can be  fast, and the high stresses region is localized in a narrow layer adjacent to the transformation front. Note also that other external loadings are possible, at which stress relaxation can restart the initially blocked reaction front.
Following these results, different perspectives could be drawn 
for coupled mechanochemistry simulations based on the chemical affinity tensor in order to be applied for more complex external loading, various geometries and towards plasticity and viscoplasticity.

% Authors must disclose all relationships or interests that 
% could have direct or potential influence or impart bias on 
% the work: 

% \section*{Conflict of interest}
%
 %The authors declare that they have no conflict of interest.

% BibTeX users please use one of
\bibliographystyle{abbrv}      % basic style, author-year citations
\bibliography{mybibfile}   % name your BibTeX data base

\end{document}